\documentclass[11pt]{article}
\usepackage{graphicx}
\usepackage{epsfig}

\newcommand{\be}{\begin{equation}}
\newcommand{\ee}{\end{equation}}
\newcommand{\bea}{\begin{eqnarray}}
\newcommand{\eea}{\end{eqnarray}}

\usepackage{amsmath}
\usepackage{amsfonts}
\usepackage{amssymb}
\usepackage{amsxtra}
\newcommand\sminus{\boxminus}
\usepackage{supertabular}
\title{ \vspace{1cm} 
How far has so far the Spin-Charge-Family theory 
succeeded to offer the explanation for the observed phenomena in
elementary particle physics and cosmology}

\author{N.S.\ Manko\v c Bor\v stnik$^{1}$ 
\\
$^1$Department of Physics, University of Ljubljana\\
SI-1000 Ljubljana, Slovenia,\\
norma.mankoc@fmf.uni-lj.si } 
\begin{document}
\maketitle

\begin{abstract}
Abstract:This talk discusses the achievements of the {\it spin-charge-family} theory. 
The project started in the year 1993 when trying to understand the internal spaces 
of fermions and bosons with the Grassmann algebra~\cite{norma93}.
Recognizing that the Grassmann algebra suggests the existence of the
anticommuting fermion states  with the integer spin (and commuting boson states
with integer spin) and that Grassmann algebra is expressible with two kinds of the 
Clifford algebras~\cite{norma93,n2019PRD,pikanorma,pikanorma2005}, both 
offering a description of the anticommuting half spin fermion 
states~(\cite{nh2021RPPNP} and references therein), it became obvious that the 
Clifford odd algebra (the superposition of odd products of $\gamma^a$) offers a 
way for describing the internal spaces for fermions, while $\tilde{\gamma}^a$ 
can be chosen to determine quantum numbers of families of
fermions~(\cite{nh2021RPPNP} and references therein). Three years ago, it
become evident that the Clifford even algebra (the superposition of even products
of $\gamma^a$) offers a way to describe the internal spaces of the
corresponding boson gauge fields~(\cite{n2023NPB,n2023MDPI} and references
therein). In odd dimensional spaces the  properties 
of the internal space of fermions and bosons differ essentially from those in even 
dimensional spaces, manifesting as the Fadeev-Popov ``ghosts''.\\ 
Can the {\it spin-charge-family} theory, extending the point fields to
strings, be related to {\it string theories}? 

\vspace{3mm}

Povzetek: Prispevek predstavi dose\v zke teorije {\it spin-charge-family}. Projekt 
te\v ce od leta 1993 s poskusom opisati notranje prostore fermionov in bozonov z 
Grassmannovo algebro~\cite{norma93}. Ob spoznanju, da Grassmannova algebra 
ponudi opis antikomutirajo\v cih fermionskih stanj s celo\v stevil\v cnim spinom (in 
komutirajo\v cih bozonskih stanj s celo\v stevil\v cnim spinom) in da je mogo\v ce 
izraziti Grassmannovo algebro z dvema vrstama Cliffordovih algeber%
~\cite{norma93,n2019PRD,pikanorma,pikanorma2005}, ki obe ponujata opis
antikomutirajo\v cih fermionskih stanj s pol\v stevilskim 
spinom~(\cite{nh2021RPPNP} in v vsebovanih referencah), je postalo o\v citno, 
da Cliffordova liha algebra (superpozicija lihih produktov $\gamma^a$) ponuja 
opis notranjih prostorov za fermione, $\tilde{\gamma}^a$ pa kvantna \v stevila
za dru\v zine fermionov~\cite{nh2021RPPNP}. Pred tremi leti je postalo o\v citno, 
da ponuja Cliffordova soda algebra (superpozicija sodih produktov $\gamma^a$) 
opis notranjih prostorov ustreznih bozonskih umeritvenih 
polj~(\cite{n2023NPB,n2023MDPI} in v vsebovanih referencah). V liho razse\v znih 
prostorih demostrirajo stanja fermionskih in bozonskih polj lastnosti Fadejevih in 
Popovih ``duhov''. Ali lahko teorijo 
{\it spin-charge-family}, ki opisuje doslej to\v ckasta polja, \v ce jih raz\v sirimo v
strune, pove\v zemo s {\it teorijami strun}?
\end{abstract}
\noindent Keywords: Second quantization of fermion and boson fields with Clifford 
algebra; Beyond the standard model; Kaluza-Klein-like theories in higher dimensional 
spaces; Clifford algebra in odd dimensional spaces; Ghosts in quantum field theories
\section {Introduction}
\label{introduction}
\vspace{2mm}

The {\it standard model} (corrected with the right-handed neutrinos) has
been experimentally confirmed without raising any severe doubts so far on its
assumptions, which, however, remain unexplained.

The {\it standard model} assumptions have several explanations in the literature, 
mostly with several new, not explained assumptions. The most popular are
the grand unifying theories~(\cite{Geor,FritzMin,PatiSal,GeorGlas,Cho} 
and many others). 

\vspace{2mm}

In a long series of works~(\cite{norma93,pikanorma2005,
n2014matterantimatter,JMP2013}, and the references there in) 
the author has found, together with the collaborators~(\cite{nh2021RPPNP,nh02,gmdn2008,gn2009,
nd2017,nh2018,
2020PartIPartII,nh2021RPPNP} and the references therein), 
the phenomenological success with the model named the
{\it spin-charge-family} theory with the properties: \\

{\bf a.}
In $d\ge (13+1)$ the creation operators manifest in $d=(3+1)$ the properties of 
all the observed quarks and leptons, with the families included, and of their gauge  
boson fields, with the scalar fields included, making several predictions. \\
{\bf a.i.} The internal space of fermions are in each even dimension ($d=2(2n+1), 
d=4n$) described by the ``basis vectors'', $\hat{b}^{m \dagger}_{f}$, Subsects.~(\ref{grassmannandclifford}, \ref{basisvectors}), $\hat{b}^{m \dagger}_{f}$, which are 
superposition of odd products of anti-commuting objects (operators) $\gamma^a$'s, 
appearing in $2^{\frac{d}{2}-1}$ families, each family with $2^{\frac{d}{2}-1}$ 
members. 
%
Correspondingly the ``basis vectors'' of one Lorentz irreducible representation in 
internal space of fermions, together with their Hermitian conjugated partners, 
anti-commute, fulfilling (on the vacuum state) all the requirements for the second 
quantized fermion fields~(\cite{prd2018,2020PartIPartII,
nh02,
nh2021RPPNP} and references therein). \\
{\bf a.i.{\it i.}} The second kind of anti-commuting objects, $\tilde{\gamma}^a$, 
Subsect.~\ref{grassmannandclifford}, equip each irreducible representation of odd 
``basis vectors'' with the family quantum number~%
\cite{2020PartIPartII,nh02}. Correspondingly each of $2^{\frac{d}{2}-1}$ families 
carries  $2^{\frac{d}{2}-1}$ members. \\
{\bf a.i.{\it ii.}} Creation operators for single fermion states, inheriting 
anti-commutativity of ``basis vectors'', are tensor products, $*_{T}$, of a finite 
number of odd ``basis vectors'', and the (continuously) infinite 
momentum/coordinate basis. Applying on the vacuum state~\cite{2020PartIPartII,
nh2021RPPNP}, the creation operators and their Hermitian conjugated partners 
fulfil the anticommutation relations for the second quantized fermion states.\\
{\bf a.i.{\it iii.}} 
The Hilbert space of second quantized fermion field is represented by the tensor
products, $*_{T_{H}}$, of all possible numbers of creation operators, from zero to
infinity~\cite{nh2021RPPNP}, applying on a vacuum state.\\
{\bf a.i.{\it iv}.} Spins from higher dimensions, $d>(3+1)$, described by the 
eigenvectors of the superposition of the Cartan subalgebra of $S^{ab}$, manifest 
in $d=(3+1)$ all the charges of the {\it standard model} quarks and leptons and 
antiquarks and antileptons,  as the reader can see in Table~\ref{Table so13+1.} 
for one, anyone, of families.\\ 
Let be pointed out that in even $d=(13+1)$ one irreducible 
representation of the Clifford odd ``basis vectors'' contains fermions and 
antifermions, quarks and leptons and antiquarks and antileptons.\\
{\bf a.ii.}  The internal space of bosons are in each even dimension ($d=2(2n+1), 
d=4n$) described by the ``basis vectors'' which 
are superposition of even products of anti-commuting objects (operators) 
$\gamma^a$'s: They appear in two (orthogonal) groups, each with 
$2^{\frac{d}{2}-1} \times 2^{\frac{d}{2}-1}$ members, called 
${}^{I}{\cal A}^{m \dagger}_{f}$ and ${}^{II}{\cal A}^{m \dagger}_{f}$, 
respectively~\footnote{
The Clifford odd ``basis vectors'', appearing in $2^{\frac{d}{2}-1}$ families, each 
family having $2^{\frac{d}{2}-1}$ members, contain together with their Hermitian 
conjugated partners twice $2^{\frac{d}{2}-1} \times 2^{\frac{d}{2}-1}$ objects, 
the same as the two groups of the Clifford even ``basis vectors''.}.\\
{\bf a.ii.{\it i.}} The Clifford even ``basis vectors'' have the properties of the gauge 
fields of the Clifford odd ``basis vectors''; They do not appear in families, and have 
their Hermitian conjugated partners within the same group.\\
{\bf a.ii.{\it ii.}} One group of the Clifford even ``basis vectors'' transforms, when 
applying algebraically on the Clifford odd ``basis vector'', this Clifford odd ``basis 
vector'' into other members of the same family. The other group of the Clifford even 
``basis vectors'' transform, when applying algebraically on the Clifford odd ``basis 
vector'', this Clifford odd ``basis vector'' into the same member of another family. 
These two groups have properties of the spin connection fields, 
$\omega_{ab \alpha}$ and $\tilde{\omega}_{ab \alpha}$, presented in the action, 
Eq.~(\ref{wholeaction}), respectively.\\ 
{\bf a.ii.{\it iii.}} Creation operators for boson states, inheriting commutativity 
of ``basis vectors'', are tensor products, $*_{T}$, of a finite number of even 
``basis vectors'', and the (continuously) infinite momentum/coordinate basis.\\
{\bf a.ii.{\it iv}.} Spins from higher dimensions, $d>(3+1)$, described by the 
eigenvectors of the superposition of the Cartan subalgebra of ${\cal S}^{ab}=
S^{ab} + \tilde{S}^{ab}$~\footnote{The definition of $S^{ab}=\frac{i}{4} 
(\gamma^a \gamma^b - \gamma^b \gamma^a)$, and  $\tilde{S}^{ab}=
\frac{i}{4} (\tilde{\gamma}^a \tilde{\gamma}^b - \tilde{\gamma}^b
\tilde{\gamma}^a) $.}, 
manifest as charges (in adjoint representations) of 
the gauge fields --- the vector ones with the space index $\alpha=(0,1,2,3)$,
and the scalar ones with the space index $\alpha \ge 5)$. \\
{\bf a.ii.{\it v}.} Besides the internal space, the Clifford even ``basis vectors''
must carry also the space index $\alpha$, to  be able to represent 
the vector and scalar gauge fields; all the scalar fields carry in the 
{\it spin-charge-family} theory the space index, assumed as
${}^{i}{\cal A}^{m \dagger}_{f} {}^{i} {\cal C}^{m}_{f \alpha}, i=(I,II)$%
\footnote{The reader can find the demonstration of the properties of the 
fermion and boson gauge fields for $d=(5+1)$ case in Subsect. 2.3 of 
Ref.~\cite{n2023NPB}}. Both groups,  
${}^{I}{\cal A}^{m \dagger}_{f} {}^{I} {\cal C}^{m}_{f \alpha}$ and
${}^{II}{\cal A}^{m \dagger}_{f} {}^{II} {\cal C}^{m}_{f \alpha}$ can 
be related to $\omega_{ab \alpha}$ and $\tilde{\omega}_{ab \alpha}$, 
presented in Eq.~(\ref{wholeaction}), respectively. These relations can be 
found in Eq.~(37) of Ref.~\cite{n2023NPB}.
\\
{\bf a.iii.}  In odd dimensional spaces, $d=(2n +1)$, the  properties of the
internal space of fermions and bosons differ essentially from those in even 
dimensional spaces; one half of the ``basis vectors'' have the properties as 
those from $d=2n$, the second half, following from those of $d=2n$
by the application of $S^{0 2n+1}$, behave as the Fadeev-Popov ghost --- 
anticommuting appear in two orthogonal groups with the Hermitian 
conjugated partners within the same group, commuting appear in families
and have their Hermitian conjugated partners in a separate 
group~\cite{n2023MDPI}.\\
{\bf a.iii.{\it i}.} The theory offers a new understanding of the second 
quantized fermion fields, as mentioned in  {\bf a.} and it is explained in 
Refs.~\cite{2020PartIPartII,nh2021RPPNP}, it also enables a new 
understanding of the second quantization  of boson fields as presented in 
Refs.~\cite{n2021SQ,n2022epjc,n2023NPB,n2023MDPI}. \\
The properties of the Clifford odd and even ``basis
vectors'' wait to be studied.\\

{\bf b.} 
In a simple starting action, Eq.~(\ref{wholeaction}), 
in  $d=2(2n+1)$-dimensional space 
\begin{eqnarray}
{\cal A}\,  &=& \int \; d^dx \; E\;\frac{1}{2}\, (\bar{\psi} \, \gamma^a p_{0a} \psi) 
+ h.c. +
\nonumber\\  
               & & \int \; d^dx \; E\; (\alpha \,R + \tilde{\alpha} \, \tilde{R})\,,
\nonumber\\
           p_{0\alpha} &=&  p_{\alpha}  - \frac{1}{2}  S^{ab} \omega_{ab \alpha} - 
                    \frac{1}{2}  \tilde{S}^{ab}   \tilde{\omega}_{ab \alpha} \,,
                    \nonumber\\  
           p_{0a } &=& f^{\alpha}{}_a p_{0\alpha} + \frac{1}{2E}\, \{ p_{\alpha},
E f^{\alpha}{}_a\}_- \,,\nonumber\\                  
R &=&  \frac{1}{2} \, \{ f^{\alpha [ a} f^{\beta b ]} \;(\omega_{a b \alpha, \beta} 
- \omega_{c a \alpha}\,\omega^{c}{}_{b \beta}) \} + h.c. \,, \nonumber \\
\tilde{R}  &=&  \frac{1}{2} \, \{ f^{\alpha [ a} f^{\beta b ]} 
\;(\tilde{\omega}_{a b \alpha,\beta} - \tilde{\omega}_{c a \alpha} \,
\tilde{\omega}^{c}{}_{b \beta})\} + h.c.\,,              
\label{wholeaction}
\end{eqnarray}
with~\footnote{$f^{\alpha}{}_{a}$ are inverted vielbeins to 
$e^{a}{}_{\alpha}$ with the properties $e^a{}_{\alpha} f^{\alpha}{\!}_b = 
\delta^a{\!}_b,\; e^a{\!}_{\alpha} f^{\beta}{\!}_a = \delta^{\beta}_{\alpha} $, 
$ E = \det(e^a{\!}_{\alpha}) $.
Latin indices  
$a,b,..,m,n,..,s,t,..$ denote a tangent space (a flat index),
while Greek indices $\alpha, \beta,..,\mu, \nu,.. \sigma,\tau, ..$ denote an Einstein 
index (a curved index). Letters  from the beginning of both the alphabets
indicate a general index ($a,b,c,..$   and $\alpha, \beta, \gamma,.. $ ), 
from the middle of both the alphabets   
the observed dimensions $0,1,2,3$ ($m,n,..$ and $\mu,\nu,..$), indexes from 
the bottom of the alphabets
indicate the compactified dimensions ($s,t,..$ and $\sigma,\tau,..$). 
We assume the signature $\eta^{ab} =
diag\{1,-1,-1,\cdots,-1\}$.} 
$f^{\alpha [a} f^{\beta b]}= f^{\alpha a} f^{\beta b} - f^{\alpha b} f^{\beta a}$~
\footnote{The vielbeins, $f^a_{\alpha}$, and the two kinds of the spin connection fields, 
$\omega_{ab \alpha}$ (the gauge fields of $S^{ab}$) and $\tilde{\omega}_{ab \alpha}$  
(the gauge fields of $\tilde{S}^{ab}$), manifest in $d=(3+1)$ as the known vector 
gauge fields and the scalar gauge fields taking care of masses of quarks and leptons and 
antiquarks and antileptons and of the weak boson fields~\cite{nd2017,%
n2014matterantimatter,JMP2013,%
nh2018}}~\footnote{
Since the multiplication with either $\gamma^a$'s or $\tilde{\gamma}^a$'s  changes 
the Clifford odd ``basis vectors'' into the Clifford even  objects, 
and even ``basis vectors'' commute, the action for fermions can not include an odd 
numbers of $\gamma^a$'s or $\tilde{\gamma}^a$'s, what the simple starting action 
of Eq.~(\ref{wholeaction}) does not. In the starting action $\gamma^a$'s and 
$\tilde{\gamma}^a$'s appear as $\gamma^0 \gamma^a \hat{p}_{0a}$  or as 
$\gamma^0 \gamma^c \, S^{ab}\omega_{abc}$  and  as 
$\gamma^0 \gamma^c \,\tilde{S}^{ab}\tilde{\omega}_{abc} $.},
massless fermions carry only 
spins and interact with only gravity --- with the vielbeins and the two kinds of spin 
connection fields (the gauge fields of momenta, of 
$S^{ab}=\frac{i}{4}(\gamma^a \gamma^b- \gamma^b \gamma^a)$ and of 
$\tilde{S}^{ab}=\frac{1}{4} (\tilde{\gamma}^a \tilde{\gamma}^b -
\tilde{\gamma}^b \tilde{\gamma}^a)$, respectively~%
\footnote{
If no fermions are present, the two kinds of spin connection fields are uniquely 
expressible by the vielbeins.}). The starting action includes only even products 
of $\gamma^a$'s and $\tilde{\gamma^a}$'s~(\cite{nh2021RPPNP} and
references therein). \\
{\bf b.i.} Vielbeins can be expressed by 
${}^{i}{\cal A}^{m \dagger}_{f} {}^{i} {\cal C}^{m}_{f \alpha}, i=(I,II)$.
Also this way of representing vielbeins needs further studies.\\%
{\bf b.ii.} Gravity fields in $d$, which are  the gauge fields of $S^{ab}$, 
($(a,b)=(5,6,....,d)$), with the space index $\alpha =m=(0,1,2,3)$, manifest as 
the {\it standard model} vector gauge fields~\cite{nd2017}, those with the
space index $ \sigma =5,6,7,...,d$, manifest as the {\it standard model} scalar 
gauge fields~\cite{nd2017,nh2021RPPNP}, those with $(a,b)=(0,1,2,3)$), and
with the space index $\alpha=(0,1,2,3)$ manifest as ordinary gravity. 
The supersymmetric transformations for all three kinds of boson fields deserve
further studies. In the gravity case the expression of graviton with superposition
of two ${}^{i}{\cal A}^{m \dagger}_{f} {}^{i} {\cal C}^{m}_{f \alpha}, 
i=(I,II)$, and gravitino with the superposition of
${}^{i}{\cal A}^{m \dagger}_{f} {}^{i} {\cal C}^{m}_{f \alpha}, i=(I,II)$ 
and $\hat{b}^{m \dagger}_{f}$ seems promising.\\


{\bf b.ii.{\it i.}} The scalar gauge fields of $\tilde{S}^{ab}$, and of particular
superposition of $S^{ab}$, with the space index $s=(7,8)$ manifest 
as the scalar higgs and Yukawa couplings~\cite{JMP2013,nh2021RPPNP},
determining mass matrices (of $ \widetilde{SU}(2) \times 
\widetilde{SU}(2) \times U(1)$ symmetry) and correspondingly 
the masses of quarks and leptons (predicting the fourth families to the 
observed three) and of the weak boson fields after (some of) the scalar 
fields with the space index $(7,8)$ gain constant values.\\
The theory predicts at low energy two groups with four families. To the 
lower group of four families the so far  observed three belong%
~\cite{mdn2006,gmdn2007,gmdn2008,gn2013,gn2014}, and the stable 
of the upper four families, the fifth family of (heavy) quarks and leptons, 
offers the explanation for the appearance of dark matter. Due to the 
heavy masses of the fifth family quarks, the nuclear interaction among 
hadrons of the fifth family members is very different than the ones so 
far observed~\cite{gn2009,nm2015}. \\ 
{\bf b.ii.{\it ii.}} The scalar gauge fields of $\tilde{S}^{ab}$ and of
$S^{ab}$ with the space index $s=(9,10,...,14)$ and $(a,b)=(5,6,....,d)$ 
offer the explanation for the observed matter/antimatter 
asymmetry~\cite{n2014matterantimatter,JMP2013,nh2018,nh2021RPPNP} 
in the universe.

\vspace{2mm}

The theory seems very promising to offer a new insight into the second 
quantization of fermion and boson fields and to show the next step beyond 
the {\it standard model}.

The more work is put into the theory, the more phenomena the theory can
explain.


Let me add: Other references use a different approach by trying to make 
the next step with Clifford algebra to the second quantized fermion field%
~\cite{MPavsic,MP2017}.

\vspace{2mm}


\noindent
In Sect.~\ref{creationannihilation}, creation and annihilation operators 
for fermions and bosons in even and odd dimensional spaces are 
presented. \\
Subsect.~\ref{grassmannandclifford} starts with relating the Grassmann 
algebra with the two Clifford subalgebras. \\
In Subsect.~\ref{basisvectors}, ``basis vectors''  in even and 
odd-dimensional spaces are presented. \\
In Subsect.~\ref{secondquantizedfermionsbosonsdeven}, creation and 
annihilation operators are described as tensor products of the ``basis 
vectors'' and basis in ordinary space.\\
In Sect.~\ref{achievements}, the achievements of the {\it spin-charge-family} 
theory so far are shortly overviewed.\\
Sect.~\ref{conclusions} presents what the reader could learn new from
 this article. \\
In App.~\ref{secondgroupofbosons}, the properties of the Clifford even
``basis vectors''  are demonstrated in the toy model in $d=(5+1)$. \\
 In App.~\ref{basis3+1}, the reader can find concrete examples for $d=(3+1)$, taken 
 from Ref.~\cite{n2023NPB}. \\
  In App.~\ref{A}, some useful formulas and relations are presented.\\
 In App.~\ref{13+1representation} one irreducible representation (one family) of
 $SO(13,1)$, group,  analysed with respect to $SO(3,1)$, $SU(2)_{I}$,  $SU(2)_{II}$,
  $SU(3)$, and $U(1)$, representing ``basis vectors'' of  quarks and leptons and antiquarks 
  and antilepons is discussed.

\vspace{2mm}
\section{Creation and annihilation operators for fermions and bosons in even and 
odd dimensional spaces}
\label{creationannihilation}
\vspace{2mm}

This section reviews the Refs.~\cite{n2023NPB,norma93,2020PartIPartII,nh2021RPPNP}.

The assumed simple action, Eq.~(\ref{wholeaction}), in which fermions interact 
with only the gravitational fields in $d=(13+1)$-dimensional space, contains from
the point of view of  $d=(3+1)$  all the vector gauge fields and the scalar gauge
fields, assumed by the {\it standard  model}~\cite{gmdn2007,nd2017,nh2017,%
gn2013,gn2014,nh2021RPPNP}. This simple action explains also several observed
phenomena in cosmology~\cite{gn2009,n2014matterantimatter,NHD,nm2015}.

Although it was expected all the time,  it became evident only three years ago
that while the Clifford odd algebra  offers the description of the internal space
of fermions, explaining the second quantisation o fermion fields, offers Clifford 
even algebra the explanation of the internal space of the corresponding boson 
gauge fields, offering the understanding of  the second quantized boson fields~\cite{Dirac,BetheJackiw,Weinberg}.

In even dimensional spaces, the number of ``basis vectors'' and their Hermitian 
conjugated partners is the same for fermion and boson fields: 
 $2^{\frac{d}{2}-1} \times 2^{\frac{d}{2}-1}\times 2 $.
 
 Fermion ``basis vectors'' appear in  $2^{\frac{d}{2}-1}$ families (irreducible
 representations), each family has  $2^{\frac{d}{2}-1} $ members. Their
 Hermitian conjugated partners appear in a separate group.

Boson ``basis vectors'' appear in  two groups, with the Hermitian conjugated
partners within the same group, each group has $2^{\frac{d}{2}-1}\times $ 
  $2^{\frac{d}{2}-1} $ members. 
  
  \vspace{2mm}

\subsection{Grassmann and Clifford algebras and representations of Clifford subalgebras}
\label{grassmannandclifford}
%
This part is a short overview of several references,
cited in Ref.~(\cite{nh2021RPPNP}, Subsects. 3.2,3.3),  also appearing in 
Ref.~\cite{nIARD2022,2020PartIPartII,n2023MDPI,n2023NPB}.

In Grassmann space the infinitesimal generators of the Lorentz transformations 
${\cal {\bf S}}^{ab}$ are expressible with anticommuting coordinates 
$\theta^a$ and their conjugate momenta $p^{\theta a} =
 i \frac{\partial}{\partial \theta_a}$%
~\cite{norma93},
\begin{eqnarray}
\{\theta^a, \theta^b\}_{+} &=& 0\,, \quad \{p^{\theta a}, 
p^{\theta b}\}_{+} = 0\, , \quad 
\{p^{\theta a}, \theta^b\}_{+} = i \,\eta^{a b}\, ,\nonumber\\ 
\label{thetarelcom}
{\cal {\bf S}}^{ab} &=& \theta^a p^{\theta b} -  \theta^b p^{\theta a}\,. 
\end{eqnarray}
Grassmann space offers the description of the internal degrees of freedom 
of fermions and bosons in the second quantized procedure~\cite{n2019PRD}. 
In both cases there exist the creation and annihilation operators, which fulfil 
the anticommutation relations required for fermions, and commutation 
relations for bosons~\cite{n2019PRD}.

Making a choice~\cite{nh2018} 
\begin{eqnarray}
(\theta^{a})^{\dagger} &=& \eta^{a a} \frac{\partial}{\partial \theta_{a}}\,,\quad
{\rm leads  \, to} \quad
(\frac{\partial}{\partial \theta_{a}})^{\dagger}= \eta^{a a} \theta^{a}\,,
\label{thetaderher0}
\end{eqnarray}
with $\eta^{a b}=diag\{1,-1,-1,\cdots,-1\}$.

$ \theta^{a}$ and $ \frac{\partial}{\partial \theta_{a}}$ are, up to the sign, Hermitian 
conjugated to each other. The identity is the self adjoint member of the algebra.
The choice for the following complex properties of $\theta^a$ 
 \begin{small}
\begin{eqnarray}
\label{complextheta}
\{\theta^a\}^* &=&  (\theta^0, \theta^1, - \theta^2, \theta^3, - \theta^5,
\theta^6,...,- \theta^{d-1}, \theta^d)\,, 
\end{eqnarray}
\end{small}
 correspondingly requires $\;\;\, $ 
$\{\frac{\partial}{\partial \theta_{a}}\}^* = (\frac{\partial}{\partial \theta_{0}},
\frac{\partial}{\partial \theta_{1}}, - \frac{\partial}{\partial \theta_{2}},
\frac{\partial}{\partial \theta_{3}}, - \frac{\partial}{\partial \theta_{5}}, 
\frac{\partial}{\partial \theta_{6}},..., - \frac{\partial}{\partial \theta_{d-1}}, 
\frac{\partial}{\partial \theta_{d}})\,. $
%

There are $2^d$ superposition of products of  $\theta^{a}$, 
the Hermitian conjugated partners of which are the corresponding $2^{d}$ 
superposition of products of $\frac{\partial}{\partial \theta_{a}}$.

There exist two kinds of the Clifford algebra elements (operators), $\gamma^{a}$ and 
$\tilde{\gamma}^{a}$, expressible with $\theta^{a}$'s and their conjugate momenta 
$p^{\theta a}= i \,\frac{\partial}{\partial \theta_{a}}$~\cite{norma93,nh2021RPPNP}, 
\begin{eqnarray}
\label{clifftheta1}
\gamma^{a} &=& (\theta^{a} + \frac{\partial}{\partial \theta_{a}})\,, \quad 
\tilde{\gamma}^{a} =i \,(\theta^{a} - \frac{\partial}{\partial \theta_{a}})\,,\nonumber\\
&& {\rm from \;where\;it\; follows}\nonumber\\
\theta^{a} &=&\frac{1}{2} \,(\gamma^{a} - i \tilde{\gamma}^{a})\,, \quad 
\frac{\partial}{\partial \theta_{a}}= \frac{1}{2} \,(\gamma^{a} + i \tilde{\gamma}^{a})\,,
\nonumber\\
\end{eqnarray}
offering together  $2\cdot 2^d$  operators: $2^d$ are superposition of products of 
$\gamma^{a}$  and  $2^d$ superposition of products of $\tilde{\gamma}^{a}$.
It is easy to prove if taking into account Eqs.~(\ref{thetaderher0}, \ref{clifftheta1}),
 that they form two anti-commuting Clifford subalgebras, 
$\{\gamma^{a}, \tilde{\gamma}^{b}\}_{+} =0$, Refs.~(\cite{nh2021RPPNP} and 
references therein)
\begin{eqnarray}
\label{gammatildeantiher0}
\{\gamma^{a}, \gamma^{b}\}_{+}&=&2 \eta^{a b}= \{\tilde{\gamma}^{a}, 
\tilde{\gamma}^{b}\}_{+}\,, \nonumber\\
\{\gamma^{a}, \tilde{\gamma}^{b}\}_{+}&=&0\,,\quad
 (a,b)=(0,1,2,3,5,\cdots,d)\,, \nonumber\\
(\gamma^{a})^{\dagger} &=& \eta^{aa}\, \gamma^{a}\, , \quad 
(\tilde{\gamma}^{a})^{\dagger} =  \eta^{a a}\, \tilde{\gamma}^{a}\,.
\end{eqnarray}
The Grassmann algebra offers the description of the ``anti-commuting integer spin
second quantized fields'' and of the ``commuting integer spin second quantized 
fields'', the reader is invited to read~\cite{n2019PRD,2020PartIPartII,nh2021RPPNP},
which offer the representations and equations of motion when using Grassmann
algebra to describe internal spaces of fermions and bosons.

Each of the two Clifford algebras which are superposition of odd products of either 
$\gamma^a$'s or $\tilde{\gamma}^a$'s offers the description of the second 
quantized half integer spin fermion fields 
in the fundamental representations of the group and subgroups, 
Table~\ref{Table so13+1.}.

\noindent
The superposition of even products of either $\gamma^a$'s or 
$\tilde{\gamma}^a$'s offer the description of the commuting second quantized 
boson fields with integer spins~\cite{n2021SQ,n2022epjc} which from the point 
of the subgroups of the $SO(d-1,1)$ group manifest spins and charges in the 
adjoint representations of the group and subgroups, Table~(\ref{Table so13+1.},
Tables~(1, 2, 3, 4) in Ref.~\cite{n2023NPB}, Table~\ref{Table so13+1.}.

\vspace{2mm}

There is no fermions with the integer spin observed and there is only one kind of 
half spin fermions and their integer spin gauge fields observed so far.

The {\it postulate}, which determines how does the operator $\tilde{\gamma}^{a}$
operate on polynomial of $\gamma^{a}$ \\
$A=\sum_{k=0}^{d}\, a_{a_{1} a_{2} \dots a_{k}}\,
\gamma^{a_1}\gamma^{a_2} \dots \gamma^{a_k} $, $ a_{i}\le a_{i+1}$%
~\cite{nh02,norma93,JMP2013,nh2018}
\begin{eqnarray}
\tilde{\gamma}^a A &=&(-)^{A}\, i \, A \gamma^a\,,
\label{tildegammareduced0}
\end{eqnarray}
with $(-)^A = -1$, if $A$ is (a function of) odd products of $\gamma^a$'s,  otherwise 
$(-)^A = 1$, reduces the two Clifford sub algebras to only one.
$\tilde{\gamma}^{a}$, indeed $\tilde{S}^{ab}=\frac{i}{4} (\tilde{\gamma}^{a}
\tilde{\gamma}^{b}- \tilde{\gamma}^{b}\tilde{\gamma}^{a})$, equip each 
irreducible representation with the family quantum numbers.

The subalgebra, determined by $\tilde{\gamma}^{a}$'s, looses its meaning.\\

The ``basis vectors'' for either fermions or bosons will be defined in 
Subsect.~\ref{basisvectors} as products of eigenvectors of each of the chosen 
Cartan subalgebra member 
\begin{small}
\begin{eqnarray}
&&S^{03}, S^{12}, S^{56}, \cdots, S^{d-1 \;d}\,, \nonumber\\
&&\tilde{S}^{03}, \tilde{S}^{12}, \tilde{S}^{56}, \cdots,  \tilde{S}^{d-1\; d}\,, 
\nonumber\\
&&{\cal {\bf S}}^{ab} = S^{ab} +\tilde{S}^{ab}=
 i \, (\theta^{a} \frac{\partial}{\partial \theta_{b}} - 
 \theta^{b} \frac{\partial}{\partial \theta_{a}})\,.
\label{cartangrasscliff0}
\end{eqnarray}
\end{small}
Each eigenvector of $S^{ab}, \tilde{S}^{ab} {\rm or\,\,}{\cal {\bf S}}^{ab}$,
chosen to be Cartan subalgebra members of the Lorentz algebra in the internal
space of fermions and bosons, Eq.~(\ref{cartangrasscliff0}), can be superposition 
of either an odd or an even number of $\gamma^{a}$'s: either as 
($\alpha\gamma^a+\beta \gamma^b$) or as ($\alpha+ 
\beta \gamma^a\gamma^b$), respectively.

\vspace{2mm}
\subsection{``Basis vectors'' describing internal spaces of fermions and bosons 
in even and odd dimensional spaces}
\label{basisvectors}
%


We use the technique~\cite{norma93,nh02} which makes ``basis vectors''   
products of nilpotents and projectors which are eigenvectors of the chosen
Cartan subalgebra members, Eq.~(\ref{cartangrasscliff0}), of the Lorentz 
algebra in the space of $\gamma^{a}$'s, either  in the case of the Clifford odd 
or in the case of the Clifford even products of  $\gamma^{a}$'s. \\
There  are  in  even-dimensional spaces $\frac{d}{2}$ members of the Cartan 
subalgebra, Eq.~(\ref{cartangrasscliff0}). In odd-dimensional spaces there are 
$\frac{d-1}{2}$ members of the Cartan subalgebra.

In even dimensional spaces, one can define for any of the $\frac{d}{2}$ Cartan 
subalgebra members $S^{ab}$ or $\tilde{S}^{ab}$ or of both kinds the nilpotent 
 $\stackrel{ab}{(k)}$ and the projector $\stackrel{ab}{[k]}$
\begin{small}
\begin{eqnarray}
\label{nilproj}
\stackrel{ab}{(k)}:&=&\frac{1}{2}(\gamma^a + 
\frac{\eta^{aa}}{ik} \gamma^b)\,, \;\;\; (\stackrel{ab}{(k)})^2=0\, , \nonumber \\
\stackrel{ab}{[k]}:&=&
\frac{1}{2}(1+ \frac{i}{k} \gamma^a \gamma^b)\,, \;\;\;(\stackrel{ab}{[k]})^2=
\stackrel{ab}{[k]}\,.
\end{eqnarray}
\end{small}
It follows, if taking into account the relations, Eqs.~(\ref{gammatildeantiher0}, 
\ref{tildegammareduced0})
\begin{small}
\begin{eqnarray}
\label{signature0}
S^{ab} \,\stackrel{ab}{(k)} = \frac{k}{2}  \,\stackrel{ab}{(k)}\,,\quad && \quad
\tilde{S}^{ab}\,\stackrel{ab}{(k)} = \frac{k}{2}  \,\stackrel{ab}{(k)}\,,\nonumber\\
S^{ab}\,\stackrel{ab}{[k]} =  \frac{k}{2}  \,\stackrel{ab}{[k]}\,,\quad && \quad 
\tilde{S}^{ab} \,\stackrel{ab}{[k]} = - \frac{k}{2}  \,\,\stackrel{ab}{[k]}\,,
\end{eqnarray}
\end{small}
with  $k^2=\eta^{aa} \eta^{bb}$~\footnote{Let us prove one of the relations in 
Eq.~(\ref{signature0}): $S^{ab}\, \stackrel{ab}{(k)}= \frac{i}{2} \gamma^a 
\gamma^b \frac{1}{2} (\gamma^a +\frac{\eta^{aa}}{ik} \gamma^b)=
\frac{1}{2^2}\{ -i (\gamma^a)^2 \gamma^b + i (\gamma^b)^2 \gamma^a 
\frac{\eta^{aa}}{ik}\}= \frac{1}{2} \frac{\eta^{aa}\eta^{bb}}{k}
\frac{1}{2} \{\gamma^a + \frac{k^2}{\eta^{bb} ik}\gamma^b\}$. For 
$k^2 = \eta^{aa} \eta^{bb}$ the first relation follows.}, 
demonstrating that the eigenvalues of $S^{ab}$ (determining the spin) on nilpotents 
and projectors expressed with $\gamma^a$ differ from the eigenvalues of 
$\tilde{S}^{ab}$ (determining the family quantum number) on nilpotents and 
projectors expressed with $\gamma^a$~\footnote{
 The reader can find the proof of Eq.~(\ref{signature0})  also in
  Ref.~\cite{nh2021RPPNP},  App.~(I).}

Let us point out that eigenvalues of $S^{ab}$, determining  the spin, and the 
eigenvalues of $\tilde{S}^{ab}$, determining the family quantum numbers,  
 are half integer,  $\pm \frac{1}{2}$ or $\pm \frac{i}{2}$, while the eigenvalues of 
 ${\cal {\bf S}}^{ab}$, expressible with ($S^{ab} + \tilde{S}^{ab}$) are integers,
$\pm 1$ and  zero or $\pm i$ and zero.  \\

\vspace{2mm}

{\it In even dimensional spaces, the ``basis vectors'' can be defined as algebraic,
$*_A$, products of nilpotents and projectors so that each product is the eigenvector
of all the $\frac{d}{2}$ Cartan subalgebra members of} Eq.(\ref{cartangrasscliff0}).\\

The fermion ``basis vectors''  can be chosen as the algebraic, $*_A$, products of 
an odd number of the nilpotents and the rest of the projectors; each of them is 
the eigenvector of one of the Cartan subalgebra members.

The boson ``basis vectors'' are the algebraic, $*_{A}$ products of an even number 
of nilpotents and the rest of the projectors. 
(In App.~\ref{basis3+1}, the reader can find concrete examples.)  

It follows that the Clifford odd ``basis vectors'', which are the superposition of odd 
products of $\gamma^{a}$, must include an odd number of nilpotents, at least one, 
while the superposition of an even products of $\gamma^{a}$, that is Clifford even 
``basis vectors'', must include an even number of nilpotents or only projectors. 

Correspondingly the Clifford odd ``basis vectors'' have in even $d$ properties 
appropriate to describe the internal space of the second quantized fermion fields 
while the Clifford even ``basis vectors'' have properties appropriate to describe the 
internal space of the second quantized boson fields. 

\vspace{2mm}

Taking into account Eqs.~(\ref{gammatildeantiher0}, \ref{tildegammareduced0}) 
one finds~\cite{n2023NPB}
\begin{small}
\begin{eqnarray}
\label{usefulrel}
 \gamma^a \stackrel{ab}{(k)}&=& \eta^{aa}\stackrel{ab}{[-k]},\; \quad
 \gamma^b \stackrel{ab}{(k)}= -ik \stackrel{ab}{[-k]}, \; \quad 
\gamma^a \stackrel{ab}{[k]}= \stackrel{ab}{(-k)},\;\quad \;\;
 \gamma^b \stackrel{ab}{[k]}= -ik \eta^{aa} \stackrel{ab}{(-k)}\,,\nonumber\\
 \tilde{\gamma^a} \stackrel{ab}{(k)} &=& - i\eta^{aa}\stackrel{ab}{[k]},\quad
 \tilde{\gamma^b} \stackrel{ab}{(k)} =  - k \stackrel{ab}{[k]}, \;\qquad  \,
 \tilde{\gamma^a} \stackrel{ab}{[k]} =  \;\;i\stackrel{ab}{(k)},\; \quad
 \tilde{\gamma^b} \stackrel{ab}{[k]} =  -k \eta^{aa} \stackrel{ab}{(k)}\,, 
%
\end{eqnarray}
\end{small}
More relations are presented in App.~\ref{A}. 

 The relations of Eq.~(\ref{usefulrel}) demonstrate that the properties of ``basis 
 vectors'' which include an odd number of nilpotents, differ essentially from the
``basis vectors'', which include an even number of nilpotents. One namely 
recognizes~\cite{n2023NPB}:\\
{\bf i.} The Hermitian conjugated partner of a nilpotent 
$\stackrel{ab}{(k)}^{\dagger}$ is $\eta^{aa}\stackrel{ab}{(-k)}$; correspondingly
neither $S^{ab}$ nor $\tilde{S}^{ab}$ nor both can transform odd products of
nilpotents to be one of the $2^{\frac{d}{2}-1}$ members of one of
$2^{\frac{d}{2}-1}$ irreducible representations (families). The Hermitian conjugated 
partners of the Clifford odd ``basis vectors'' must belong to a different group of 
$2^{\frac{d}{2}-1}$ members of $2^{\frac{d}{2}-1}$ families.\\
Since $S^{ac}$ transforms $\stackrel{ab}{(k)} *_A \stackrel{cd}{(k')}$ into
$\stackrel{ab}{[-k]} *_A \stackrel{cd}{[-k']}$, and $\tilde{S}^{ab}$ transforms
$\stackrel{ab}{(k)} *_A \stackrel{cd}{(k')}$ into $\stackrel{ab}{[k]} *_A $
$ \stackrel{cd}{[k']}$, 
%
 the Hermitian conjugated partners of the Clifford even ``basis
vectors'' must belong to the same group of $2^{\frac{d}{2}-1}\times $
$2^{\frac{d}{2}-1}$ members. Projectors are self-adjoint. \\
{\bf ii.} While an odd product of $\gamma^{a}$ anti-commute with another odd 
product of $\gamma^{a}$, the Clifford odd ``basis vectors'' anti-commute.
An even product of $\gamma^{a}$ commute with another even (or odd)
product of $\gamma^{a}$, therefore  the Clifford even ``basis vectors'' commute.\\
In the tensor product, $*_{T}$, with basis in ordinary space, Clifford odd and 
Clifford even ``basis vectors'', form creation and annihilation operators that 
inherit anti-commutation or commutation relations from basis vectors.
The Clifford odd creation operators  manifest, together with their Hermitian 
conjugated annihilation operators on the vacuum state, Eq.~(\ref{vaccliffodd0}),
 the properties of the anti-commutation relations postulated by Dirac for the 
 second quantized fermion fields. The Clifford even creation operators  manifest
correspondingly the commutation relations for the second quantized boson fields.\\
{\bf iii.} The Clifford odd ``basis vectors'' have all the eigenvalues of the Cartan
subalgebra members equal to either $\pm \frac{1}{2}$ or to $\pm \frac{i}{2}$.\\
The Clifford even ``basis vectors'' have all the eigenvalues of the Cartan subalgebra
members ${\bf {\cal S}}^{ab}=S^{ab} + \tilde{S}^{ab}$ equal to either 
$\pm 1$ and zero or to $\pm i$ and zero.\\

\vspace{2mm}

 \noindent
{\it In odd dimensional spaces, $d=(2n +1)$, the properties of the Clifford odd and the 
 Clifford even ``basis vectors'' differ essentially from their properties in even 
dimensional spaces}~(\cite{n2023MDPI} in Subsect.~\ref{dodd}, \cite{n2023NPB}
in Sect.~2.2.2. ).\\

%
\noindent
Half of the ``basis vectors'' have properties as those Clifford odd and even ``basis
vectors'' of $d'=2n$. The second half, appearing by applying on these half of ``basis 
vectors'' by $S^{0 \, 2n+1}$, although anti-commuting, the Clifford odd ``basis 
vectors'' manifest properties of the Clifford even ``basis vectors'' in even 
dimensional spaces; they appear in two separate groups, each group having their 
Hermitian conjugated within their own group. And the Clifford even ``basis vectors'', 
although commuting, manifest properties of the Clifford odd ``basis vectors'' in even 
dimensional spaces; they appear in $2^{\frac{d}{2}-1}$ families, each family with
$2^{\frac{d}{2}-1}$ members, their Hermitian conjugated partners appear in a
separate group~\cite{n2023MDPI}.

\vspace{2mm}

\subsubsection{Clifford odd and even ``basis vectors'' in even $d$}
\label{deven}
This subsection shortly reviews  Subsect.~2.2.1 of Ref.~\cite{n2023NPB}.\\

Let us start with the {\it Clifford odd ``basis vectors''  }.\\


The Clifford odd  ``basis vectors''  must be products of an odd number of nilpotents, 
and the rest, up to $\frac{d}{2}$,  of projectors,  each nilpotent and each projector 
must be the ``eigenvector'' of one of the members of the Cartan subalgebra, 
Eq.~(\ref{cartangrasscliff0}), correspondingly are the ``basis vectors'' eigenvectors 
of all the members of the Cartan subalgebra of the Lorentz algebra: $S^{ab}$'s 
determine $2^{\frac{d}{2}-1}$ members of one family, $\tilde{S}^{ab}$'s 
transform  each member of one family to the same member of the rest of 
$2^{\frac{d}{2}-1}$ families. 

Let us call the Clifford odd ``basis vectors''  $\hat{b}^{m \dagger}_{f}$,  if it is 
the $m^{th}$ membership  of the family $f$. The Hermitian conjugated partner 
of $\hat{b}^{m \dagger}_{f}$ is called $\hat{b}^{m}_{f} \,(=
(\hat{b}^{m \dagger}_{f})^{\dagger}$.

In  $d=2(2n+1)$ the ``basis vector'' $\hat{b}^{1 \dagger}_{1}$ is chosen to be
the product of only nilpotents, all the rest members belonging to the $f=1$ 
family follow by the application of $S^{01}$, $S^{03}$, $ \dots, S^{0d}, S^{15}$,
$\dots, S^{1d}, S^{5 d}\dots, S^{d-2\, d}$. They are presented on the left-hand 
side of Eq.~(\ref{allcartaneigenvec}).
Their Hermitian conjugated partners~\footnote{
Taking into account that $(\gamma^a)^{\dagger} =\eta^{aa} \gamma^{a}, k^{2}=
 \eta^{aa}\eta^{bb}$, and $k^{\dagger} k=1$, it follows:\\
 $(\stackrel{ab}{(k)})^{\dagger}= \frac{1}{2}\eta^{aa} (\gamma^{a} + 
 \frac{\eta^{bb} k k}{i k^{\dagger}k (- k)}\gamma^{b})=\frac{1}{2}\eta^{aa} 
 (\gamma^{a} +  \frac{\eta^{aa} }{i (- k)}\gamma^{b})=\stackrel{ab}{(-k)}$, 
 and  $(\stackrel{ab}{[k]})^{\dagger}= \frac{1}{2} (1 +\frac{i}{k} \gamma^{a} 
\gamma^{b})^{\dagger}=\frac{1}{2} (1 +\frac{-i}{k^{\dagger}}\eta^{aa}
\eta^{bb} \gamma^{b} \gamma^{a})=\stackrel{ab}{[k]}$}
 are presented on the right-hand side. 
The algebraic product mark $*_{A}$ among nilpotents and projectors is skipped.
\begin{small}
\begin{eqnarray}
\label{allcartaneigenvec}
&& \qquad  \qquad \qquad \qquad \qquad \qquad    d=2(2n+1)\, ,\nonumber\\
&& \hat{b}^{1 \dagger}_{1}=\stackrel{03}{(+i)}\stackrel{12}{(+)} \stackrel{56}{(+)}
\cdots \stackrel{d-1 \, d}{(+)}\,,\qquad  \qquad \qquad \quad \quad
\hat{b}^{1}_{1}=\stackrel{03}{(-i)}\stackrel{12}{(-)}\cdots \stackrel{d-1 \, d}{(-)}\,,
\nonumber\\
&&\hat{b}^{2 \dagger}_{1} = \stackrel{03}{[-i]} \stackrel{12}{[-]} 
\stackrel{56}{(+)} \cdots \stackrel{d-1 \, d}{(+)}\,,\qquad \qquad \qquad \qquad\;\;
\hat{b}^{2 }_{1} = \stackrel{03}{[-i]} \stackrel{12}{[-]} 
\stackrel{56}{(-)} \cdots \stackrel{d-1 \, d}{(-)}\,,\nonumber\\
&& \cdots \qquad  \qquad \qquad \qquad \qquad  \qquad \qquad \qquad \qquad\;
\cdots \nonumber\\
&&\hat{b}^{2^{\frac{d}{2}-1} \dagger}_{1} = \stackrel{03}{[-i]} \stackrel{12}{[-]} 
\stackrel{56}{(+)} \dots \stackrel{d-3\,d-2}{[-]}\;\stackrel{d-1\,d}{[-]}\,, \qquad
\hat{b}^{2^{\frac{d}{2}-1} \dagger}_{1} = \stackrel{03}{[-i]} \stackrel{12}{[-]} 
\stackrel{56}{(-)} \stackrel{78}{[-]} \dots \stackrel{d-3\,d-2}{[-]}\;\stackrel{d-1\,d}{[-]}\,,
\nonumber\\
&& \cdots\,, \qquad \qquad  \qquad \qquad \qquad \qquad \qquad \qquad  \cdots\,.
\end{eqnarray}
\end{small}
In $d=4n$  the choice of the starting ``basis vector''  with maximal number of nilpotents
must have one projector, all the rest follows in equivalent way as in the case of 
$d=2(2n+1)$.

The reader can notice that all the ``basis vectors''   within any family, as well as  the 
``basis vectors'' among families, are orthogonal; that is, their mutual algebraic products 
are zero. The same is true within their Hermitian conjugated partners.
\begin{eqnarray}
\hat{b}^{m \dagger}_f *_{A} \hat{b}^{m `\dagger }_{f `}&=& 0\,, 
\quad \hat{b}^{m}_f *_{A} \hat{b}^{m `}_{f `}= 0\,, \quad \forall m,m',f,f `\,. 
\label{orthogonalodd}
\end{eqnarray}
Choosing the vacuum state equal to
\begin{eqnarray}
\label{vaccliffodd0}
|\psi_{oc}>= \sum_{f=1}^{2^{\frac{d}{2}-1}}\,\hat{b}^{m}_{f}{}_{*_A}
\hat{b}^{m \dagger}_{f} \,|\,1\,>\,,
\end{eqnarray}
for one of members $m$, anyone of the odd irreducible representations $f$,
it follows that the Clifford odd ``basis vectors''  obey the relations
\begin{eqnarray}
\label{almostDirac}
\hat{b}^{m}_{f} {}_{*_{A}}|\psi_{oc}>&=& 0.\, |\psi_{oc}>\,,\nonumber\\
\hat{b}^{m \dagger}_{f}{}_{*_{A}}|\psi_{oc}>&=&  |\psi^m_{f}>\,,\nonumber\\
\{\hat{b}^{m}_{f}, \hat{b}^{m'}_{f `}\}_{*_{A}+}|\psi_{oc}>&=&
 0.\,|\psi_{oc}>\,, \nonumber\\
\{\hat{b}^{m \dagger}_{f}, \hat{b}^{m' \dagger}_{f  `}\}_{*_{A}+}|\psi_{oc}>
&=& 0. \,|\psi_{oc}>\,,\nonumber\\
\{\hat{b}^{m}_{f}, \hat{b}^{m' \dagger}_{f `}\}_{*_{A}+}|\psi_{oc}>
&=& \delta^{m m'} \,\delta_{f f `}|\psi_{oc}>\,.
\end{eqnarray}
In above equation the normalization 
$<\psi_{oc}| (\hat{b}^{m \dagger}_{f})^{\dagger}\, *_{A}\,\hat{b}^{m \dagger}_{f}
*_{A}|\psi_{oc}> = 1$  is used. The anti-commutation relation
is defined: $\{\hat{b}^{m \dagger}_{f}, \hat{b}^{m' \dagger}_{f  `}\}_{*_{A}+}=$
$\hat{b}^{m \dagger}_{f} \,*_A\, \hat{b}^{m' \dagger}_{f  `}+
\hat{b}^{m' \dagger}_{f `} \,*_A \,\hat{b}^{m \dagger}_{f }$.


The reader can find in tables 1. and 2. of Ref.~{\cite{n2023NPB} the properties of the 
Clifford odd ``basis vectors'' for $d=(5+1)$ case. Assuming that this case represents
$SO(6)$ broken into $SU(3)\times U(1)$, Fig.~1 manifests the triplet and singlet of
any of  $2^{\frac{6}{2}-1}=4$ families, the same figure is presented in Fig.%
~\ref{FigSU3U1odd} in this paper.\\

\vspace{3mm} 

We shall see that the {\it Clifford even ``basis vectors'' } manifest in even 
dimensional spaces properties of the internal spaces of the gauge fields of the 
corresponding {\it Clifford odd  ``basis vectors'' }, which manifest properties of the 
fermion second quantized fields~\cite{n2022epjc,n2023NPB}.

\vspace{2mm}

The Clifford even ``basis vectors'' must be products of an even number of 
nilpotents and the rest, up to $\frac{d}{2}$, of projectors. Again, like in the 
Clifford odd ``basis vectors'', each nilpotent and each projector is  the 
``eigenvector'' of one of the members of the Cartan subalgebra of the Lorentz 
algebra, ${\bf {\cal S}}^{ab}= S^{ab} + \tilde{S}^{ab}$, 
Eq.~(\ref{cartangrasscliff0}). Correspondingly the ``basis vectors'' are the 
eigenvectors of all the members of the Cartan subalgebra of the Lorentz algebra.

The Clifford even ``basis vectors'', called  ${}^{i}\hat{\cal A}^{m \dagger}_{f}$,
where $i=(I,II)$, appear in two groups, each group has  $2^{\frac{d}{2}-1}
\times $ $2^{\frac{d}{2}-1}$ members, each member is either self adjoint or 
has his Hermitian conjugated partner within the same group. 


$S^{ab}$ and $\tilde{S}^{ab}$ generate from the starting ``basis vector'' of
each group all the $2^{\frac{d}{2}-1} \times$ $2^{\frac{d}{2}-1}$ members.
Let us present the members in $d=2(2n+1)$ dimensions. In $d=4n$ we can start 
with only nilpotents, in $d=2(2n+1)$ case we  must start with one projector.
\begin{eqnarray}
\label{allcartaneigenvecevenI} 
d&=&2(2n+1)\nonumber\\
{}^I\hat{{\cal A}}^{1 \dagger}_{1}=\stackrel{03}{(+i)}\stackrel{12}{(+)}\cdots 
\stackrel{d-1 \, d}{[+]}\,,\qquad &&
{}^{II}\hat{{\cal A}}^{1 \dagger}_{1}=\stackrel{03}{(-i)}\stackrel{12}{(+)}\cdots 
\stackrel{d-1 \, d}{[+]}\,,\nonumber\\
{}^I\hat{{\cal A}}^{2 \dagger}_{1}=\stackrel{03}{[-i]}\stackrel{12}{[-]} 
\stackrel{56}{(+)} \cdots \stackrel{d-1 \, d}{[+]}\,, \qquad  && 
{}^{II}\hat{{\cal A}}^{2 \dagger}_{1}=\stackrel{03}{[+i]}\stackrel{12}{[-]} 
\stackrel{56}{(+)} \cdots \stackrel{d-1 \, d}{[+]}\,,
\nonumber\\ 
{}^I\hat{{\cal A}}^{3 \dagger}_{1}=\stackrel{03}{(+i)} \stackrel{12}{(+)} 
\stackrel{56}{(+)} \cdots \stackrel{d-3\,d-2}{[-]}\;\stackrel{d-1\,d}{(-)}\,, \qquad &&
{}^{II}\hat{{\cal A}}^{3 \dagger}_{1}=\stackrel{03}{(-i)} \stackrel{12}{(+)} 
\stackrel{56}{(+)} \cdots \stackrel{d-3\,d-2}{[-]}\;\stackrel{d-1\,d}{(-)}\,,  \nonumber\\
\dots \qquad && \dots 
\end{eqnarray}
%
There are $2^{\frac{d}{2}-1}\times \,2^{\frac{d}{2}-1}$ Clifford  even ``basis 
vectors'' of the kind ${}^{I}{\hat{\cal A}}^{m \dagger}_{f}$ and  the same 
number of the Clifford  even  ``basis vectors'' of the kind 
${}^{II}{\hat{\cal A}}^{m \dagger}_{f}$.\\
 
 

Table~1, presented in Ref.~(\cite{n2023NPB}, Subsect.~2.3) illustrates 
properties of the Clifford odd and Clifford even ``basis vectors'' on the case of 
$d=(5+1)$. 
We shall discuss in here only the general case  by carefully inspecting 
properties of both kinds of ``basis vectors''.
%

The Clifford even ``basis vectors''  belonging to two different groups are 
orthogonal due to the fact that they differ in the sign of one nilpotent or one 
projector, or the algebraic product of a member of one group with a member 
of another group gives zero according to the first two lines of 
Eq.~(\ref{graficcliff0}): $\stackrel{ab}{(k)}\stackrel{ab}{[k]} =0$, 
$\stackrel{ab}{[k]}\stackrel{ab}{(-k)} =0$, 
$\stackrel{ab}{[k]}\stackrel{ab}{[-k]} =0$.
\begin{eqnarray}
\label{AIAIIorth}
{}^{I}{\hat{\cal A}}^{m \dagger}_{f} *_A {}^{II}{\hat{\cal A}}^{m \dagger}_{f} 
&=&0={}^{II}{\hat{\cal A}}^{m \dagger}._{f} *_A 
{}^{I}{\hat{\cal A}}^{m \dagger}_{f}\,.
\end{eqnarray}
The members of each of these two groups have the property 
\begin{eqnarray}
\label{ruleAAI}
{}^{i}{\hat{\cal A}}^{m \dagger}_{f} \,*_A\, {}^{i}{\hat{\cal A}}^{m' \dagger}_{f `}
\rightarrow  \left \{ \begin{array} {r}
 {}^{i}{\hat{\cal A}}^{m \dagger}_{f `}\,, i=(I,II) \\
{\rm or \,zero}\,.
\end{array} \right.
\end{eqnarray}
For a chosen ($m, f, f `$) there is only one $m'$ (out of  $2^{\frac{d}{2}-1}$)
which gives nonzero contribution. The reader should pay attention on the repetition of 
the index $m$ and $f `$ on the left and on the right hand side of Eq.~(\ref{ruleAAI}).

%

Two ``basis vectors'', ${}^{i}{\hat{\cal A}}^{m \dagger}_{f}$  and 
${}^{i}{\hat{\cal A}}^{m' \dagger}_{f '}$, the algebraic product, $*_{A}$, of which 
 gives non zero contribution, ``scatter'' into the third one 
 ${}^{i}{\hat{\cal A}}^{m \dagger}_{f `}$ of the same kind, for $i=(I,II)$. 

\vspace{2mm}

It remains to evaluate the algebraic application, $*_{A}$, of the Clifford even ``basis vectors'' 
${}^{I,II}{\hat{\cal A}}^{m \dagger}_{f }$ on the Clifford odd ``basis vectors'' 
$ \hat{b}^{m' \dagger}_{f `} $. One finds, taking into account Eq.~(\ref{graficcliff0}),
   for ${}^{I}{\hat{\cal A}}^{m \dagger}_{f }$
\begin{eqnarray}
\label{calIAb1234gen}
{}^{I}{\hat{\cal A}}^{m \dagger}_{f } \,*_A \, \hat{b}^{m' \dagger }_{f `}
\rightarrow \left \{ \begin{array} {r} \hat{b }^{m \dagger}_{f `}\,, \\
{\rm or \,zero}\,,
\end{array} \right.
\end{eqnarray}

The reader should pay attention on the repetition of the index $m$ and $f `$ on 
the left and on the right hand side in the above equation.

For each ${}^{I}{\hat{\cal A}}^{m \dagger}_{f}$  there are among 
$2^{\frac{d}{2}-1}\times \;2^{\frac{d}{2}-1}$ members of the Clifford odd 
``basis vectors'' (describing the internal space of fermion fields) 
$2^{\frac{d}{2}-1}$ members, $\hat{b}^{m' \dagger}_{f `}$, fulfilling the
relation of Eq.~(\ref{calIAb1234gen}). All the rest ($2^{\frac{d}{2}-1}\times 
\, (2^{\frac{d}{2}-1}-1)$)  Clifford odd ``basis vectors'' give zero contributions.
 Or equivalently, there are 
$ 2^{\frac{d}{2}-1}$ pairs of quantum numbers $(f,m')$ for which 
$\hat{b }^{m \dagger}_{f `}\ne 0$~\footnote{
\begin{small}
Let us treat a particular case in $d=2(2n+1)$-dimensional space:\\
${}^{I}{\hat{\cal A}}^{m \dagger}_{f} (\equiv \stackrel{03}{(+i)}\stackrel{12}{(+)} 
\stackrel{56}{(+)}\dots \stackrel{d-3\, d-2}{(+)}\stackrel{d-1\, d}{[+]} ) *_{A}$
 $\hat{b}^{m' \dagger }_{f `} (\equiv \stackrel{03}{(-i)}\stackrel{12}{(-)} 
\stackrel{56}{(-)}\dots \stackrel{d-3\, d-2}{(-)}\stackrel{d-1\, d}{(+)}) \rightarrow
\hat{b}^{m \dagger }_{f `} (\equiv \stackrel{03}{[+i]}\stackrel{12}{[+]} 
\stackrel{56}{[+]}\dots \stackrel{d-3\, d-2} {[+]}\stackrel{d-1\, d}{(+)}$. 
The ${\cal {\bf{S}}}^{ab}$ (meaning ${\cal {\bf{S}}}^{03}, {\cal {\bf{S}}}^{12},
{\cal {\bf{S}}}^{56}, \dots {\cal {\bf{S}}}^{d-1\,d}$) say for  the above case
that the boson field with the quantum numbers $(i, 1, 1, \dots, 1, 0)$ when ``scattering''
on the fermion field with the Cartan subalgebra quantum numbers ($S^{03}, S^{1,2}, 
S^{56}\dots S^{d-3\, d-2}, S^{d-1\,d}$) $=(-\frac{i}{2}, -\frac{1}{2},  -\frac{1}{2},
\dots, -\frac{1}{2}, \frac{1}{2})$, and the family quantum numbers $(-\frac{i}{2}, 
-\frac{1}{2},  -\frac{1}{2}, \dots,$ $-\frac{1}{2}, \frac{1}{2})$ transfers to the fermion 
field its quantum numbers $(i, 1, 1, \dots, 1, 0)$, transforming fermion family members 
quantum numbers to   $ (\frac{i}{2}, \frac{1}{2}, \frac{1}{2},
\dots, \frac{1}{2}, \frac{1}{2})$, leaving family quantum numbers unchanged. 
\end{small}
}. \\

\vspace{2mm}

Taking into account Eq.~(\ref{graficcliff0}) one finds
\begin{eqnarray}
\label{calbIA1234gen}
 \hat{b}^{m \dagger }_{f } *_{A} {}^{I}{\hat{\cal A}}^{m'  \dagger}_{f `} = 0\,, \quad
 \forall (m, m`, f, f `)\,.
\end{eqnarray}

\vspace{2mm}


Eqs.~(\ref{calIAb1234gen}, \ref{calbIA1234gen}) demonstrates that 
${}^{I}{\hat{\cal A}}^{m \dagger}_{f}$, 
applying on $\hat{b}^{m' \dagger }_{f `} $, transforms the Clifford odd ``basis vector''
into another Clifford  odd   ``basis vector'' of the same family, transferring to the
   Clifford odd ``basis vector''  integer spins, or gives zero.
   
 For  ``scattering'' the Cifford even ``basis vectors'' 
${}^{II}{\hat{\cal A}}^{m \dagger}_{f }$ on the Clifford odd ``basis vectors'' 
$ \hat{b}^{m' \dagger}_{f `} $ it follows 
%
\begin{eqnarray}
\label{calIIAb1234gen}
{}^{II}{\hat{\cal A}}^{m \dagger}_{f } \,*_A \, \hat{b}^{m' \dagger }_{f `}= 0\,,\;\;
\forall (m,m',f,f `)\,,
\end{eqnarray}
while we get 
\begin{eqnarray}
\label{calbIIA1234gen}
\hat{b}^{m \dagger }_{f } *_{A} {}^{II}{\hat{\cal A}}^{m' \dagger}_{f `} \,
\rightarrow \left \{ \begin{array} {r} \hat{b }^{m \dagger}_{f ``}\,, \\
{\rm or \,zero}\,,
\end{array} \right.
\end{eqnarray}
For each $\hat{b}^{m \dagger}_{f}$ 
 there are among $2^{\frac{d}{2}-1}$
$\times \;2^{\frac{d}{2}-1}$ members of the Clifford even ``basis vectors'' 
(describing the internal space of boson  fields) \,,  
${}^{II}{\hat{\cal A}}^{m' \dagger}_{f `}$,  
$2^{\frac{d}{2}-1}$ members (with appropriate $f `$ and $m'$) 
fulfilling the relation of Eq.~(\ref{calbIIA1234gen}) while  $f ``$ runs over
$(1 - 2^{\frac{d}{2}-1})$.

All the rest ($2^{\frac{d}{2}-1}\times \;(2^{\frac{d}{2}-1}-1)$)  Clifford even``basis 
vectors'' give zero contributions.

 Or equivalently, there are 
$ 2^{\frac{d}{2}-1}$ pairs of quantum numbers $(f  ',m')$ for which 
$\hat{b }^{m \dagger}_{f }$ and ${}^{II}{\hat{\cal A}}^{m' \dagger}_{f `}$ give 
non zero contribution~\footnote{
\begin{small}
Let us treat a particular case in $d=2(2n+1)$-dimensional space:\\
 $\hat{b}^{m \dagger }_{f } (\equiv \stackrel{03}{(-i)}\stackrel{12}{(-)} 
\stackrel{56}{(-)}\dots \stackrel{d-3\, d-2}{(-)}\stackrel{d-1\, d}{(+)})*_{A}$
${}^{II}{\hat{\cal A}}^{m `\dagger}_{f `} (\equiv \stackrel{03}{(+i)}\stackrel{12}{(+)} 
\stackrel{56}{(+)}\dots \stackrel{d-3\, d-2}{(+)}\stackrel{d-1\, d}{[-]} ) \rightarrow$
 $\hat{b}^{m \dagger }_{f  `'} (\equiv \stackrel{03}{[-i]}\stackrel{12}{[-]} $
$\stackrel{56}{[-]}\dots \stackrel{d-3\, d-2}{[-]}\stackrel{d-1\, d}{(+)}) $
 When the fermion field with the Cartan subalgebra family members quantum numbers 
 ($S^{03}, S^{12}, S^{56} \dots $ $S^{d-3\, d-2}, S^{d-1\,d}$) 
 $=(-\frac{i}{2}, -\frac{1}{2}, 
  -\frac{1}{2}, \dots, -\frac{1}{2}, \frac{1}{2})$  and family quantum numbers 
  ($\tilde{S}^{03}, \tilde{S}^{12}, \tilde{S}^{56}\dots \tilde{S}^{d-3\, d-2},
  \tilde{S}^{d-1\,d}$) $ (-\frac{i}{2}, -\frac{1}{2},  -\frac{1}{2}, \dots, -\frac{1}{2}, 
  \frac{1}{2})$ 
 ``absorbs'' a boson field  with the Cartan subalgebra  quantum numbers 
 ${\cal {\bf{S}}}^{ab}$ (meaning ${\cal {\bf{S}}}^{03}, {\cal {\bf{S}}}^{12},
{\cal {\bf{S}}}^{56}, \dots {\cal {\bf{S}}}^{d-1\,d}$)  equal to 
 $(i, 1, 1, \dots, 1, 0)$,  the fermion field changes the family quantum numbers 
  ($\tilde{S}^{03}, \tilde{S}^{1,2}, \tilde{S}^{56}\dots \tilde{S}^{d-3\, d-2},
  \tilde{S}^{d-1\,d}$)  to $ (\frac{i}{2}, \frac{1}{2},  \frac{1}{2}, \dots, \frac{1}{2}, 
  \frac{1}{2})$, keeping family members quantum numbers unchanged. \\ 
 \end{small}
}.
\vspace{2mm}

Eqs.~(\ref{calIIAb1234gen}, \ref{calbIIA1234gen}) demonstrate that 
${}^{II}{\hat{\cal A}}^{m' \dagger}_{f'}$, ``absorbed'' by 
$\hat{b}^{m \dagger }_{f } $, transforms the Clifford odd ``basis vector''
into the Clifford  odd   ``basis vector'' of the same family member and of 
another family, or gives zero.

Tables 1, 2, 3, presented in Subsect.~2.3 in Ref.~\cite{n2023NPB}, and
Table~\ref{Cliff basis5+1even II.}, presented in App.~\ref{secondgroupofbosons}
illustrate properties of the Clifford odd and Clifford even ``basis vectors'' on the
case of $d=(5+1)$. Looking at this case, the reader can easily evaluate properties of
either even or odd ``basis vectors''. We discuss in this subsection the
general case by carefully inspecting the properties of both kinds of ``basis vectors''.\\

Assuming that this case represents
$SO(6)$ broken into $SU(3)\times U(1)$, Fig.~2 manifests one octet (the sextet 
with two self-adjoint members with all quantum numbers equal zero) and the triplet 
and the anti-triplet triplet with two 
selfadjoint members with all the quantum numbers equal zero.\\


\vspace{2mm}


While the Clifford odd ``basis vectors'', $\hat{b}^{m \dagger}_{f}$, offer the
description of the internal space of the second quantized anti-commuting fermion
fields, appearing in families, the Clifford even ``basis vectors'',
${}^{I,II}{\hat{\cal A}}^{m \dagger}_{f }$, offer the description of the internal
space of the second quantized commuting boson fields, having no families and
appearing in two groups. One of the two groups,
${}^{I}{\hat{\cal A}}^{m \dagger}_{f }$, transferring their integer quantum
numbers to the Clifford odd ``basis vectors'', $\hat{b}^{m \dagger}_{f}$,
changes the family members quantum numbers leaving the family quantum
numbers unchanged. The second group, ${}^{II}{\hat{\cal A}}^{m \dagger}_{f }$,
transferring their integer quantum numbers to the Clifford odd ``basis vector'', 
changes the family quantum numbers leaving the family members quantum 
numbers unchanged.\\

{\it Both groups of Clifford even ``basis vectors'' manifest as the gauge fields
of the corresponding fermion fields: One concerning the family members
quantum numbers, the other concerning the family quantum numbers.}\\

One can find in Fig.~\ref{FigSU3U1even} in this article the demonstration of
the properties of both gauge fields, ${}^{I}{\hat{\cal A}}^{m \dagger}_{f }$
and ${}^{II}{\hat{\cal A}}^{m \dagger}_{f }$, to the fermion fields presented
in Fig.~\ref{FigSU3U1odd} for the case of $d=(5+1)$, assuming that in all three
cases $SO(6)$ is broken into $SU(3)\times U(1)$. Although ``basis vectors''
of the two groups are different, and they differently affect the fermion fields,
their quantum numbers are the same. \\ 



%
\subsubsection{Clifford odd and even ``basis vectors'' in $d$ odd}
\label{dodd}
 %
 %

Let us shortly overview properties of the fermion and boson ``basis vectors'' in odd
dimensional spaces, as presented in Ref.~\cite{n2023MDPI}, Subsect.~2.2.

In even dimensional spaces the Clifford odd ``basis vectors'' fulfil the postulates for 
the second quantized fermion fields, Eq.~(\ref{almostDirac}), and the Clifford even 
''basis vectors'' have the properties of the internal spaces of their corresponding gauge 
fields, Eqs.~(\ref{ruleAAI}, \ref{calIAb1234gen}, \ref{calbIIA1234gen}). In odd 
dimensional spaces, the Clifford odd and even ''basis vectors'' have unusual properties 
resembling properties of the internal spaces of the Faddeev--Popov ghosts, as we 
described in~\cite{n2023MDPI}.

In $d=(2n+1)$-dimensional cases, $n=1,2,\dots$,  half of the ``basis vectors'', 
$2^{\frac{2n}{2}-1}$ $\times \,2^{\frac{2n}{2}-1}$, can be taken from the 
$2n$-dimensional part of space, presented in Eqs.~(\ref{allcartaneigenvec}, 
\ref{allcartaneigenvecevenI}, \ref{ruleAAI}).  

The rest  of  the ``basis vectors'' in odd dimensional spaces, $2^{\frac{2n}{2}-1}$ 
$\times \, 2^{\frac{2n}{2}-1}$, follow if  $S^{0 \,2n+1}$  is applied on these half of the 
``basis vectors''. Since $S^{0 \,2n+1}$ are Clifford even operators,  they do not change 
the oddness or evenness of the ``basis vectors''.

For the Clifford odd ``basis vectors'',  the $2^{\frac{d-1}{2}-1}$ members appearing in 
$2^{\frac{d-1}{2}-1}$ families and representing the part which is the same as in even,
$d=2n$, dimensional space are present on the left-hand side of 
Eq.~(\ref{allcartaneigenvecbdgen}), the part obtained by  applying $S^{0 \,2n+1}$ on 
the one of the left-hand side is presented on the right hand side.  Below  the ``basis 
vectors'' and their Hermitian conjugated partners are presented.
\begin{small}
\begin{eqnarray}
\label{allcartaneigenvecbdgen}
 d=&&2(2n+1)+1\, \nonumber\\
 \hat{b}^{1 \dagger}_{1}=\stackrel{03}{(+i)}\stackrel{12}{(+)} \stackrel{56}{(+)}
\cdots \stackrel{d-2 \, d-1}{(+)} \,,\quad 
&& \hat{b}^{1 \dagger}_{2^{\frac{d-1}{2}-1}+1}=\stackrel{03}{[-i]}
\stackrel{12}{(+)} \stackrel{56}{(+)} \cdots \stackrel{d-2 \, d-1}{(+)} \gamma^{d}\,,
\nonumber\\
%
\cdots \quad 
&& \cdots\nonumber\\
\hat{b}^{2^{\frac{d-1}{2}-1} \dagger}_{1} = \stackrel{03}{[-i]} \stackrel{12}{[-]} 
\stackrel{56}{(+)} \dots \stackrel{d-2\,d-1}{[-]}\;, \quad
&&\hat{b}^{2^{\frac{d-1}{2}-1} \dagger}_{2^{{d-1}{2}-1}+1} = \stackrel{03}{(+i)} \stackrel{12}{[-]} 
\stackrel{56}{(+)} \dots \stackrel{d-2\,d-1}{[-]}\; \gamma^{d}\,, \nonumber\\
\cdots \quad
&& \cdots \,,\nonumber\\
&& \cdots \,,\nonumber\\
 \hat{b}^{1}_{1}=\stackrel{03}{(-i)}\stackrel{12}{(-)} \stackrel{56}{(-)}
\cdots \stackrel{d-2 \, d-1}{(-)} \,,\quad 
&& \hat{b}^{1}_{2^{\frac{d-1}{2}-1}+1}=\stackrel{03}{[+i]}\stackrel{12}{(-)} \stackrel{56}{(-)}
\cdots \stackrel{d-2 \, d-1}{(-)} \gamma^{d}\,,\nonumber\\
\cdots \quad
&& \cdots \,.
\end{eqnarray}
\end{small}

The application of $S^{0d}$ or $\tilde{S}^{0d}$ on the left-hand side of the 
``basis vectors'' (and the Hermitian conjugated partners of both) generate the whole 
set of $2\times 2^{d-2}$ members of the Clifford odd ``basis vectors'' and their 
Hermitian conjugated partners in $d=(2n+1)$- dimensional space
 appearing on the left-hand side and the right-hand sides of 
 Eq.~(\ref{allcartaneigenvecbdgen}).\\
 

It is not difficult to see that $ \hat{b}^{m \dagger}_{2^{\frac{d-1}{2}-1}+k}$ and
$ \hat{b}^{m'}_{2^{\frac{d-1}{2}-1}+k'}$ on the right-hand side of
Eq.~(\ref{allcartaneigenvecbdgen}) obtain properties of the two groups (they are
orthogonal to each other; the algebraic products, $*_{A}$, of a member from one
group, and any member of another group give zero)
with the Hermitian conjugated partners within the same group;
they have properties of the Clifford even ``basis vectors'' from the point of view of
the Hermiticity property: The operators $\gamma^a$ are up to a constant the self-adjoint
operators, while $S^{0 d}$ transforms one nilpotent into a projector. 

$S^{ab}$ do not change the Clifford oddness of $ \hat{b}^{m \dagger}_{f}$, and
$ \hat{b}^{m}_{f}$;  $ \hat{b}^{m \dagger}_{f}$ remain to be Clifford odd objects,
however, with the properties of boson fields.\\

Let us find the  Clifford even ``basis vectors'' in odd dimensional space $d=2(2n+1) +1$.
\begin{small}
\begin{eqnarray}
\label{allcartaneigenvecAdgen}
 d=&&2(2n+1)+1\, \nonumber\\
 {}^{I}{\bf {\cal A}}^{1 \dagger}_{1} =\stackrel{03}{(+i)}\stackrel{12}{(+)} \stackrel{56}{(+)} \cdots \stackrel{d-2 \, d-1}{[+]} \,,\quad 
&& {}^{I}{\bf {\cal A}}^{1 \dagger}_{2^{{d-1}{2}-1}+1}=\stackrel{03}{[-i]}\stackrel{12}{(+)} \stackrel{56}{(+)}
\cdots \stackrel{d-2 \, d-1}{[+]} \gamma^{d}\,,\nonumber\\
%
\cdots \quad 
&& \cdots\nonumber\\
 {}^{I}{\bf {\cal A}}^{2^{\frac{d-1}{2}-1} \dagger}_{1} = \stackrel{03}{[-i]} 
 \stackrel{12}{[-]} \stackrel{56}{[-]} \dots \stackrel{d-2\,d-1}{[+]}\;, \quad
&& {}^{I}{\bf {\cal A}}^{2^{\frac{d-1}{2}-1} \dagger}_{2^{{d-1}{2}-1}+1} = 
\stackrel{03}{(+i)} \stackrel{12}{[-]} 
\stackrel{56}{[-]} \dots \stackrel{d-2\,d-1}{[+]}\; \gamma^{d}\,, \nonumber\\
\cdots \quad
&& \cdots \,,\nonumber\\
\cdots \quad 
&& \cdots\nonumber\\
 {}^{II}{\bf {\cal A}}^{1 \dagger}_{1} =\stackrel{03}{(-i)}\stackrel{12}{(+)} \stackrel{56}{(+)} 
 \cdots \stackrel{d-2 \, d-1}{[+]} \,,\quad 
&& {}^{II}{\bf {\cal A}}^{1 \dagger}_{2^{{d-1}{2}-1}+1}=\stackrel{03}{[+i]}\stackrel{12}{(+)} 
\stackrel{56}{(+)}
\cdots \stackrel{d-2 \, d-1}{[+]} \gamma^{d}\,,\nonumber\\
\cdots \quad
&& \cdots \,.
\end{eqnarray}
\end{small}
The right hand side of Eq.~(\ref{allcartaneigenvecbdgen}), although anti-commuting, is resembling
the properties of the Clifford even ``basis vectors'' on the left hand side of 
Eq.~(\ref{allcartaneigenvecAdgen}), while the right-hand side of 
Eq.~(\ref{allcartaneigenvecAdgen}), although commuting, resembles the properties of the Clifford 
odd  ``basis vectors'', from the left hand side of Eq.~(\ref{allcartaneigenvecbdgen}): $\gamma^a$ 
are up to a constant the self adjoint operators, while $S^{0 d}$ transform one nilpotent into a 
projector (or one projector into a nilpotent). However, $S^{ab}$ do not change Clifford eveness 
of  ${}^{I}{\bf {\cal A}}^{m \dagger}_{f}, i=(I,II)$. 

The reader can see the  illustration with the special case for $d=(4+1)$ in Subsect.~3.2.2. of 
 Ref.~\cite{n2023MDPI}.

\vspace{1mm}

\subsection{Second quantized fermion and boson fields with internal spaces
described by Clifford ``basis vectors'' in even dimensional spaces}
\label{secondquantizedfermionsbosonsdeven}
%

\vspace{1mm}

We learn in Subsect.~\ref{basisvectors} (See also Sect.~2.3 in 
Ref.~\cite{n2023NPB}) that in even dimensional spaces the Clifford odd and 
the Clifford even ``basis vectors'', which are the superposition of the Clifford 
odd and the Clifford even products of $\gamma^a$'s, respectively, offer the 
description of the internal spaces of fermion and boson fields:\\
The Clifford odd ``basis vectors'' explain the second quantization postulates 
for fermions~(\cite{n2023NPB,nh2021RPPNP} and the references therein), the 
Clifford even ``basis vectors'' explain the second quantization postulates for 
bosons, the gauge fields of fermions.

The Clifford odd algebra offers $2^{\frac{d}{2}-1}$ ``basis vectors''
$\hat{b}^{m \dagger}_{f}$, appearing in $2^{\frac{d}{2}-1}$ families,
with the family quantum numbers determined by $\tilde{S}^{ab}= 
\frac{i}{4} \{ \tilde{\gamma}^a, \tilde{\gamma}^b\}_{-}$, which, together 
with their $2^{\frac{d}{2}-1}\times$ $2^{\frac{d}{2}-1}$ Hermitian conjugated 
partners $\hat{b}^{m}_{f}$ fulfil the postulates for the second quantized fermion 
fields, Eq.~(\ref{almostDirac}) in this paper, Eq.(26) in Ref.~\cite{nh2021RPPNP},
explaining the second quantization postulate of Dirac.

The Clifford even algebra offers $2^{\frac{d}{2}-1}\times$ $2^{\frac{d}{2}-1}$
``basis vectors'' of ${}^{I}{\hat{\cal A}}^{m \dagger}_{f}$, and the same number
of ${}^{II}{\hat{\cal A}}^{m \dagger}_{f}$, with the properties of the second
quantized boson fields manifesting the gauge fields of fermion fields described
by the Clifford odd ``basis vectors'' $\hat{b}^{m \dagger}_{f}$, as we can see
in Figs.~(\ref{FigSU3U1odd}, \ref{FigSU3U1even}). 

The first figure represents the ``basis vectors'' of fermions in 
$d=(5+1)$-dimensional space for any of the $2^{\frac{d}{2}-1}$ families, 
analysed with respect to the subgroups $SU(3)$ and $U(1)$ of the group 
$SO(5,1)$, their ``basis vectors'' are presented in Table 2 of Ref.~\cite{n2023NPB}. 
The reader can see in the figure three members of a colour charged triplet, and 
one colourless singlet, manifesting the colour part of quarks and the colourless
leptons. The same figure represent any of the families.

The second figure represents the ``basis vectors'' for anyone of the two 
corresponding Clifford even groups, ${}^{I}{\hat{\cal A}}^{m \dagger}_{f}$,
or ${}^{II}{\hat{\cal A}}^{m \dagger}_{f}$. The corresponding ``basis vectors'' 
for ${}^{I}{\hat{\cal A}}^{m \dagger}_{f}$ are presented in Table 3 of 
Ref.~\cite{n2023NPB}, the ``basis vectors'' of 
${}^{II}{\hat{\cal A}}^{m \dagger}_{f}$ can be found in 
App.~\ref{secondgroupofbosons} in Table~\ref{Cliff basis5+1even II.}. The 
reader can see in Fig.~\ref{FigSU3U1even} four ``basis vectors'' in the centre of the
coordinate system, one sextet, one triplet and one  antitriplet. These Clifford 
even ``basis vectors'' manifest the colour 
octet (the sextet and two of  four ``singlets''), the gauge fields of the Clifford 
odd triplet ``basis vectors''. The triplets and antitriplets, if applied as presented
in Eq.~(\ref{calbIIA1234gen}) on the Clifford odd ``basis vector'', representing 
the singlet, transform the singlet to one member of the triplet.\\
\begin{figure}
  \centering
   \includegraphics[width=0.45\textwidth]{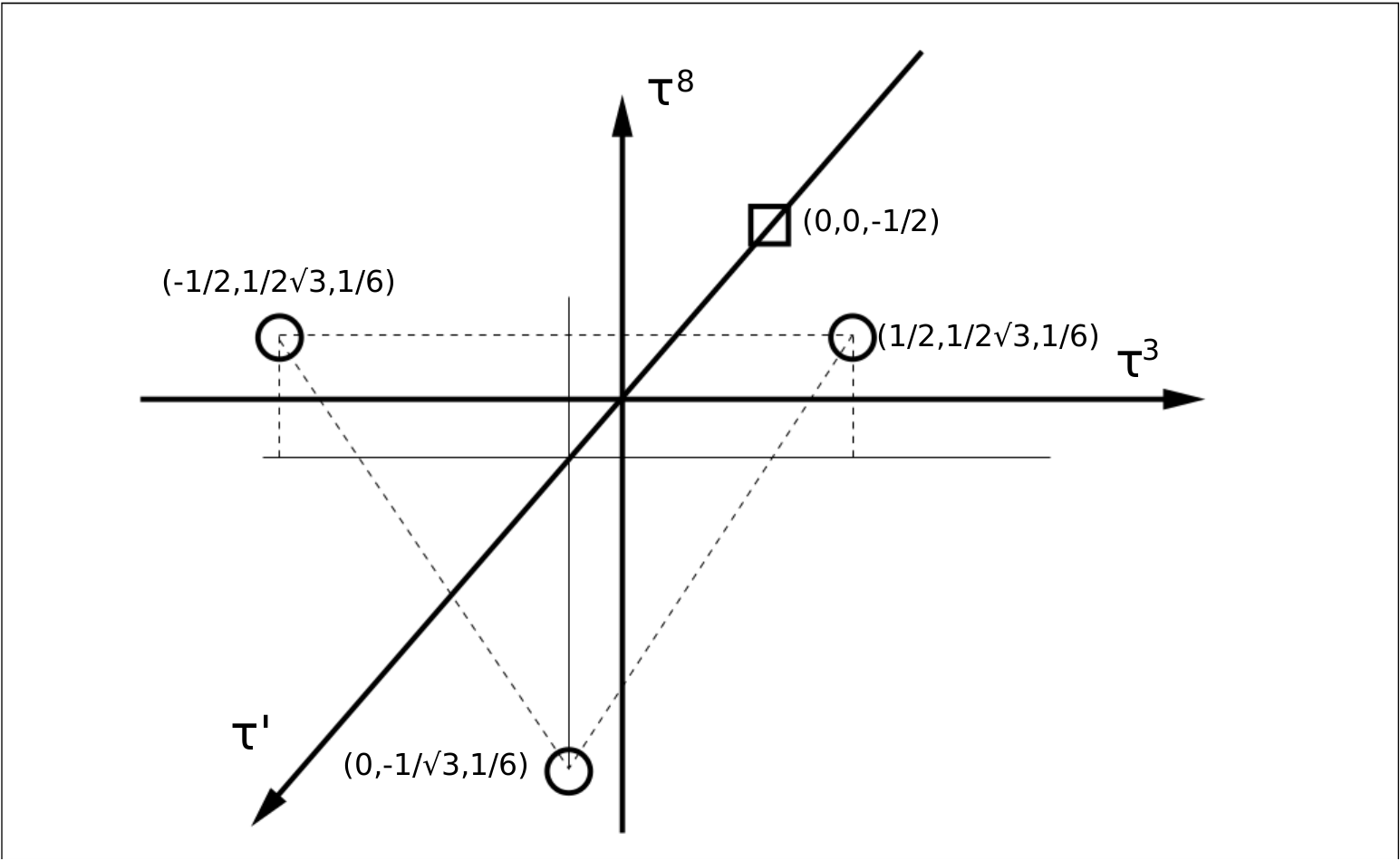}
  \caption{\label{FigSU3U1odd} 
The representations for the Clifford odd ``basis vectors'' of the subgroups 
$SU(3)$ and $U(1)$ of the group $SO(5,1)$, the properties of which appear in 
Tables 1 and 2 of Ref.~\cite{n2023NPB}, 
 are presented, taken from Ref.~\cite{n2023NPB} ($\tau^3=\frac{1}{2} 
 (- S^{12} -i S^{03})$, $\tau^8=\frac{1}{2\sqrt{3}} 
 ( S^{12} -i S^{03}- 2 S^{56})$, $\tau'=- \frac{1}{3} (S^{12} -i S^{03}
  +S^{56})$.
On the abscissa axis, on the ordinate axis and on the third axis, are the 
eigenvalues of $\tau^3$, $\tau^8$ and  $\tau'$. One notices one triplet,
denoted by ${\bf \bigcirc}$ with the values $\tau'=\frac{1}{6}$, 
($\tau^3=-\frac{1}{2}, \tau^8=\frac{1}{2\sqrt{3}}, \tau'=\frac{1}{6})$,
($\tau^3= \frac{1}{2}, \tau^8=\frac{1}{2\sqrt{3}}, \tau'=\frac{1}{6}$), 
($\tau^3=0, \tau^8=-\frac{1}{\sqrt{3}}, \tau'=\frac{1}{6}$), 
respectively, and one singlet denoted by  $\Box$
($\tau^3=0, \tau^8=0, \tau'=-\frac{1}{2}$).
The triplet and the singlet appear in four families, with the family 
quantum numbers presented in the last three columns of Table  2
of Ref.~\cite{n2023NPB}}. 
\end{figure}
%
%

%
\begin{small}
\begin{figure}
  \centering
   \includegraphics[width=0.45\textwidth]{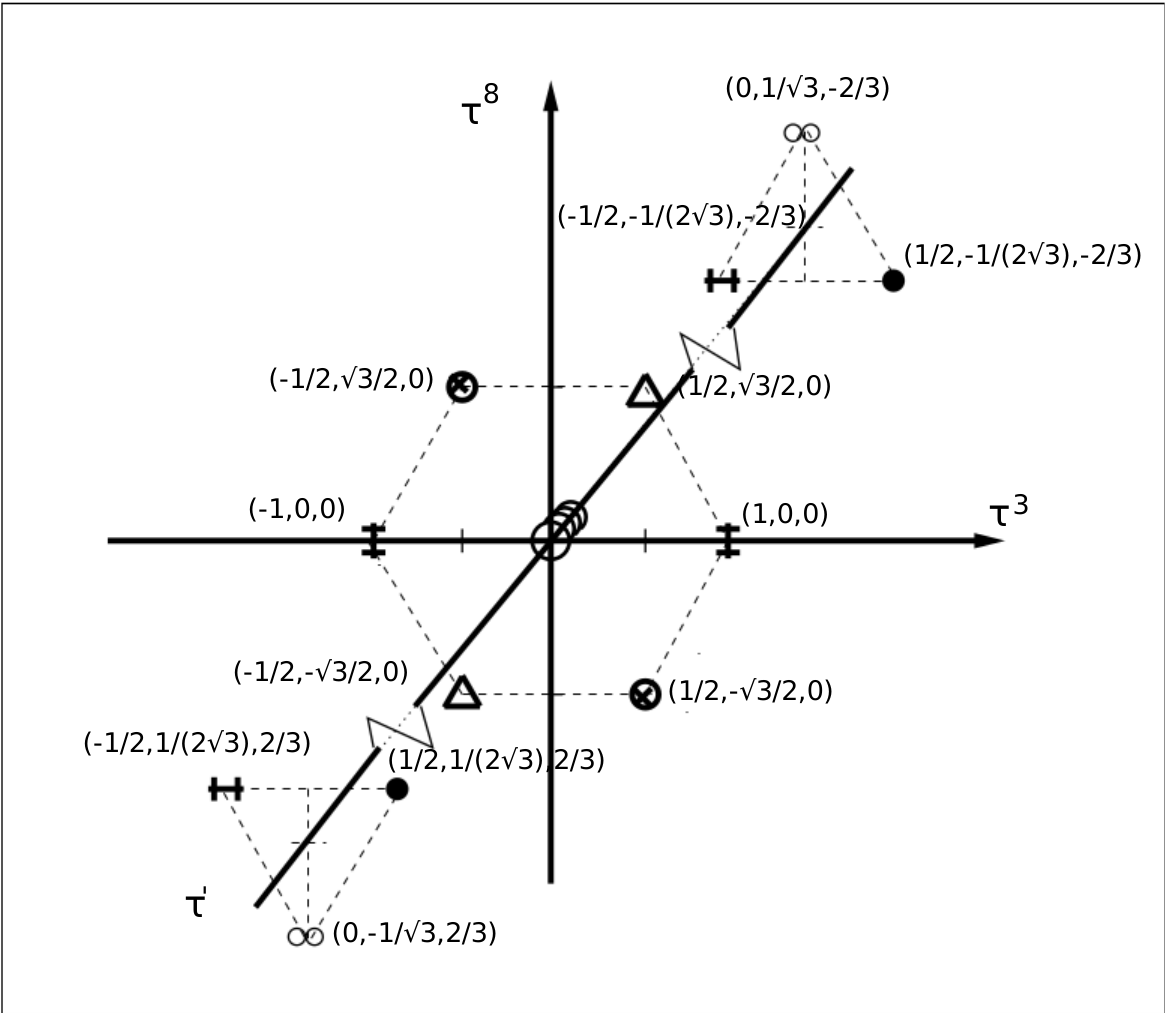}
  \caption{\label{FigSU3U1even} 
The Clifford even ''basis vectors'' ${}^{II}{\hat{\cal A}}^{m\dagger}_{f}$ in the case that
$d=(5+1)$ are presented, with respect to the eigenvalues of the commuting operators
of the subgroups $SU(3)$ and $U(1)$ of the group $SO(5,1)$,  
(${\cal \bf{\tau}}^3 =\frac{1}{2}(- {\bf {\cal S}}^{12} - 
i {\bf {\cal S}}^{03})$, ${\cal \bf{\tau}}^8 =\frac{1}{2\sqrt{3}} ({\bf {\cal S}}^{12} -i
{\bf {\cal S}}^{03}-  2 {\bf {\cal S}}^{56})$ and ${\cal \bf{\tau}}' =- \frac{1}{3} 
({\bf {\cal S}}^{12} -i {\bf {\cal S}}^{03} +  {\bf {\cal S}}^{56})$).
Their properties can be found in Table~\ref{Cliff basis5+1even II.}. The abscissa axis carries the 
eigenvalues of ${\cal \bf{\tau}}^3$, the ordinate axis carries the eigenvalues of 
${\cal \bf{\tau}}^8$, and the third axis carries the eigenvalues of ${\cal \bf{\tau}}'$.
One notices four ``singlets'' in the origin of the coordinate system with 
(${\cal \bf{\tau}}^3=0, {\cal \bf{\tau}}^8=0, {\cal \bf{\tau}}'=0$),
denoted by $\bigcirc$, representing four self adjoint Clifford even ``basis vectors''
${}^{II}{\hat{\cal A}}^{m \dagger}_{f}$, with ($f=1, m=4$), ($f=2, m=3$),
($f=3, m=1$), ($f=4, m=2$)\,, one sextet of three pairs, Hermitian conjugated to each 
other, with $ {\cal \bf{\tau}}'=0$, denoted by $\bigtriangleup$ 
(${}^{II}{\hat{\cal A}}^{2\dagger}_{1}$ with ($ {\cal \bf{\tau}}'=0, {\cal \bf{\tau}}^3
=-\frac{1}{2}, {\cal \bf{\tau}}^8=-\frac{3}{2\sqrt{3}}$) and 
${}^{II}{\hat{\cal A}}^{4\dagger}_{4}$ with ($ {\cal \bf{\tau}}'=0, {\cal \bf{\tau}}^3=
\frac{1}{2}, {\cal \bf{\tau}}^8=\frac{3}{2\sqrt{3}}$)), by 
$\ddagger$ (${}^{II}{\hat{\cal A}}^{3\dagger}_{1}$ with ($ {\cal \bf{\tau}}'=0, 
{\cal \bf{\tau}}^3=-1, {\cal \bf{\tau}}^8=0$) and ${}^{II}{\hat{\cal A}}^{4\dagger}_{2}$
 with $ {\cal \bf{\tau}}'=0, {\cal \bf{\tau}}^3=1, {\cal \bf{\tau}}^8=0$)), and by 
 $\otimes$ (${}^{II}{\hat{\cal A}}^{2\dagger}_{2}$ with (${\cal \bf{\tau}}'=0, 
 {\cal \bf{\tau}}^3=\frac{1}{2}, {\cal \bf{\tau}}^8=-\frac{3}{2\sqrt{3}}$) and 
 ${}^{II}{\hat{\cal A}}^{3\dagger}_{4}$ with (${\cal \bf{\tau}}'=0, {\cal \bf{\tau}}^3
 =- \frac{1}{2}, {\cal \bf{\tau}}^8=\frac{3}{2\sqrt{3}}$)), and one triplet, denoted by 
 $\star \star$  (${}^{II}{\hat{\cal A}}^{4\dagger}_{3}$ with (${\cal \bf{\tau}}'=
 \frac{2}{3}, {\cal \bf{\tau}}^3=-\frac{1}{2}, {\cal \bf{\tau}}^8=\frac{1}{2\sqrt{3}}$)),
by $\bullet$ (${}^{II}{\hat{\cal A}}^{3\dagger}_{3}$ with (${\cal \bf{\tau}}'=
\frac{2}{3}, {\cal \bf{\tau}}^3= \frac{1}{2}, {\cal \bf{\tau}}^8=\frac{1}{2\sqrt{3}}$)), 
and by $\odot \odot$ (${}^{II}{\hat{\cal A}}^{2\dagger}_{3}$ with (${\cal \bf{\tau}}'=
\frac{2}{3}, {\cal \bf{\tau}}^3=0, {\cal \bf{\tau}}^8=-\frac{1}{\sqrt{3}}$)),
as well as one antitriplet, Hermitian conjugated to triplet, denoted by $\star \star$
(${}^{II}{\hat{\cal A}}^{1 \dagger}_{1}$ with (${\cal \bf{\tau}}'=-\frac{2}{3},
{\cal \bf{\tau}}^3=-\frac{1}{2}, {\cal \bf{\tau}}^8=-\frac{1}{2\sqrt{3}}$)),
by $\bullet$ (${}^{II}{\hat{\cal A}}^{1\dagger}_{2}$ with (${\cal \bf{\tau}}'=-\frac{2}{3},
{\cal \bf{\tau}}^3= \frac{1}{2}, {\cal \bf{\tau}}^8=- \frac{1}{2\sqrt{3}}$)), and by $\odot \odot$
(${}^{II}{\hat{\cal A}}^{4 \dagger}_{1}$ with (${\cal \bf{\tau}}'=-\frac{2}{3}, {\cal \bf{\tau}}^3=0,
{\cal \bf{\tau}}^8=\frac{1}{\sqrt{3}}$)).}
\end{figure}
\end{small}
%
As we learn in Subsect.~\ref{basisvectors}, the Clifford odd and the Clifford even ``basis vectors'' are chosen to be products of
nilpotents, $\stackrel{ab}{(k)}$ (with the odd number of nilpotents if describing
fermions and the even number of nilpotents if describing bosons), and projectors,
$\stackrel{ab}{[k]}$. Nilpotents and projectors are (chosen to be) eigenvectors
of the Cartan subalgebra members of the Lorentz algebra in the internal space of
$S^{ab}$ for the Clifford odd ``basis vectors'' and of ${\bf {\cal S}}^{ab} (=
S^{ab}+ \tilde{S}^{ab}$) for the Clifford even ``basis vectors''.

\vspace{2mm}

To define the creation operators, for fermions or bosons,
besides the ``basis vectors'' defining the internal space of fermions and bosons,
the basis in ordinary space in momentum or coordinate representation is needed.
Here Ref.~\cite{nh2021RPPNP}, Subsect.~3.3 and App. J is overviewed. \\

Let us introduce the momentum part of the single-particle states. (The extended version
is presented in Ref.~\cite{nh2021RPPNP} in Subsect.~3.3 and App. J.)
\begin{eqnarray}
\label{creatorp}
|\vec{p}>&=& \hat{b}^{\dagger}_{\vec{p}} \,|\,0_{p}\,>\,,\quad
<\vec{p}\,| = <\,0_{p}\,|\,\hat{b}_{\vec{p}}\,, \nonumber\\
<\vec{p}\,|\,\vec{p}'>&=&\delta(\vec{p}-\vec{p}')=
<\,0_{p}\,|\hat{b}_{\vec{p}}\; \hat{b}^{\dagger}_{\vec{p}'} |\,0_{p}\,>\,,
\nonumber\\
&&{\rm pointing  \;out\;} \nonumber\\
<\,0_{p}\,| \hat{b}_{\vec{p'}}\, \hat{b}^{\dagger}_{\vec{p}}\,|\,0_{p}\, > &=&\delta(\vec{p'}-\vec{p})\,,
\end{eqnarray}
 with the normalization $<\,0_{p}\, |\,0_{p}\,>=1$.
While the quantized operators $\hat{\vec{p}}$ and $\hat{\vec{x}}$ commute,
$\{\hat{p}^i\,, \hat{p}^j \}_{-}=0$ and $\{\hat{x}^k\,, \hat{x}^l \}_{-}=0$,
it follows for $\{\hat{p}^i\,, \hat{x}^j \}_{-}=i \eta^{ij}$. One correspondingly
finds
\begin{small}
\begin{eqnarray}
\label{eigenvalue10}
<\vec{p}\,| \,\vec{x}>&=&<0_{\vec{p}}\,|\,\hat{b}_{\vec{p}}\;
\hat{b}^{\dagger}_{\vec{x}}
|0_{\vec{x}}\,>=(<0_{\vec{x}}\,|\,\hat{b}_{\vec{x}}\;
\hat{b}^{\dagger}_{\vec{p}} \,\,
|0_{\vec{p}}\,>)^{\dagger}\, \nonumber\\
<0_{\vec{p}}\,|\{\hat{b}^{\dagger}_{\vec{p}}\,, \,
\hat{b}^{\dagger}_{\vec{p}\,'}\}_{-}|0_{\vec{p}}\,>&=&0\,,\qquad
<0_{\vec{p}}\,|\{\hat{b}_{\vec{p}}, \,\hat{b}_{\vec{p}\,'}\}_{-}|0_{\vec{p}}\,>=0\,,\qquad
<0_{\vec{p}}\,|\{\hat{b}_{\vec{p}}, \,\hat{b}^{\dagger}_{\vec{p}\,'}\}_{-}|0_{\vec{p}}\,>=0\,,
\nonumber\\
<0_{\vec{x}}\,|\{\hat{b}^{\dagger}_{\vec{x}}, \,\hat{b}^{\dagger}_{\vec{x}\,'}\}_{-}|0_{\vec{x}}\,>&=&0\,,
\qquad
<0_{\vec{x}}\,|\{\hat{b}_{\vec{x}}, \,\hat{b}_{\vec{x}\,'}\}_{-}|0_{\vec{x}}\,>=0\,,\qquad
<0_{\vec{x}}\,|\{\hat{b}_{\vec{x}}, \,\hat{b}^{\dagger}_{\vec{x}\,'}\}_{-}|0_{\vec{x}}\,>=0\,,
\nonumber\\
<0_{\vec{p}}\,|\{\hat{b}_{\vec{p}}, \,\hat{b}^{\dagger}_{\vec{x}}\}_{-}|0_{\vec{x}}\,>&=&
e^{i \vec{p} \cdot \vec{x}} \frac{1}{\sqrt{(2 \pi)^{d-1}}}\,,\quad
<0_{\vec{x}}\,|\{\hat{b}_{\vec{x}}, \,\hat{b}^{\dagger}_{\vec{p}}\}_{-}|0_{\vec{p}}\,>=
e^{-i \vec{p} \cdot \vec{x}} \frac{1}{\sqrt{(2 \pi)^{d-1}}}\,.
\end{eqnarray}
\end{small}
The internal space of either fermion or boson fields has the finite number of ``basis
vectors'', $2^{\frac{d}{2}-1}\times 2^{\frac{d}{2}-1}$ for fermions (and the same
number of their Hermitian conjugated partners),  and twice 
$2^{\frac{d}{2}-1}\times 2^{\frac{d}{2}-1}$ for bosons. The momentum basis is
continuously infinite.\\

The creation operators for either fermions or bosons must be tensor products,
$*_{T}$, of both contributions, the ``basis vectors'' describing the internal space of
fermions or bosons and the basis in ordinary momentum or coordinate space.

The creation operators for a free massless fermion of the energy
$p^0 =|\vec{p}|$, belonging to a family $f$ and to a superposition of
family members $m$ applying on the vacuum state
$|\psi_{oc}>\,*_{T}\, |0_{\vec{p}}>$ 
can be written as~(\cite{nh2021RPPNP}, Subsect.3.3.2, and the references therein, 
\cite{n2023NPB})
\begin{eqnarray}
\label{wholespacefermions}
{\bf \hat{b}}^{s \dagger}_{f} (\vec{p}) \,&=& \,
\sum_{m} c^{sm}{}_f (\vec{p}) \,\hat{b}^{\dagger}_{\vec{p}}\,*_{T}\,
\hat{b}^{m \dagger}_{f} \, \,,
\end{eqnarray}
where the vacuum state for fermions $|\psi_{oc}>\,*_{T}\, |0_{\vec{p}}> $
includes both spaces, the internal part, Eq.(\ref{vaccliffodd0}), and the momentum
part, Eq.~(\ref{creatorp}) (in a tensor product for a starting single particle state
with zero momentum, from which one obtains the other single fermion states of the
same ''basis vector'' by the operator $\hat{b}^{\dagger}_{\vec{p}}$ which pushes
the momentum by an amount $\vec{p}$~\footnote{
The creation operators and their Hermitian conjugated annihilation
operators in the coordinate representation can be
read in~\cite{nh2021RPPNP} and the references therein:
$\hat{\bf b}^{s \dagger}_{f }(\vec{x},x^0)=
\sum_{m} \,\hat{b}^{ m \dagger}_{f} \, *_{T}\, \int_{- \infty}^{+ \infty} \,
\frac{d^{d-1}p}{(\sqrt{2 \pi})^{d-1}} \, c^{s m }{}_{f}\;
(\vec{p}) \; \hat{b}^{\dagger}_{\vec{p}}\;
e^{-i (p^0 x^0- \varepsilon \vec{p}\cdot \vec{x})}
$
~(\cite{nh2021RPPNP}, subsect. 3.3.2., Eqs.~(55,57,64) and the references therein).}.
\\
The creation operators and annihilation operators for fermion fields
fulfil the anti-commutation relations for the second quantized
fermion fields~\footnote{Let us evaluate:
 $<0_{\vec{p}}\,|
\{ \hat{\bf b}^{s' }_{f `}(\vec{p'})\,,\,
\hat{\bf b}^{s \dagger}_{f }(\vec{p}) \}_{+} \,|\psi_{oc}> |0_{\vec{p}}>=
\delta^{s s'} \delta_{f f'}\,\delta(\vec{p}' - \vec{p})\,\cdot |\psi_{oc}> =$
$ <0_{\vec{p}}\,|\hat{\bf b}^{s' }_{f `}\,\hat{\bf b}^{s \dagger}_{f }\,\hat{b}_{\vec{p}'}
\hat{b}^{\dagger}_{\vec{p}} +\, \hat{b}^{\dagger}_{\vec{p}} \hat{b}_{\vec{p}'}\,
\hat{\bf b}^{s \dagger}_{f }\, \hat{\bf b}^{s' }_{f `}\,|\psi_{oc}>|0_{\vec{p}}> =$
$<0_{\vec{p}}\,|\hat{\bf b}^{s' }_{f `}\,\hat{\bf b}^{s \dagger}_{f }\,\hat{b}_{\vec{p}'}
\hat{b}^{\dagger}_{\vec{p}} \,|\psi_{oc}> |0_{\vec{p}}>=\delta^{s s'} \delta_{f f'}\,
\delta(\vec{p}-\vec{p}')\,|\psi_{oc}>$,
since, according to Eq.~(\ref{almostDirac}), $ \hat{\bf b}^{s' }_{f ` }\,|\psi_{oc}>=0.$
}
\begin{small}
\begin{eqnarray}
<0_{\vec{p}}\,|\{ \hat{\bf b}^{s' }_{f `}(\vec{p'})\,,\,
\hat{\bf b}^{s \dagger}_{f }(\vec{p}) \}_{+} \,|\psi_{oc}> |0_{\vec{p}}>&=&
\delta^{s s'} \delta_{f f'}\,\delta(\vec{p}' - \vec{p})\,\cdot |\psi_{oc}> 
\,,\nonumber\\
\{ \hat{\bf b}^{s' }_{f `}(\vec{p'})\,,\,
\hat{\bf b}^{s}_{f }(\vec{p}) \}_{+} \,|\psi_{oc}> |0_{\vec{p}}>&=&0\, \cdot \,
|\psi_{oc}> |0_{\vec{p}}>
\,,\nonumber\\
\{ \hat{\bf b}^{s' \dagger}_{f '}(\vec{p'})\,,\,
\hat{\bf b}^{s \dagger}_{f }(\vec{p}) \}_{+}\, |\psi_{oc}> |0_{\vec{p}}>&=&0\, \cdot
\,|\psi_{oc}> |0_{\vec{p}}>
\,,\nonumber\\
\hat{\bf b}^{s \dagger}_{f }(\vec{p}) \,|\psi_{oc}> |0_{\vec{p}}>&=&
|\psi^{s}_{f}(\vec{p})>\,,\nonumber\\
\hat{\bf b}^{s}_{f }(\vec{p}) \, |\psi_{oc}> |0_{\vec{p}}>&=&0\, \cdot\,
\,|\psi_{oc}> |0_{\vec{p}}>\,, \nonumber\\
|p^0| &=&|\vec{p}|\,.
\label{Weylpp'comrel}
\end{eqnarray}
\end{small}
The creation operators $ \hat{\bf b}^{s\dagger}_{f }(\vec{p}) $ and their
Hermitian conjugated partners annihilation operators
$\hat{\bf b}^{s}_{f }(\vec{p}) $, creating and annihilating the single fermion
states, respectively, fulfil when applying the vacuum state,
$|\psi_{oc}> *_{T} |0_{\vec{p}}>$, the anti-commutation relations for the second 
quantized fermions, postulated by Dirac (Ref.~\cite{nh2021RPPNP}, Subsect.~3.3.1,
Sect.~5).~\footnote{
The anti-commutation relations of Eq.~(\ref{Weylpp'comrel}) are valid also if we
replace the vacuum state, $|\psi_{oc}>|0_{\vec{p}}>$, by the Hilbert space of the
Clifford fermions generated by the tensor products multiplication, $*_{T_{H}}$, of 
any number of the Clifford odd fermion states of all possible internal quantum
numbers and all possible momenta (that is, of any number of
$ \hat {\bf b}^{s\, \dagger}_{f} (\vec{p})$ of any
$(s,f, \vec{p})$), Ref.~(\cite{nh2021RPPNP}, Sect. 5.).}\\
%


To write the creation operators for boson fields, we must take into account that
boson gauge fields have the space index $\alpha$, describing the $\alpha$
component of the boson field in the ordinary space~\footnote{
In the {\it spin-charge-family} theory the Higgs's scalars origin
in the boson gauge fields with the vector index $(7,8)$, Ref.~(\cite{nh2021RPPNP},
Sect.~7.4.1, and the references therein).}.
We, therefore, add the space index $\alpha$ as follows~(\cite{n2023NPB} and 
references therein)
\begin{eqnarray}
\label{wholespacebosons}
{\bf {}^{i}{\hat{\cal A}}^{m \dagger}_{f \alpha}} (\vec{p}) \,&=&
\hat{b}^{\dagger}_{\vec{p}}\,*_{T}\,
{}^{i}{\cal C}^{ m}{}_{f \alpha}\, {}^{i}{\hat{\cal A}}^{m \dagger}_{f} \, \,, i=(I,II)\,.
\end{eqnarray}
We treat free massless bosons of momentum $\vec{p}$ and energy $p^0=|\vec{p}|$
and of particular ``basis vectors'' ${}^{i}{\hat{\cal A}}^{m \dagger}_{f}$'s, $i=(I.II),$ 
which are eigenvectors of all the Cartan subalgebra members~\footnote{
In the general case, the energy eigenstates of bosons are in a superposition of
${\bf {}^{i}{\hat{\cal A}}^{m \dagger}_{f}}$, for either $i=I$ or $i=II$. One example,
which uses the
superposition of the Cartan subalgebra eigenstates manifesting the $SU(3)\times U(1)$
subgroups of the group $SO(5,1)$, is presented in Fig.~\ref{FigSU3U1even}.},
${}^{i}{\cal C}^{ m}{}_{f \alpha}$ carry the space index $\alpha$ of the boson
field. Creation operators operate on the vacuum state
$|\psi_{oc_{ev}}>\,*_{T}\, |0_{\vec{p}}> $ with the internal space part
just a constant, $|\psi_{oc_{ev}}>=$ $|\,1>$, and for
a starting single boson state with zero momentum from which one obtains
the other single boson states with the same ''basis vector'' by the operators
$\hat{b}^{\dagger}_{\vec{p}}$ which push the momentum by an amount
$\vec{p}$, making also ${}^{i}{\cal C}^{ m}{}_{f \alpha}$ depending on $\vec{p}$.


For the creation operators for boson fields in a coordinate
representation one finds using Eqs.~(\ref{creatorp}, \ref{eigenvalue10})
\begin{eqnarray}
{\bf {}^{i}{\hat{\cal A}}^{m \dagger}_{f \alpha}}
(\vec{x}, x^0)& =& \int_{- \infty}^{+ \infty} \,
\frac{d^{d-1}p}{(\sqrt{2 \pi})^{d-1}} \,
{}^{i}{\hat{\cal A}}^{m \dagger}_{f \alpha} (\vec{p})\,
e^{-i (p^0 x^0- \varepsilon \vec{p}\cdot \vec{x})}|_{p^0=|\vec{p}|}\,,i=(I,II)\,.
\label{Weylbosonx}
\end{eqnarray}
\vspace{2mm}

To understand what new the Clifford algebra description of the internal space
of fermion and boson fields, Eqs.~(\ref{wholespacebosons}, \ref{Weylbosonx},
\ref{wholespacefermions}), bring to our understanding of the second quantized
fermion and boson fields and what new can we learn from this offer,
we need to relate $\sum_{ab} c^{ab} \omega_{ab \alpha}$ and
$ \sum_{m f} {}^{I}{\hat{\cal A}}^{m \dagger}_{f} \,{}^{I}{\cal C}^{m}{}_{f\alpha}$,
recognizing that ${}^{I}{\hat{\cal A}}^{m \dagger}_{f} \,{}^{I}{\cal C}^{m}{}_{f\alpha}$
are eigenstates of the Cartan subalgebra members, while $\omega_{ab \alpha}$
are not. And, equivalently, we need to relate $\sum_{ab} \tilde{c}^{ab} \tilde{\omega}_{ab \alpha}$ and
$ \sum_{m f} {}^{II}{\hat{\cal A}}^{m \dagger}_{f}\, {}^{II}{\cal C}^{m}{}_{f\alpha}$.

The gravity fields, the vielbeins and the two kinds of spin connection fields,
$f^{a}{}_{\alpha}$, $\omega_{ab \alpha}$, $\tilde{\omega}_{ab \alpha}$,
respectively, are in the {\it spin-charge-family} theory
(unifying spins, charges and families of fermions and offering not only the 
explanation for all the assumptions of the {\it standard model} but also for 
the increasing number of phenomena observed so far) the only boson fields in
$d=(13+1)$, observed in $d=(3+1)$ besides as gravity also as all the other
boson fields with the Higgs's scalars included~\cite{nd2017}.

We, therefore, need to relate: 
\begin{eqnarray}
\label{relationomegaAmf0}
\{\frac{1}{2} \sum_{ab} S^{ab}\, \omega_{ab \alpha} \}
\sum_{m } \beta^{m f}\, \hat{\bf b}^{m \dagger}_{f }(\vec{p}) &{\rm related\, \,to}&
\{ \sum_{m' f '} {}^{I}{\hat{\cal A}}^{m' \dagger}_{f '} \,
{\cal C}^{m' f '}_{\alpha} \}
\sum_{m } \beta^{m f} \, \hat{\bf b}^{m \dagger}_{f }(\vec{p}) \,, \nonumber\\
&&\forall f \,{\rm and}\,\forall \, \beta^{m f}\,, \nonumber\\
{\bf \cal S}^{cd} \,\sum_{ab} (c^{ab}{}_{mf}\, \omega_{ab \alpha}) &{\rm related\, \,to}&
{\bf \cal S}^{cd}\, ({}^{I}{\hat{\cal A}}^{m \dagger}_{f}\, {\cal C}^{m f}_{\alpha})\,, \nonumber\\
&& \forall \,(m,f), \nonumber\\
&&\forall \,\,{\rm Cartan\,\,subalgebra\, \, \, member} \,{\bf \cal S}^{cd} \,.
\end{eqnarray}
Let be repeated that ${}^{I}{\hat{\cal A}}^{m \dagger}_{f } $ are chosen to be
the eigenvectors of the Cartan subalgebra members, Eq.~(\ref{cartangrasscliff0}).
Correspondingly we can relate a particular ${}^{I}{\hat{\cal A}}^{m \dagger}_{f }
\,{}^{I}{\cal C}^{m}{}_{f \alpha}$ with such a superposition of $\omega_{ab \alpha}$'s,
which is the eigenvector with the same values of the Cartan subalgebra members as
there is a particular ${}^{I}{\hat{\cal A}}^{m \dagger}_{f } {\cal C}^{m f }_{\alpha}$.
We can do this in two ways:\\
{\bf i.} $\;\;$ Using the first relation in Eq.~(\ref{relationomegaAmf0}). On the left
hand side of this relation ${S}^{ab}$'s apply on $ \hat{b}^{m \dagger}_{f} $ part of
$ \hat{\bf b}^{m \dagger}_{f }(\vec{p}) $.
On the right hand side ${}^{I}{\hat{\cal A}}^{m \dagger}_{f }$ apply as well on the
same ``basis vector'' $ \hat{b}^{m \dagger}_{f} $. \\
{\bf ii.} $\;\;$ Using the second relation, in which ${\bf \cal S}^{cd}$ apply on
the left hand side on $\omega_{ab \alpha}$'s,
\begin{eqnarray}
\label{sonomega}
\, {\bf \cal S}^{cd} \,\sum_{ab}\, c^{ab}{}_{mf}\, \omega_{ab \alpha}
&=& \sum_{ab}\, c^{ab}{}_{mf}\, i \,(\omega_{cb \alpha} \eta^{ad}-
\omega_{db \alpha} \eta^{ac}+ \omega_{ac \alpha} \eta^{bd}-
\omega_{ad \alpha} \eta^{bc}),
\end{eqnarray}
on each $ \omega_{ab \alpha}$ separately; $c^{ab}{}_{mf}$ are constants to be
determined from the second relation, where on the right-hand side of this relation
${\bf \cal S}^{cd} (= S^{cd}+ \tilde{S}^{cd})$ apply on the ``basis vector''
${}^{I}{\hat{\cal A}}^{m \dagger}_{f }$ of the corresponding gauge field~\footnote{
The reader can find the relation of Eq.~(\ref{relationomegaAmf0}) demonstrated for the
case $d={3+1}$ in Ref.~\cite{n2022epjc} at the end of Sect.~3.}. \\

While ${}^{I}{\hat{\cal A}}^{m \dagger}_{f }\,{}^{I}{\cal C}^{m}{}_{f \alpha}$
determine the observed vector gauge fields~\cite{nd2017,nh2021RPPNP} determine 
${}^{II}{\hat{\cal A}}^{m \dagger}_{f }\,{}^{II}{\cal C}^{m}{}_{f \alpha}$ the 
observed scalar gauge fields, those which determine masses of quarks and leptons 
and antiquarks and antileptons and of weak bosons, after the 
electroweak break~(\cite{nh2021RPPNP}, Subsect.~6.2.2)~\footnote{
There are ${}^{I}{\hat{\cal A}}^{m \dagger}_{f }\,
{}^{I}{\cal C}^{m}{}_{f \alpha}$ fields, carrying the space index $\alpha =(7,8)$, 
and indices $(m,f)$ which manifest the charges $Q, Y, \tau$, Eq.~(108) of Ref.~%
\cite{nh2021RPPNP}, determining together with  
${}^{II}{\hat{\cal A}}^{m' \dagger}_{f `}\,
{}^{II}{\cal C}^{m'}{}_{f ` \alpha}$ the Higgs scalar and Yukawa couplings.}.

The fields ${}^{II}{\hat{\cal A}}^{m \dagger}_{f }\,
{}^{II}{\cal C}^{m}{}_{f \alpha}$ must correspondingly be related with the fields 
$\tilde{\omega}_{ab \alpha}$.
\\

\vspace{1mm}
\section{Short overview of some of achievements of {\it spin-charge-family} theory}
\label{achievements}
%




The {\it spin-charge-family} theory~(\cite{norma93,pikanorma2005,%
n2014matterantimatter,JMP2013,nh02,gmdn2008,gn2009,nd2017,nh2018,%
2020PartIPartII,n2023NPB,n2023MDPI}, and the references therein)
assumes in $d=(13+1)$-dimensional space a simple action,
Eq.~(\ref{wholeaction}), for massless fermions and for massless
vielbeins and two kinds of spin connection fields, with which fermions
interact. Description of the internal degrees of fermions by the odd
Clifford ``basis vectors'' $\hat{b}^{m \dagger}_{f}$,
Eq.~(\ref{allcartaneigenvec}), offer the unique
explanation of spins, charges and families from the point of view of
$d=(3+1)$, explaining the assumptions of the {\it standard model}
for fermions before the electroweak break. Defining the creation
operators for fermions as tensor products of Clifford odd ``basis
vectors'' with the basis in ordinary space, explain as well the
second quantisation postulates for fermions, assumed by
Dirac~\cite{Dirac,BetheJackiw,Weinberg}.

In the last three years~(\cite{n2023MDPI,n2023NPB}, and
references therein) the author recognised that the description of the
internal degrees of boson fields by the Clifford even ``basis vectors'',
Eq.~(\ref{allcartaneigenvecevenI}), offer the unique explanation
of spins and charges from the point of view of $d=(3+1)$ for the
vector gauge fields as assumed by the standard model.
Fig.~\ref{FigSU3U1odd} and Fig.~\ref{FigSU3U1even} in
Subsect.~\ref{basisvectors}, representing the fermion ``basis
vectors'' $\hat{b}^{m \dagger}_{f}$ of the colour triplet quarks
and the colourless singlet leptons, four members of any of four
families~\footnote{Assuming $d=(5+1)$ it follows that there
are $2^{\frac{6}{2}-1}$ members in any of $2^{\frac{6}{2}-1}$ 
families~\cite{n2023NPB}.}, 
and the ``basis vectors'' of the corresponding vector gauge fields, 
${}^{I}{\cal A}^{m \dagger}_{f }\,
{}^{I}{\cal C}^{m}{}_{f \alpha} $ with 16 members, demonstrate 
the relation between the fermion ``basis vectors'' and the ``basis 
vectors'' of the vector gauge fields. Besides the vector gauge fields 
carrying the vector index ${\alpha}=(0,1,2,3)$~\footnote{
Index $\alpha$ is explained
in Subsect.~\ref{secondquantizedfermionsbosonsdeven}.},
explaining the assumed vector gauge fields of the {\it standard
model}, there are the scalar gauge fields with the space index
$\alpha \ge 5$, explaining the scalar gauge fields of the
{\it standard model} --- scalar Higgs and Yukawa couplings. 
Defining the creation operators for bosons as tensor products
of the Clifford even ``basis vectors'' ,
${}^{i}{\cal A}^{m \dagger}_{f}\,, i=(I,II)$,
Eq.~(\ref{allcartaneigenvecevenI}), and the basis in ordinary
space, explain the second quantisation postulates for bosons.

\vspace{2mm}

{\it Extending the point second quantised fields for fermion and
their boson gauge fields to strings moving in ordinary space
promises to understand gravity as the second quantised
field.}

The presentation above introduces the next step, in
which the point second quantised fermion and boson fields are
extended by strings to assure the renormalisability of fermion
and boson second quantised fields. This next step is just starting.
\vspace{2mm}

The action in Eq.~(\ref{wholeaction}) includes besides 
vielbeins $f^{\alpha}_{a}$ two kinds of the spin connection fields,
$\omega_{ab\alpha}$ and $\tilde{\omega}_{ab \alpha}$.

It is shown in Ref.~(\cite{nd2017}, Eq.~(21)) that the vector gauge 
fields with the space index $\alpha =(0,1,2,3)$ are expressible by 
the spin connection fields $\omega_{ab\alpha}$, while spin 
connection fields can be expressed by vielbein fields $f^{\alpha}_a$
and the fermion fields, Eq.~(4) of Ref.~\cite{nd2017}~\footnote{
Varying the action of Eq.~(\ref{wholeaction})  with respect to the spin 
connection fields, the expression for the spin connection fields 
$\omega_{ab}{}^e$ follows
\begin{eqnarray}
\label{omegaabe}
\omega_{ab}{}^{e} &=& 
 \frac{1}{2E} \{   e^{e}{}_{\alpha}\,\partial_\beta(Ef^{\alpha}{}_{[a} f^\beta{}_{b]} )
      - e_{a\alpha}\,\partial_\beta(Ef^{\alpha}{}_{[b}f^{\beta e]})
{} - e_{b\alpha} \partial_\beta (Ef^{\alpha [e} f^\beta{}_{a]})\}
                     \nonumber\\
                  &+& \frac{1}{4}   \{\bar{\Psi} (\gamma^e \,S_{ab} - 
 \gamma_{[a}  S_{b]}{}^{e} )\Psi \}  \nonumber\\
                  &-& \frac{1}{d-2}  
   \{ \delta^e_{a} [
\frac{1}{E}\,e^d{}_{\alpha} \partial_{\beta}
             (Ef^{\alpha}{}_{[d}f^{\beta}{}_{b]})
                        + \bar{\Psi} \gamma_d  S^{d}{}_{b} \,\Psi ] 
{}  - \delta^{e}_{b} [\frac{1}{E} e^{d}{}_{\alpha} \partial_{\beta}
             (Ef^{\alpha}{}_{[d}f^{\beta}{}_{a]} )
            + \bar{\Psi} \gamma_{d}  S^{d}{}_{a}\, \Psi ]\}\,. 
                        \end{eqnarray}
If replacing $S^{ab}$ in Eq.~(\ref{omegaabe}) with $\tilde{S}^{ab}$, the  
expression for the spin connection fields  $\tilde{\omega}_{ab}{}^{e}$ follows.
}.
Eq.~(20) of Ref.~\cite{nd2017}) relates for a particular symmetry 
in $d\ge 5$ vielbeins with vector gauge fields.

Eq.~(\ref{relationomegaAmf0}) relates 
${}^{I}{\cal A}^{m \dagger}_{f} {}^{I}{\cal C}^{m}_{f \alpha}$  and 
${}^{II}{\cal A}^{m \dagger}_{f} {}^{II}{\cal C}^{m}_{f \alpha}$ with
$\omega_{ab\alpha}$ and $\tilde{\omega}_{ab \alpha}$, respectively, 
from where it follows 
that vielbeins are expressible by ${}^{I}{\cal A}^{m \dagger}_{f} 
{}^{I}{\cal C}^{m}_{f \alpha}$  and ${}^{II}{\cal A}^{m \dagger}_{f} 
{}^{II}{\cal C}^{m}_{f \alpha}$.

Description of the internal spaces of boson gauge fields with the Clifford 
even ''basis vectors'' suggest to replace $\omega_{ab\alpha}$ and 
$\tilde{\omega}_{ab \alpha}$ in Eq.~(\ref{wholeaction}) with 
${}^{I}{\cal A}^{m \dagger}_{f} {}^{I}{\cal C}^{m}_{f \alpha}$  and 
${}^{II}{\cal A}^{m \dagger}_{f} {}^{II}{\cal C}^{m}_{f \alpha}$.

In the case that there are no fermion fields present $\omega_{ab\alpha}$ 
and $\tilde{\omega}_{ab \alpha}$ are uniquely described by vielbeins, 
as we can see in Eq.~(\ref{omegaabe}) (after neglecting terms determined 
by fermion fields $\psi$).  In this case $\omega_{ab\alpha}$ is equal to
$\tilde{\omega}_{ab \alpha}$~\footnote{This demonstrates that there is
the relation between the coordinate dependence of   
${}^{I}{\cal A}^{m \dagger}_{f} {}^{I}{\cal C}^{m}_{f \alpha}$  and 
 ${}^{II}{\cal A}^{m \dagger}_{f} {}^{II}{\cal C}^{m}_{f \alpha}$, 
 although their ``basis vectors'' remain different.}.

 \vspace{2mm}
 
Most of the work in the {\it spin-charge-family} theory is done so far by using
the action of Eq.~(\ref{wholeaction}) with fermion fields described by tensor
products of the Clifford odd ``basis vectors'' and basis in ordinary space
applying on the vacuum state and in interaction with only gravity, the
vielbeins and two kinds of spin connection fields presented in this action.
The review of achievements so far is presented in Ref.~\cite{nh2021RPPNP}.

To manifest that the assumed simple starting action, Eq.~(\ref{wholeaction}),
in $d =(13+1)$, together with the description of the internal space of
fermions with the Clifford odd ``basis vectors'', offer the explanation of all
the assumptions of the {\it standard model} for quarks and leptons and
antiquarks and antileptons, their vector gauge fields and scalar gauge fields
(scalar Higgs and Yukawa couplings), for dark matter, for matter/antimatter
asymmetry in the universe, making several predictions, some of the
achievements of the {\it spin-charge-family} theory are very
shortly presented in what follows.

The reader can find more in Ref.~\cite{nh2021RPPNP} and the
references presented in this reference.

\vspace{2mm}

Let us present a few of the achievements.\\


\vspace{2mm}

{\bf A.} {\it Each irreducible representation of the Lorentz group} $SO(13,1)$, 
$S^{ab}$, analysed with respect to the {\it standard model} groups, 
$SO(3,1), SU(2)_I, SU(2)_{II}, SU(3), U{1}$  as presented in 
Table~\ref{Table so13+1.}, includes quarks and leptons (with the right 
handed neutrino included) and antiquarks and antileptons related to 
handedness as required by the {\it standard model}.
(The $SO(10)$ unifying theories must relate charges and handedness "by 
hand".)

{\bf A.i.}
{\it The irreducible representations of $SO(13,1)$} represent families of 
fermions, with quantum numbers determined by $\tilde{S}^{ab}$,
Eqs.~(\ref{cartangrasscliff0}, \ref{signature0}).
\\

{\bf B.} {\it Fermions interact in $d=(13 +1)$ with the gravity only}, 
manifesting in $d=(3+1)$ all the {\it observed vector and scalar gauge 
fields, as well as gravity}, Ref.~\cite{nd2017}. 

{\bf B.i.}
Gravity is represented by the {\it vielbeins} (the gauge fields of momenta) 
and the {\it two kinds of the spin connection fields} (the gauge fields of 
$S^{ab}$ and $\tilde{S}^{ab}$). 

{\bf B.ii.}
Eq.~(\ref{faction}), Subsect.~6.1 of Ref.~\cite{nh2021RPPNP}, represents 
the interaction of massless fermions with the vector gauge fields (first line 
in Eq.~(\ref{faction})), and with the scalar gauge fields carrying the space 
index $(7,8)$, the space index $(7,8)$ determines the weak charge, 
$SU(2)_I$, and the hyper charge of scalar fields~(\cite{nh2021RPPNP},
Eq.~(109,110,111)),  (second line of in Eq.~(\ref{faction})), before these 
scalar fields gain non zero vacuum expectation values causing the
electroweak break. The last line determines scalar triplets with respect
to the space index $(9,10,11,12,13,14)$ offering the explanation for 
matter/antimatter asymmetry in our universe~\cite{n2014matterantimatter}
\begin{eqnarray}
\label{faction}
{\mathcal L}_f &=&  \bar{\psi}\gamma^{m} (p_{m}- \sum_{A,i}\; g^{Ai}\tau^{Ai} 
A^{Ai}_{m}) \psi + \nonumber\\
               & &  \{ \sum_{s=7,8}\;  \bar{\psi} \gamma^{s} p_{0s} \; \psi \} +
 \nonumber\\ 
& & \{ \sum_{t=5,6,9,\dots, 14}\;  \bar{\psi} \gamma^{t} p_{0t} \; \psi \}
\,, 
\end{eqnarray}
where $p_{0s} =  p_{s}  - \frac{1}{2}  S^{s' s"} \omega_{s' s" s} - 
                    \frac{1}{2}  \tilde{S}^{ab}   \tilde{\omega}_{ab s}$, 
$p_{0t}   =    p_{t}  - \frac{1}{2}  S^{t' t"} \omega_{t' t" t} - 
                    \frac{1}{2}  \tilde{S}^{ab}   \tilde{\omega}_{ab t}$,                    
with $ m \in (0,1,2,3)$, $s \in (7,8),\, (s',s") \in (5,6,7,8)$, $(a,b)$ 
(appearing in $\tilde{S}^{ab}$) run within  either $ (0,1,2,3)$ or 
$ (5,6,7,8)$, $t$ runs  $ \in (5,\dots,14)$, $(t',t")$ run either 
$ \in  (5,6,7,8)$ or $\in (9,10,\dots,14)$. 
The appearance of the scalar condensate (so far just assumed, Sect.~6
in Ref.~\cite{nh2021RPPNP}) breaks the manifold ${\cal M}^{13,1}$ 
to ${\cal M}^{7,	1}\times {\cal M}^{6}$, brings masses of the scale 
$\propto 10^{16}$ GeV or higher to all the vector and scalar gauge 
fields, which interact with the condensate~\cite{n2014matterantimatter}, 
{\it leaving the weak, colour, electromagnetic and gravitational fields 
massless}. Fermions $\psi$ correspondingly appear in 
$2^{\frac{7+1}{2}-1}=8$ massless families, divided in two groups, 
each group interacting with different scalar fields as discussed in 
Subsect.~6.2.2 in Ref.~\cite{nh2021RPPNP}, remaining massless up 
to the electroweak break.

Table 8 and Eqs.~(108,109,110) of~\cite{nh2021RPPNP} point out 
the properties of two groups of four families of quarks and leptons, 
explaining that both groups of four families have the same symmetry 
of $4 \times 4$ mass matrices for quarks and leptons.

The {\it spin-charge-family} theory obviously predicts at observable energies
two groups of four families. To the lower four families the observed three
families belong.
The stable (at low energies) of the upper four  families of quarks, clustered
into neutrons, contributes to the dark matter.



How strong is the  influence of scalar fields on the masses of quarks and leptons,
depends on the coupling constants  and the masses of the scalar fields.
The {\it spn-charge-family} predicts  that in both groups of four families, 
the  mass matrices $4 \times 4$ have the symmetry 
$SU(2)\times SU(2)\times U(1)$ of the form~\footnote{
The symmetry $SU(2)\times SU(2)\times U(1)$  of the mass matrices, 
Eq.~(\ref{M0}), is expected to remain in all loop corrections~\cite{NA2017}.}
\begin{small}
 \begin{equation}
 \label{M0}
 {\cal M}^{\alpha} = \begin{pmatrix} 
 - a_1 - a & e     &   d & b\\ 
 e*     & - a_2 - a &   b & d\\ 
 d*     & b*     & a_2 - a & e\\
 b*     &  d*    & e*   & a_1 - a
 \end{pmatrix}^{\alpha}\,,
 \end{equation}
 \end{small}
with $\alpha$ representing family members --- quarks and leptons~%
\cite{mdn2006,gmdn2007,gmdn2008,gn2013,gn2014,NA2017}. 

The symmetry of mass matrices allows to calculate properties of the
fourth family from the known masses of the observed three families and
from the mixing matrices of quarks and leptons: Knowing the values
of the $3 \times 3$ submatrix of the unitary $4 \times 4$ matrix allows
 to calculate all the remaining matrix elements of the $4 \times 4$ 
matrix. The masses of the lower three families do not make a  problem.
There are measured elements of the $3 \times 3$ submatrix of the 
unitary $4 \times 4$ mixing matrix which are even for quarks 
far to be known accurately enough to allow prediction of the masses 
of the fourth family of quarks and correspondingly to calculate the 
rest elements of the mixing matrix.
Fitting the experimental data (and the meson decays evaluations in 
the literature,  as well as our own evaluations) the authors of the 
paper~\cite{gn2014} very roughly  estimate that the fourth family quarks 
masses might be pretty above $1$ TeV. 


Since the matrix elements of the $3 \times 3$ submatrix of the $4 \times 4$ 
mixing matrix depend weakly on the fourth family masses, the 
calculated mixing matrix
offers the prediction to what values will more accurate measurements move
the present experimental data and also the fourth family mixing matrix 
elements in dependence of the fourth family masses, Eq.~(\ref{vudoldexp}): 


In Eq.~(\ref{vudoldexp}) the matrix elements, taken from Ref.~\cite{gn2014}, 
of the $4\times 4$ mixing matrix for quarks obtained when the $4\times 4$ 
mass matrices respect the symmetry of Eq.~(\ref{M0}) while the parameters 
of the mass matrices are fitted to the  ($exp$) experimental data~%
\cite{datanew} are presented for two choices of the fourth family quark 
masses:  $m_{u_4}= m_{d_4}=700$ GeV  ($scf_{1}$) and 
$m_{u_4}= m_{d_4}=$ $1\,200$ GeV ($scf_{2}$). 
In parentheses, $(\;)$ and $[\;\,]$, the  changes of the matrix elements are 
presented, which are due to the changes of the top mass
within the experimental inaccuracies: with the $m_{t} =$ $(172 + 3\times 0.76)$ 
GeV and $m_{t} =$ $(172 - 3\times 0.76)$, respectively  (if there are one, 
two or more numbers in parentheses the last one or more numbers 
are different, if there is no parentheses no numbers are different) .


\vspace{0.1cm}

\begin{small}
\begin{equation}
\label{vudoldexp}
      |V_{(ud)}|= \begin{pmatrix}
    %
     exp  &    0.97425 \pm 0.00022    &  0.2253 \pm 0.0008 
&  0.00413 \pm 0.00049&   \\
     \hline
     scf_1  &    0.97423(4)            &  0.22539(7)          
&  0.00299  &     0.00776(1)\\  
      scf_2  &    0.97423[5]            &  0.22538[42]        &  0.00299  
&  0.00793[466]\\ 
     \hline 
     exp  &  0.225   \pm 0.008      &  0.986  \pm 0.016     
&  0.0411  \pm 0.0013&   \\
    \hline
    scf_1  &  0.22534(3)       &  0.97335              &  0.04245(6) 
&   0.00349(60) \\  
    scf_2  &  0.22531[5]       &  0.97336[5]          &  0.04248     
&   0.00002[216] \\ 
    \hline
    exp  &  0.0084  \pm 0.0006     &  0.0400 \pm 0.0027    
&  1.021   \pm 0.032&     \\
   \hline
    scf_1  &  0.00667(6)            &  0.04203(4)         &  0.99909   
&     0.00038\\  
     scf_2  &  0.00667                &  0.04206[5]         &  0.99909   
&     0.00024[21] \\
   \hline
    scf_1   & 0.00677(60) & 0.00517(26)    & 0.00020    & 0.99996\\
   scf_2   & 0.00773      & 0.00178   & 0.00022  & 0.99997[9]
     \end{pmatrix}\,.
     \end{equation}
\end{small}
Let us conclude that according to Ref.~\cite{gn2014} the masses of the  fourth 
family lie  much above the known three.
%
The larger are  masses of the fourth family the larger are $V_{u_1 d_4}$
in comparison with $V_{u_1 d_3}$ and the more is valid that 
$V_{u_2 d_4} <V_{u_1 d_4}$, $V_{u_3 d_4}<V_{u_1 d_4}$. 
The flavour changing neutral currents are correspondingly weaker. 

Although the results of Ref.~\cite{gn2014} are old and that new evaluations
are needed, the accuracy of the measured mixing matrices for quarks has not 
improve in the meantime enough to predict masses of the fourth families of quarks.\\

\vspace{2mm}


 The {\it spin-charge-family} {\it theory 
predicts the existence} of besides the lower group of four families of 
quarks and leptons and antiquarks and antileptons also {\it the upper group 
of four families}; quarks and leptons carry the same charges
as the lower group of four families, the members of one family of which
are presented Table~\ref{Table so13+1.}. There are, however, different
scalar fields with the space index $(7,8)$, which determine mass matrices
of the upper four families, although demonstrating the same
$\widetilde{SU}(2)\times \widetilde{SU}(2)\times U(1)$ symmetry,
discussed in Subsect.~6.2.2 of Ref.~\cite{nh2021RPPNP}, in particular in
Eq.~(108,111).

Different scalar fields are responsible for much higher masses of quarks
and leptons than those of the lower four families. Correspondingly, the
"nuclear" force among the baryons and mesons of these quarks and
antiquarks differ a lot from the nuclear force of the baryons and mesons
of the lower four families~\footnote{
In Ref.~\cite{nm2015} the weak and "nuclear"   scattering  of such 
very  heavy baryons by ordinary nucleons is studied, showing that the 
cross-section for such scattering is very small and therefore consistent 
with the observation of experiments so far, provided that the quark mass
of this  baryon is about 100 TeV or above.}.

The stable of the upper four families offers an explanation for the 
appearance of the {\it dark matter} in our universe~\footnote{
In Ref.~\cite{gn2009} a simple hydrogen-like model is used to evaluate 
properties of baryons of these heavy quarks, with one gluon  exchange
determining the force among the constituents of the fifth family baryons.
The weak force and the electromagnetic force start to be at small 
distances due to heavy masses of quarks of the same order of 
magnitude as the colour force.}.

A rough estimation of properties of baryons of the stable fifth family
members, of their behaviour during the evolution of the universe and when
scattering on the ordinary matter, as well as a study of possible limitations
on the family properties due to the cosmological and direct experimental
evidences are done in Ref.~\cite{gn2009}.


The authors of Ref.~\cite{gn2009} study the freeze out procedure of 
the fifth family quarks and anti-quarks and the formation of  baryons and 
anti-baryons up to the temperature  $ k_b T= 1$ GeV,  when the colour 
phase transition starts which depletes almost all the fifth family quarks 
and antiquarks, while the colourless fifth family neutrons with very small 
scattering cross section decouples long before (at $ k_b T= 100$ GeV), 
Fig.~\ref{DiagramI.}.
\begin{figure}[h]
\begin{center}
\includegraphics[width=15cm,angle=0]{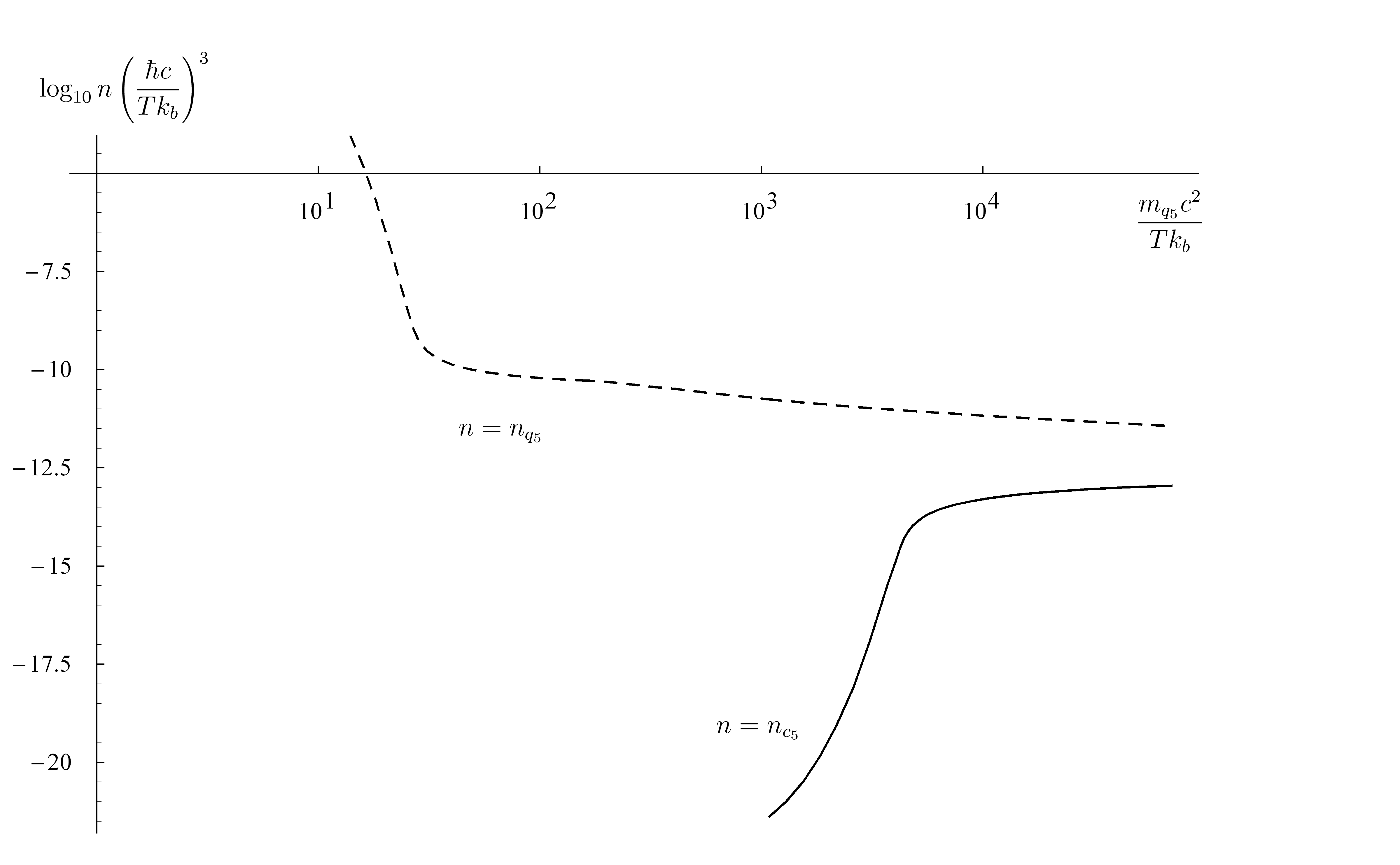}
\caption{The dependence of the two number densities, $n_{q_5}$ of 
the fifth family  quarks and $n_{c_5}$ of the fifth family clusters of 
quarks, as functions of $\frac{m_{q_5} \, c^2}{ k_b \, T}$ is presented 
for the special values $m_{q_5} = 71 \,{\rm TeV}$. The estimated 
scattering cross sections, entering into Boltzmann equation, are 
presented in Ref.~\cite{gn2009}, Eqs.~(2,3,4,5). In the treated energy 
(temperature $ k_b T$) interval the one gluon exchange gives the  main 
contribution to the scattering cross sections entering into the Boltzmann 
equations for $n_{q_5}$ and $n_{c_5}$. The figure is taken from 
Ref.~\cite{gn2009}.
}
\end{center}
\label{DiagramI.}
\end{figure}
The cosmological evolution 
suggests for the mass limits the range $10$ TeV $< m_{q_5}  < 
{\rm a \, few} \cdot 10^2$ TeV 
and for the  scattering cross sections 
$ 10^{-8}\, {\rm fm}^2\, < \sigma_{c_5}\, <   10^{-6} \,{\rm fm}^2  $. 
The measured density of  the  dark matter 
does not put much limitation on the properties of heavy enough clusters %
\footnote{   
In the case that the weak interaction determines the  cross section of  the 
neutron $n_5$, the interval for the  fifth family quarks would be 
$10\; {\rm TeV} < m_{q_5} \, c^2< 10^5$ TeV. }.   

\vspace{2mm}

The DAMA/LIBRA experiments~\cite{Rita2022} limit (provided that they 
measure the heavy fifth family clusters) the quark mass in the interval:
 $ 200 \,{\rm TeV} < m_{q_{5}} < 10^5\, {\rm TeV}$, Ref.~\cite{gn2009}.
 
Masses of the stable fifth family of quarks and leptons are much above 
the fourth family members~\footnote{
Although the upper four families carry the weak (of two kinds) and the 
colour charge, these group of four families are completely decoupled from 
the lower four families up to $<10^{16}$  GeV, unless the breaks 
of symmetries recover.}.

\vspace{2mm}

The masses of quarks and leptons of two groups of four families are 
spread from $10^{-11}$GeV ($\nu$ of the first family) to $10^{15}$GeV 
($u$ and $d$ of the fourth family of the upper four families). \\ 

\vspace{2mm}

{\it There are additional scalar fields} in the {\it spin-charge-family} 
theory, carrying the space index $s=(9, 10, 11, 12, $ $13, 14)$: 
The reader can find the properties of these scalar fields in 
Eqs.~(113,114)  and Table 9 in Subsect.~6.2.2 of 
Ref.~\cite{nh2021RPPNP}. They are triplets or antitriplets with respect 
to the space index $s$, carrying additional quantum numbers in adjoint 
representations determined by $S^{ab}$ or 
$\tilde{S}^{ab}$,  causing transitions of antileptons into quarks and 
back and leptons into antiquarks and back, what might be responsible 
in the expanding universe for the matter/antimatter asymmetry and 
also for the proton decay~\cite{n2014matterantimatter}.

\vspace{2mm}

If the antiquark $ \bar{u}_{L}^{\bar{c2}}$, from the line $43$ 
presented in Table~\ref{Table so13+1.},  with the "fermion" charge 
$\tau^{4}=-\frac{1}{6}$, the weak charge $\tau^{13} =0$, the 
second $SU(2)_{II}$ charge $\tau^{23} =-\frac{1}{2}$, the colour 
charge $(\tau^{33},\tau^{38})=(\frac{1}{2},-\frac{1}{2\sqrt{3}})$, 
the hyper charge $Y(=\tau^{4}+\tau^{23})=$ $-\frac{2}{3}$ 
and the electromagnetic charge $Q (\,=Y + \tau^{13})=$
$ -\frac{2}{3}$ emits the scalar field 
$A^{2 \sminus}_{\scriptscriptstyle{\stackrel{9\,10}{(\oplus)}}} $
with $\tau^{4}=2\times(-\frac{1}{6})$, $\tau^{13} =0$, $\tau^{23} 
=-1$, $(\tau^{33},\tau^{38})=(\frac{1}{2},\frac{1}{2\sqrt{3}})$, 
$Y=-\frac{4}{3}$ and $Q= -\frac{4}{3}$, 
it transforms into $u_{R}^{c3}$ from the line $17$ of 
Table~\ref{Table so13+1.}, carrying the quantum numbers 
$\tau^{4}=\frac{1}{6}$, $\tau^{13} =0$, 
$\tau^{23} =\frac{1}{2}$, $(\tau^{33},\tau^{38})=
(0,-\frac{1}{\sqrt{3}})$, $Y=\frac{2}{3}$ and $Q=\frac{2}{3} $. 
If this scalar field 
$A^{2 \sminus}_{\scriptscriptstyle{\stackrel{9\,10}{(\oplus)}}} $
is absorbed by the colourless antielectron, $\bar{e}_{L}$, 
presented in Table~\ref{Table so13+1.} in the line $57$, carrying 
the "fermion" charge $\tau^{4}=\frac{1}{2}$, the weak charge 
$\tau^{13} =0$, the second $SU(2)_{II}$ charge $\tau^{23} =
\frac{1}{2}$, $Y=1, Q=1$, this antielectron $\bar{e}_{L}$
transforms into $ \bar{d}_{R}^{\bar{c1}}$ quark from the line $3$ 
in Table~\ref{Table so13+1.},  with the "fermion" charge 
$\frac{1}{6}$, the weak charge $\tau^{13} =0$, the 
second $SU(2)_{II}$ charge $\tau^{23} =-\frac{1}{2}$, the colour 
charge $(\tau^{33},\tau^{38})=(\frac{1}{2},\frac{1}{2\sqrt{3}})$, 
the hyper charge $Y=-\frac{1}{3}$ and the electromagnetic charge 
$Q = -\frac{1}{3}$.

These two quarks, $d_{R}^{c1} $ and $u_{R}^{c3}$ can bind 
together with $u_{R}^{c2}$ from the $9^{th}$ line of the same 
table (at low enough energy, after the electroweak transition), 
into the colour chargeless baryon - a proton.
This transition is presented in Fig.~\ref{proton is born1.}.

The opposite transition at low energies would make the proton decay.
\begin{figure}
\includegraphics{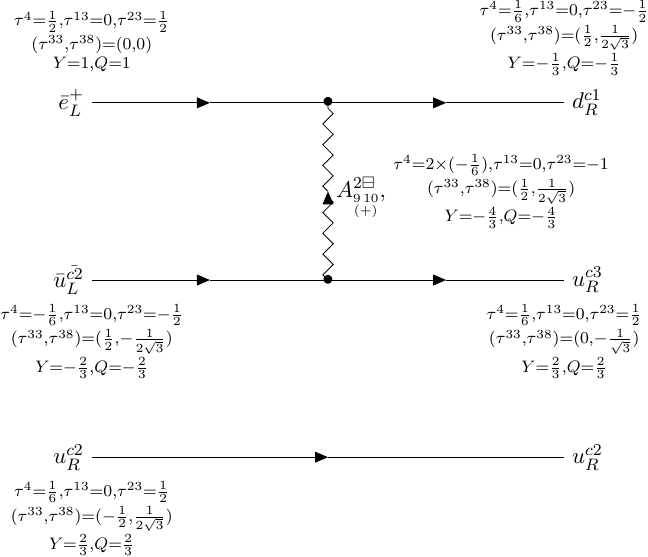}
\centering
\caption{\label{proton is born1.} The birth of a "right handed proton" 
out of an positron%
~$\bar{e}^{\,\,+}_{L}$, 
antiquark $\bar{u}_L^{\bar{c2}}$ and quark (spectator) $u_{R}^{c2}$.  
This can happen for any of the members of the lower four families, 
presented in Table~\ref{Table III.}.}
\end{figure}
\vspace{2mm}

Table~\ref{Table III.} presents the two groups of four families of ``basis vectors''
for two of the members of quarks and leptons and antiquarks and antileptons,
 presented in Table~\ref{Table so13+1.}, that is for $\hat{u}^{c1 \dagger}_{R}$ --- 
the right handed $u$-quark with spin $\frac{1}{2}$ and the colour charge 
$(\tau^{33}=1/2$, $\tau^{38}=1/(2\sqrt{3}))$, appearing in the first line of 
Table~\ref{Table so13+1.} --- and of  the colourless right handed neutrino 
$\hat{\nu}^{\dagger}_{R}$ of spin $\frac{1}{2}$, appearing 
in the $25^{th}$ line of Table~\ref{Table so13+1.}. The reader can notice
that the  part $SO(7,1)$ of ``basis vectors''   are completely the same for any of 
the eight families for quarks and neutrinos (leptons indeed). They differ in the 
$SU(3)$ and $U(1)$ part of the $SO(13,1)$. 

%
\begin{table}
 \begin{center}
   \begin{tiny}
\begin{minipage}[t]{16.5 cm}
\caption{Eight families of the ''basis vectors'', 
$\hat{u}^{c1 \dagger}_{R}$ --- the right 
handed $u$-quark with spin $\frac{1}{2}$ and the colour charge $(\tau^{33}=1/2$, 
$\tau^{38}=1/(2\sqrt{3}))$, appearing in the first line of Table~\ref{Table so13+1.} --- 
and of  the colourless right handed neutrino $\hat{\nu}^{\dagger}_{R}$ 
of spin $\frac{1}{2}$ , appearing 
in the $25^{th}$ line of Table~\ref{Table so13+1.} ---   
are presented in the  left and in the right part of this table, respectively. Table is taken 
from~\cite{normaJMP2015}. 
Families belong to two groups of four families, the group ($I$) is a doublet with respect to 
($\vec{\tilde{N}}_{L}$ and  $\vec{\tilde{\tau}}^{(1)}$) and  a singlet with respect 
to ($\vec{\tilde{N}}_{R}$ and  $\vec{\tilde{\tau}}^{(2)}$), Eqs.~(85,86) in 
Subsect.~4.3.2 of Ref.~\cite{nh2021RPPNP}), 
the other ($II$) is a singlet with respect to ($\vec{\tilde{N}}_{L}$ and  
$\vec{\tilde{\tau}}^{(1)}$) and  a doublet with respect to 
($\vec{\tilde{N}}_{R}$ and  $\vec{\tilde{\tau}}^{(2)}$), Eqs.~(85,86), in 
Subsect.~4.3.2 of Ref.~\cite{nh2021RPPNP}).
All the families follow from the starting one by the application of the operators 
($\tilde{N}^{\pm}_{R,L}$, $\tilde{\tau}^{(2,1)\pm}$). } 
\label{Table III.}
\end{minipage}
\end{tiny}
\end{center}
\begin{center}
\begin{tiny}
 \begin{tabular}{|r|c|c|c|c|c c c c c|}
 \hline
 &&&&&$\tilde{\tau}^{13}$&$\tilde{\tau}^{23}$&$\tilde{N}_{L}^{3}$&$\tilde{N}_{R}^{3}$&
 $\tilde{\tau}^{4}$\\
 \hline
 $I$&$\hat{u}^{c1 \dagger}_{R\,1}$&
   $ \stackrel{03}{(+i)}\,\stackrel{12}{[+]}|\stackrel{56}{[+]}\,\stackrel{78}{(+)} ||
   \stackrel{9 \;10}{(+)}\;\;\stackrel{11\;12}{[-]}\;\;\stackrel{13\;14}{[-]}$ & 
    $\hat{\nu}^{\dagger}_{R\,1}$&
   $ \stackrel{03}{(+i)}\,\stackrel{12}{[+]}|\stackrel{56}{[+]}\,\stackrel{78}{(+)} ||
   \stackrel{9 \;10}{(+)}\;\;\stackrel{11\;12}{(+)}\;\;\stackrel{13\;14}{(+)}$ 
  &$-\frac{1}{2}$&$0$&$-\frac{1}{2}$&$0$&$-\frac{1}{2}$ 
 \\
  $I$&$\hat{u}^{c1 \dagger}_{R\,2}$&
   $ \stackrel{03}{[+i]}\,\stackrel{12}{(+)}|\stackrel{56}{[+]}\,\stackrel{78}{(+)} ||
   \stackrel{9 \;10}{(+)}\;\;\stackrel{11\;12}{[-]}\;\;\stackrel{13\;14}{[-]}$ & 
   $\hat{\nu}^{\dagger}_{R\,2}$&
   $ \stackrel{03}{[+i]}\,\stackrel{12}{(+)}|\stackrel{56}{[+]}\,\stackrel{78}{(+)} ||
   \stackrel{9 \;10}{(+)}\;\;\stackrel{11\;12}{(+)}\;\;\stackrel{13\;14}{(+)}$ 
  &$-\frac{1}{2}$&$0$&$\frac{1}{2}$&$0$&$-\frac{1}{2}$
 \\
  $I$&$\hat{u}^{c1 \dagger}_{R\,3}$&
   $ \stackrel{03}{(+i)}\,\stackrel{12}{[+]}|\stackrel{56}{(+)}\,\stackrel{78}{[+]} ||
   \stackrel{9 \;10}{(+)}\;\;\stackrel{11\;12}{[-]}\;\;\stackrel{13\;14}{[-]}$ & 
    $\hat{\nu}^{\dagger}_{R\,3}$&
   $ \stackrel{03}{(+i)}\,\stackrel{12}{[+]}|\stackrel{56}{(+)}\,\stackrel{78}{[+]} ||
   \stackrel{9 \;10}{(+)}\;\;\stackrel{11\;12}{(+)}\;\;\stackrel{13\;14}{(+)}$ 
  &$\frac{1}{2}$&$0$&$-\frac{1}{2}$&$0$&$-\frac{1}{2}$
 \\
 $I$&$\hat{u}^{c1 \dagger}_{R\,4}$&
  $ \stackrel{03}{[+i]}\,\stackrel{12}{(+)}|\stackrel{56}{(+)}\,\stackrel{78}{[+]} ||
  \stackrel{9 \;10}{(+)}\;\;\stackrel{11\;12}{[-]}\;\;\stackrel{13\;14}{[-]}$ & 
   $\hat{\nu}^{\dagger}_{R\,4}$&
  $ \stackrel{03}{[+i]}\,\stackrel{12}{(+)}|\stackrel{56}{(+)}\,\stackrel{78}{[+]} ||
  \stackrel{9 \;10}{(+)}\;\;\stackrel{11\;12}{(+)}\;\;\stackrel{13\;14}{(+)}$ 
  &$\frac{1}{2}$&$0$&$\frac{1}{2}$&$0$&$-\frac{1}{2}$
  \\
  \hline
  $II$& $\hat{u}^{c1 \dagger}_{R\,5}$&
        $ \stackrel{03}{[+i]}\,\stackrel{12}{[+]}|\stackrel{56}{[+]}\,\stackrel{78}{[+]}||
        \stackrel{9 \;10}{(+)}\;\;\stackrel{11\;12}{[-]}\;\;\stackrel{13\;14}{[-]}$ & 
         $\hat{\nu}^{\dagger}_{R\,5}$&
        $ \stackrel{03}{[+i]}\,\stackrel{12}{[+]}|\stackrel{56}{[+]}\,\stackrel{78}{[+]}|| 
        \stackrel{9 \;10}{(+)}\;\;\stackrel{11\;12}{(+)}\;\;\stackrel{13\;14}{(+)}$ 
        &$0$&$-\frac{1}{2}$&$0$&$-\frac{1}{2}$&$-\frac{1}{2}$
 \\ 
  $II$& $\hat{u}^{c1 \dagger}_{R\,6}$&
      $ \stackrel{03}{(+i)}\,\stackrel{12}{(+)}|\stackrel{56}{[+]}\,\stackrel{78}{[+]}||
      \stackrel{9 \;10}{(+)}\;\;\stackrel{11\;12}{[-]}\;\;\stackrel{13\;14}{[-]}$ & 
    $\hat{\nu}^{\dagger}_{R\,6}$&
      $ \stackrel{03}{(+i)}\,\stackrel{12}{(+)}|\stackrel{56}{[+]}\,\stackrel{78}{[+]}|| 
      \stackrel{9 \;10}{(+)}\;\;\stackrel{11\;12}{(+)}\;\;\stackrel{13\;14}{(+)}$ 
      &$0$&$-\frac{1}{2}$&$0$&$\frac{1}{2}$&$-\frac{1}{2}$
 \\ 
 $II$&$\hat{u}^{c1 \dagger}_{R\,7}$&
 $ \stackrel{03}{[+i]}\,\stackrel{12}{[+]}|\stackrel{56}{(+)}\,\stackrel{78}{(+)}||
 \stackrel{9 \;10}{(+)}\;\;\stackrel{11\;12}{[-]}\;\;\stackrel{13\;14}{[-]}$ & 
      $ \hat{\nu}^{\dagger}_{R\,7}$&
      $ \stackrel{03}{[+i]}\,\stackrel{12}{[+]}|\stackrel{56}{(+)}\,\stackrel{78}{(+)}|| 
      \stackrel{9 \;10}{(+)}\;\;\stackrel{11\;12}{(+)}\;\;\stackrel{13\;14}{(+)}$ 
    &$0$&$\frac{1}{2}$&$0$&$-\frac{1}{2}$&$-\frac{1}{2}$
  \\
   $II$& $\hat{u}^{c1 \dagger}_{R\,8}$&
    $ \stackrel{03}{(+i)}\,\stackrel{12}{(+)}|\stackrel{56}{(+)}\,\stackrel{78}{(+)}||
    \stackrel{9 \;10}{(+)}\;\;\stackrel{11\;12}{[-]}\;\;\stackrel{13\;14}{[-]}$ & 
    $\hat{\nu}^{\dagger}_{R\,8}$&
    $ \stackrel{03}{(+i)}\,\stackrel{12}{(+)}|\stackrel{56}{(+)}\,\stackrel{78}{(+)}|| 
    \stackrel{9 \;10}{(+)}\;\;\stackrel{11\;12}{(+)}\;\;\stackrel{13\;14}{(+)}$ 
    &$0$&$\frac{1}{2}$&$0$&$\frac{1}{2}$&$-\frac{1}{3}$
 \\ 
 \hline 
 \end{tabular}
 \end{tiny}
 \end{center}
 \end{table}

\vspace{3mm}

{\it To reproduce the observed fermion and boson fields}  
{\it the spontaneous breaks are needed},
first from $SO(13,1)$ (and $\widetilde{SO}(13,1)$) to $SO(7,1) \times 
SU(3)\times U(1)$ (and to $\widetilde{SO}(7,1) \times 
\widetilde{SU}(3)\times \widetilde{U}(1)$)~\footnote{
The assumed break from the starting symmetry $SO(13,1)$ to 
$SO(7,1) \times SU(3)\times U(1)$ is supposed to  be caused by the 
appearance of the condensate of two right handed neutrinos with the 
family quantum numbers of the upper four families (that is of the four  
families, which do not contain the three so far observed families) at the 
energy of  $\ge 10^{16}$ GeV.
This condensate is presented in Table~6 in Subsect.~4.3.2 of 
Ref.~\cite{nh2021RPPNP}. 
The condensate  makes massive the $SU(2)_{II}$ 
vector  gauge fields and $U(1)_{\tau^4}$ vector gauge fields, as well 
as all the scalar gauge fields, leaving massless only the $SU(3)$ colour, 
$U(1)_{Y}$ and $SU(2)$ weak vector gauge fields, as well as the
gravity in $d=(3+1)$.},
and then further to  $SO(3,1) \times SU(2) \times SU(3)
\times U(1)$ (and to $\widetilde{SO}(3,1) \times \widetilde{SU}(2) \times$ 
$\widetilde{SU}(3) \times \widetilde{U}(1)$)~\footnote{
The reader can notice in Table~\ref{Table so13+1.} that the $SO(7,1)$ 
content of $SO(13,1)$ is identical for quarks and leptons and identical for 
antiquarks and antileptons. Quarks distinguish from leptons and antiquarks 
from antileptons only in the $SO(6)$ content of $SO(13,1)$, that is in the 
colour charge and in the "fermion" charge.}.
%

In Refs.~\cite{nh2008,NHD,NHDJMP,NHD2010,ND2012Disk} we study the 
toy model of $d=(5+1)$, in which the spin connection fields with the space 
index $s=(5,6)$ force the infinite surface in the fifth and the sixth dimension 
to form an almost $S^2$ sphere, keeping rotational symmetry of the 
surface around one point, while fermions in $d=(3+1)$ of particular 
handedness keep their masslessness. These happens for all the families of 
fermions of particular handedness~\cite{ND2012Disk}. 

We have not yet successfully repeated the $d=(5+1)$ toy model of the 
spontaneous "compactification" in the case of $SO(13,1)$ 
(we put compactification into quotation marks, since an almost $S^{n}$ 
sphere, as also $S^{2}$ sphere in the toy model of $d=(5+1)$, has 
the singular points in all infinities). 

\vspace{2mm}

Fermions --- quarks and leptons and antiquarks and antileptons --- remain 
massless and mass protected, with
the spin, handedness, $SU(3)$ triplet or singlet charges, weak $SU(2)$ charge,
hyper charge and family charge as presented in Tables~\ref{Table III.},
\ref{Table so13+1.}, "waiting for" spontaneous break of mass protection at the
electroweak break.
The simple starting action of the {\it spin-charge-family} theory, 
Eqs.~(\ref{wholeaction}, \ref{faction}),  offers three singlet and twice two triplet
scalar gauge fields,  
with the space index $s=(7,8)$ (carrying with respect to the space index the 
weak and the hyper charge as required for Higgs's scalar in the {\it standard 
model})  which break the mass protection of fermions and cause the 
electroweak break.

\section{Conclusions}
\label{conclusions}

\vspace{2mm}

This conclusions reviews the similar Sect.~3  of the Ref.~\cite{n2023NPB},
adding new recognitions. 

In the {\it spin-charge-family} theory~\cite{norma93,pikanorma2005,nh02,%
gmdn2008,gn2009,n2014matterantimatter,nd2017,JMP2013,nh2018,%
2020PartIPartII,nh2021RPPNP},  and the references therein,
the Clifford odd algebra describes the internal space of fermion fields. 

The Clifford odd ``basis vectors'', they are the superposition of odd 
products of $\gamma^a$'s, in a tensor product with the basis in ordinary 
space form the creation and annihilation operators, in which the 
anti-commutativity of the ``basis vectors'' is transferred to the creation 
and annihilation operators for fermions, explaining the second 
quantization postulates for fermion fields.

The Clifford odd ``basis vectors'' have all the properties of fermions: 
Half integer spins concerning the Cartan subalgebra members of the 
Lorentz algebra in the internal space of fermions in even dimensional 
spaces ($d=2(2n+1)$ or $d=4n$),  Subsects.~(\ref{basisvectors}, 
\ref{secondquantizedfermionsbosonsdeven}, App~\ref{basis3+1}).
The  Clifford odd ``basis vectors'' appear in families and have their 
Hermitian conjugated partners in a different group.

The Clifford even ``basis vectors'' are offering the description of the 
internal space of boson fields. The Clifford even ``basis vectors'' are 
the superposition of even products of $\gamma^a$'s. In a tensor 
product with the basis in ordinary space the Clifford even ``basis 
vectors'' form the creation and annihilation operators which manifest 
the commuting properties of the second quantized boson fields, 
offering the explanation for the second quantization postulates for 
boson fields~\cite{n2021SQ,n2022epjc}.

The Clifford even ``basis vectors'' have all the properties of boson 
fields: Integer spins for the Cartan subalgebra members of the 
Lorentz algebra in the internal space of bosons, as discussed in 
Subsects.~\ref{basisvectors}.

With respect to the subgroups of the $SO(d-1, 1)$ group the Clifford 
even ``basis vectors'' manifest the adjoint representations, as 
illustrated in Figs.~(\ref{FigSU3U1odd}, \ref{FigSU3U1even}).
The  Clifford even``basis vectors''  appear in two groups. All the
members of each group have their Hermitian conjugated partners 
within the same group, or they are self adjoint. (They do not form 
families.)  There are the same number of  the Clifford odd ``basis 
vectors'', appearing in families, and of their Hermitian conjugated 
partners, as there are the sum of the members of the two groups
of  the Clifford even ``basis vectors''.

There are two kinds of anti-commuting algebras~\cite{norma93}: The Grassmann
algebra, offering in $d$-dimensional space $2\,.\, 2^d$ operators ($2^d$ $\,\theta^a$'s
and $2^d$ $\frac{\partial}{\partial \theta_a}$'s, Hermitian conjugated to each other,
Eq.~(\ref{thetaderher0})), and the two Clifford subalgebras, each with $2^d$ operators
named $\gamma^a$'s and $\tilde{\gamma}^a$'s, respectively, \cite{norma93,nh02},
Eq.~(\ref{gammatildeantiher0})~\footnote{
The operators in each of the two Clifford subalgebras appear in even-dimensional spaces
in two groups of $2^{\frac{d}{2}-1}\times $ $2^{\frac{d}{2}-1}$ of the Clifford odd
operators (the odd products of either $\gamma^a$'s in one subalgebra or of
$\tilde{\gamma}^a$'s in the other subalgebra), which are Hermitian conjugated
to each other: In each Clifford odd group of any of the two subalgebras, there appear
$2^{\frac{d}{2}-1}$ irreducible representation each with the $2^{\frac{d}{2}-1}$
members and the group of their Hermitian conjugated partners.

There are as well the Clifford even operators (the even products of either
$\gamma^a$'s in one subalgebra or of $\tilde{\gamma}^a$'s in another
subalgebra) which again appear in two groups of $2^{\frac{d}{2}-1}\times $
$2^{\frac{d}{2}-1}$ members each. In the case of the Clifford even objects, the
members of each group of $2^{\frac{d}{2}-1}\times $ $2^{\frac{d}{2}-1}$
members have the Hermitian conjugated partners within the same
group, Subsect.~\ref{basisvectors}.

The Grassmann algebra operators are expressible with the operators of the two
Clifford subalgebras and opposite,~Eq.~(\ref{clifftheta1}). The two Clifford sub-algebras
are independent of each other, Eq.~(\ref{gammatildeantiher0}), forming two independent
spaces.}.

Either the Grassmann algebra~\cite{nh2018} or the two Clifford
subalgebras can be used to describe the internal space of anti-commuting objects,
if the superposition of odd products of operators
($\theta^a$'s or $\gamma^a$'s, or $ \tilde{\gamma}^a$'s) are used to describe the
internal space of these objects. The commuting objects must be a superposition of
even products of operators ($\theta^a$'s or $ \gamma^a$'s or $\tilde{\gamma}^a$'s).

\vspace{1mm}

No integer spin anti-commuting objects have been observed so far, and to describe the
internal space of the so far observed fermions only one of the two Clifford odd
subalgebras are needed.


The problem can be solved by reducing the two Clifford subalgebras to only one, the one
(chosen to be) determined by $\gamma^{a}$'s, 
as presented in Eq.~(\ref{tildegammareduced0}),
what enables that $2^{\frac{d}{2}-1}$ irreducible representations
of $S^{ab}= \frac{i}{4}\, \{\gamma^a\,,\, \gamma^b\}_{-}$ (each with the
$2^{\frac{d}{2}-1}$ members) obtain the family quantum numbers determined by
$\tilde{S}^{ab}= \frac{i}{4}\, \{\tilde{\gamma}^a\,,\,\tilde{\gamma}^b\}_{-}$.

\vspace{2mm}

The decision to use in the {\it spin-charge-family} theory in $d=2(2n +1)$, $n\ge 3$
($d\ge (13+1)$ indeed),
the superposition of the odd products of the Clifford algebra elements $\gamma^{a}$'s
to describe the internal space of fermions which interact with gravity only
(with the vielbeins, the gauge fields of momenta, and the two kinds of the spin
connection fields, the gauge fields of $S^{ab}$ and $\tilde{S}^{ab}$, respectively),
Eq.~(\ref{wholeaction}), offers not
only the explanation for all the assumed properties of fermions and bosons in
the {\it standard model}, with the appearance of the families of quarks and leptons
and antiquarks and antileptons~(\cite{nh2021RPPNP} and the references therein) and
of the corresponding vector gauge fields and the Higgs's scalars included~\cite{nd2017},
but also for the appearance of the dark matter~\cite{gn2009} in the universe, for the
explanation of the matter/antimatter asymmetry in the
universe~\cite{n2014matterantimatter}, and for several other observed phenomena,
making several predictions~\cite{pikan2006,gmdn2007,gmdn2008,gn2013}.\\
This decision offers also the understanding of the second quantized postulates for 
fermion fields.

\vspace{2mm}

The recognition that the use of the superposition of the even products of the Clifford 
algebra elements $\gamma^{a}$'s to describe the internal space of boson fields, what 
appears to manifest all the properties of the observed boson fields, as demonstrated 
in this article, makes clear that the Clifford algebra offers not only the explanation 
for the postulates of the second quantized anti-commuting fermion fields but also 
for the postulates of the second quantized boson fields. 

This recognition, however, offers the possibility to relate
\begin{small}
\begin{eqnarray}
\label{relationomegaAmf01}
\{\frac{1}{2} \sum_{ab} S^{ab}\, \omega_{ab \alpha} \}
\sum_{m } \beta^{m f}\, \hat{\bf b}^{m \dagger}_{f }(\vec{p}) &{\rm \,\, to}&
\{ \sum_{m' f '} {}^{I}{\hat{\cal A}}^{m' \dagger}_{f '} \,
{}^{I}{\cal C}^{m'}{}_{f `\alpha} \}
\sum_{m } \beta^{m f} \, \hat{\bf b}^{m \dagger}_{f }(\vec{p}) \,, \nonumber\\
&&\forall f \,{\rm and}\,\forall \, \beta^{m f}\,, \nonumber\\
{\bf \cal S}^{cd} \,\sum_{ab} (c^{ab}{}_{mf}\, \omega_{ab \alpha}) &{\rm \,\, to}&
{\bf \cal S}^{cd}\, ({}^{I}{\hat{\cal A}}^{m \dagger}_{f}\, {}^{I}{\cal C}^{m}{}_{f \alpha})\,, 
\nonumber\\
&& \forall \,(m,f), \nonumber\\
&&\forall \,\,{\rm Cartan\,\,subalgebra\, \, \, member} \,{\bf \cal S}^{cd} \,,
\nonumber
\end{eqnarray}
\end{small}
and equivalently for ${}^{II}{\hat{\cal A}}^{m \dagger}_{f}\,{}^{II}{\cal C}^{m}{}_{f \alpha}$ 
and $\tilde{\omega}_{ab \alpha}$, what offers the possibility to replace the covariant derivative
$ p_{0 \alpha }$
\begin{small}
$$p_{0\alpha} = p_{\alpha} - \frac{1}{2} S^{ab} \omega_{ab \alpha} -
\frac{1}{2} \tilde{S}^{ab} \tilde{\omega}_{ab \alpha}
\quad \quad \quad\quad\;$$
\end{small}
in Eq.~(\ref{wholeaction}) with
\begin{small}
$$ p_{0\alpha} = p_{\alpha} -
\sum_{m f} {}^{I}{ \hat {\cal A}}^{m \dagger}_{f}\,
{}^{I}{\cal C}^{m}{}_{f \alpha} -
\sum_{m f} {}^{II}{\hat{\cal A}}^{m \dagger}_{f}\,
{}^{II}{\cal C}^{m}{}_{f \alpha}\,, $$
\end{small}
\noindent
where the relations among ${}^{I}{\hat{\cal A}}^{m \dagger}_{f}
{}^{I}{\cal C}^{m}_{f \alpha}$ and
${}^{II}{\hat{\cal A}}^{m \dagger}_{f}\,
{}^{II}{\cal C}^{m}_{f \alpha}$ with respect to $\omega_{ab \alpha}$
and $\tilde{\omega}_{ab \alpha}$, 
need additional study. 

But let us point out that the ``basis vectors'' describing the internal spaces 
of either fermion or boson fields remain the same independent of the choice
of the basis of ordinary space. Transformation of the  basis in ordinary space
\begin{eqnarray}
\hat{\bf b}^{s \dagger}_{f }(\vec{x},x^0)&=&
\sum_{m} \,\hat{b}^{ m \dagger}_{f} \, *_{T}\, \int_{- \infty}^{+ \infty} \,
\frac{d^{d-1}p}{(\sqrt{2 \pi})^{d-1}} \, c^{s m }{}_{f}\;
(\vec{p}) \; \hat{b}^{\dagger}_{\vec{p}}\;
e^{-i (p^0 x^0- \varepsilon \vec{p}\cdot \vec{x})}|_{p^0=|\vec{p}|}\,,\nonumber\\
\label{Weylfermionx}
{\bf {}^{i}{\hat{\cal A}}^{s \dagger}_{g \alpha}}
(\vec{x}, x^0) &=& \sum_{m, f} {}^{I}{ \hat {\cal A}}^{m \dagger}_{f}\, *_{T}\,
\int_{- \infty}^{+ \infty} \,
\frac{d^{d-1}p}{(\sqrt{2 \pi})^{d-1}} \,
{}^{i}{\cal C}^{m \dagger}_{f \alpha} (\vec{p})\,
e^{-i (p^0 x^0- \varepsilon \vec{p}\cdot \vec{x})}|_{p^0=|\vec{p}|}\,,i=(I,II)\,,
\end{eqnarray}
does not influence the internal spaces of either  the Clifford odd or the Clifford 
even ``basis vectors''.

\vspace{2mm}

Let us add that in odd dimensional spaces, $d=(2n+1), n=$ integer, the properties of
the internal
spaces of fermion and boson fields differ essentially from the properties of the internal
spaces of fermion and boson fields in even dimensional spaces, $d=2(2n+1), d=4n$~%
\cite{n2023MDPI}: One half of the ``basis vectors'' have the properties of those of
$d=2n$, the other half, following from the first half by the application of $S^{0 \, 2n+1}$,
behave as the Fadeev-Popov ghosts. The anticommuting ones, remaining the
superposition of odd products of $\gamma^a$'s appear in two orthogonal groups with
their Hermitian conjugated partners within the same group. The commuting one,
still remaining superposition of even products of $\gamma^a$'s appear in families and
have their Hermitian conjugated partners in a separate group, suggesting that taking into account odd dimensional spaces might help to make the theory renormalisable.\\

We need to make the {\it spin-charge-family} theory renormalisable. This weak
point of the proposal is shared with all the Kaluza-Klein-like theories~\cite{mil}.\\

The {\it strings theories}~\cite{string} seem promising to solve this problem,
suggesting that the {\it spin-charge-family} --- describing the internal spaces
with the Clifford odd (for fermions) and the Clifford even (for bosons) ``basis 
vectors'' while fermion and boson fields are presented in  ordinary space with 
the points --- should extend the points in ordinary space to strings.


We see in the two above equations, Eq.~(\ref{Weylfermionx}), that the internal
spaces described by the ``basis vectors'' remain the same if transforming
the external basis from momentum to coordinate representation. This means that
we must relate the {\it spin-charge-family} theory with the {\it string theories}
by extension of the points in the coordinate space to strings, while keeping the
``basis vectors'' as the odd products of nilpotents and the rest of projectors
for fermions, and as the even products of nilpotents and the rest of projectors
for bosons. 

Let us recognize as the first step that multiplying algebraically
$\hat{b}^{m }_{f}$ by $\hat{b}^{m'\dagger}_{f `} $ one reproduces
${}^{II}\hat{\cal A}^{m'' \dagger}_{f ``}$, what the {\it strings theories}
call the left and the right movers forming the boson strings while
multiplying algebraically
$\hat{b}^{m \dagger}_{f}$ by $\hat{b}^{m'}_{f `} $ one reproduces
${}^{II}\hat{\cal A}^{m'' \dagger}_{f''}$, what the {\it strings theories}
could call the right and the left movers forming the boson strings.

But the {\it strings theories}  usually do not present two kinds of boson
fields.

This recognition is the very starting trial to extend the {\it spin-charge-family} theory
from the point second quantised fields to strings in collaboration with
Holger beck Nielsen. We hope to write the first trial for this proceeding.


\appendix

\section{Properties of boson fields}
\label{secondgroupofbosons}

\vspace{2mm}

In Ref.~(\cite{n2023NPB}, Table~3), the Clifford even ``basis vectors'' for 
one kind of  ${}^{i}{\hat{\cal A}}^{m \dagger}_{f}, i=(I,II)$ is presented, 
namely  for $i=I$ in the case that $d=(5+1)$. 
Table~\ref{Cliff basis5+1even II.} represents the second kind of the Clifford 
even ``basis vectors'', ${}^{II}{\hat{\cal A}}^{m \dagger}_{f}$, for the 
same particular case $d=(5+1)$. Comparing both tables we see that the 
Clifford even ``basis vector'', which are products of even number of nilpotents,
and the rest of projectors, contain different nilpotents and projectors.
Correspondingly, their application on the Clifford odd ``basis vectors'' 
$\hat{b}^{m \dagger}_{f}$ differ from the application of 
${}^{I}{\hat{\cal A}}^{m \dagger}_{f}$, as we notice in 
Eqs.~(\ref{calIAb1234gen}, \ref{calbIA1234gen}, \ref{calIIAb1234gen}, 
\ref{calbIIA1234gen}). However, to both the same Fig.~\ref{FigSU3U1even}.

\begin{table}
\begin{tiny}
\caption{The  Clifford even ''basis vectors''  ${}^{II}{\hat{\cal A}}^{m \dagger}_{f}$,  
each of them is the product of projectors and an even number of nilpotents, and each is 
the eigenvector of all the Cartan subalgebra members, ${\cal S}^{03}$, 
${\cal S}^{12}$, ${\cal S}^{56}$, Eq.~(\ref{cartangrasscliff0}), are presented for 
$d= (5+1)$-dimensional case. Indexes $m$ and $f$ determine  
$2^{\frac{d}{2}-1}\times 2^{\frac{d}{2}-1}$ different members  
${}^{II}{\hat{\cal A}}^{m \dagger}_{f}$. 
In the third column  the  ''basis vectors'' ${}^{II}{\hat{\cal A}}^{m \dagger}_{f}$  
which are  Hermitian conjugated partners to each other are pointed out with the 
same symbol. For example, with $\star \star$ 
are equipped the first member with $m=1$ and $f=1$ and the last member with $m=4$ 
and $f=3$.
The sign $\bigcirc$ denotes the  Clifford even ''basis vectors'' which are self adjoint  
($({}^{II}{\hat{\cal A}}^{m \dagger}_{f})^{\dagger}$ 
$={}^{II}{\hat{\cal A}}^{m' \dagger}_{f `}$). It is obvious that ${}^{\dagger}$
has no meaning, since ${}^{II}{\hat{\cal A}}^{m \dagger}_{f}$ are self adjoint or 
are Hermitian conjugated partner to another ${}^{II}{\hat{\cal A}}^{m' \dagger}_{f `}$.
This table represents also the  eigenvalues of the three commuting operators 
${\cal N}^3_{L,R}$ and ${\cal S}^{56 }$ of the subgroups
 $SU(2)\times SU(2)\times U(1)$ of the group
$SO(5,1)$ and the eigenvalues of the three 
commuting operators ${\tau}^3, {\tau}^8$ and ${ \tau'}$ of the 
subgroups  $SU(3)\times U(1)$.
\vspace{3mm}
}
%
%
\label{Cliff basis5+1even II.}
 \begin{center}
 \begin{tabular}{|r| r|r|r|r|r|r|r|r|r|r|r|}
 \hline
$\, f $&$m $&$*$&${}^{II}\hat{\cal A}^{m \dagger}_f$
&${\cal S}^{03}$&$ {\cal S}^{1 2}$&${\cal S} ^{5 6}$&
${\cal N}^3_L$&${\cal N}^3_R$&
${\cal \tau}^3$&${\cal \tau}^8$&${\cal \tau}'$
\\
\hline
%
$I$&$1$&$\star \star$
&$
\stackrel{03}{[-i]}\,\stackrel{12}{(+)} \stackrel{56}{(+)}$&
$0$&$1$&$1$
&$\frac{1}{2}$&$\frac{1}{2}$&$-\frac{1}{2}$&$-\frac{1}{2\sqrt{3}}$&$-\frac{2}{3}$
\\
$$ &$2$&$\bullet$&$
\stackrel{03}{(+i)}\,\stackrel{12}{[-]}\,\stackrel{56}{(+)}$&
$ i$&$0$&$1$
&$-\frac{1}{2}$&$\frac{1}{2}$&$\frac{1}{2}$&$-\frac{3}{2\sqrt{3}}$&$-\frac{2}{3}$
\\
$$ &$3$&$\odot \odot$&$
\stackrel{03}{(+i)}\,\stackrel{12}{(+)}\,\stackrel{56}{[-]}$&
$i$&$ 1$&$0$
&$0$&$1$&$0$&$\frac{1}{\sqrt{3}}$&$-\frac{2}{3}$
\\
$$ &$4$&$\bigcirc$&$
\stackrel{03}{[-i]}\,\stackrel{12}{[-]}\,\stackrel{56}{[-]}$&
$0$&$0$&$0$
&$0$&$0$&$0$&$0$&$0$
\\
\hline 
$II$&$1$&$\bigtriangleup$&$
\stackrel{03}{(-i)}\,\stackrel{12}{[+]}\, \stackrel{56}{(+)}$&
$-i$&$0$&$1$&
$\frac{1}{2}$&$-\frac{1}{2}$&$-\frac{1}{2}$&$-\frac{3}{2\sqrt{3}}$&$0$\\
$$ &$2$&$\otimes$&$
\stackrel{03}{[+i]}\,\stackrel{12}{(-)}\,\stackrel{56}{(+)}$&
$0$&$-1$&$1$
&$-\frac{1}{2}$&$-\frac{1}{2}$&$\frac{1}{2}$&$-\frac{3}{2\sqrt{3}}$&$0$
\\
$$ &$3$&$\bigcirc$&$
\stackrel{03}{[+i]}\,\stackrel{12}{[+]}\,\stackrel{56}{[-]}$&
$0$&$ 0$&$0$
&$0$&$0$&$0$&$0$&$0$
\\
$$ &$4$&$\odot \odot$&$
\stackrel{03}{(-i)}\, \stackrel{12}{(-)}\,\stackrel{56}{[-]}$&
$-i$&$-1$&$0$
&$-0$&$-1$&$0$&$-\frac{1}{\sqrt{3}}$&$\frac{2}{3}$
\\ 
%
%
 \hline
$III$&$1$&$\bigcirc$&$
\stackrel{03}{[-i]}\,\stackrel{12}{[+]}\, \stackrel{56}{[+]}$&
$0$&$0$&$0$&
$0$&$0$&$0$&$0$&$0$\\
$$ &$2$&$\ddagger$&$
\stackrel{03}{(+i)}\,\stackrel{12}{(-)}\,\stackrel{56}{[+]}$&
$i$&$-1$&$0$
&$-1$&$0$&$1$&$0$&$0$\\
$$ &$3$&$\bigtriangleup$&$
\stackrel{03}{(+i)}\,\stackrel{12}{[+]}\,\stackrel{56}{(-)}$&
$i$&$ 0$&$-1$
&$-\frac{1}{2}$&$\frac{1}{2}$&$\frac{1}{2}$&$\frac{3}{2\sqrt{3}}$&$0$
\\
$$ &$4$&$\star \star$&$
\stackrel{03}{[-i]} \stackrel{12}{(-)}\,\stackrel{56}{(-)}$&
$0$&$- 1$&$- 1$
&$-\frac{1}{2}$&$-\frac{1}{2}$&$\frac{1}{2}$&$\frac{1}{2\sqrt{3}}$&$\frac{2}{3}$
\\
\hline
$IV$&$1$&$\ddagger$&$
\stackrel{03}{(-i)}\,\stackrel{12}{(+)}\, \stackrel{56}{[+]}$&
$-i$&$1$&$0$&
$1$&$0$&$-1$&$0$&$0$
\\
$$ &$2$&$\bigcirc$&$
\stackrel{03}{[+i]}\,\stackrel{12}{[-]}\,\stackrel{56}{[+]}$&
$0$&$0$&$0$
&$0$&$0$&$0$&$0$&$0$
\\
$$ &$3$&$\otimes$&$
\stackrel{03}{[+i]}\,\stackrel{12}{(+)}\,\stackrel{56}{(-)}$&
$0$&$ 1$&$-1$
&$\frac{1}{2}$&$\frac{1}{2}$&$-\frac{1}{2}$&$\frac{3}{2\sqrt{3}}$&$0$
\\
$$ &$4$&$\bullet$&$
\stackrel{03}{(-i)}\, \stackrel{12}{[-]}\,\stackrel{56}{(-)}$&
$-i$&$0$&$-1$
&$\frac{1}{2}$&$-\frac{1}{2}$&$-\frac{1}{2}$&$\frac{1}{2\sqrt{3}}$&$\frac{2}{3}$\\ 
\hline 
 \end{tabular}
 \end{center}
\end{tiny}
\end{table}

Comparing Table~3 in Ref.~\cite{n2023NPB} and Table~\ref{Cliff basis5+1even II.} we see
that both Clifford even ''basis vectors'',  ${}^{I}{\hat{\cal A}}^{m \dagger}_{f }$ and
${}^{II}{\hat{\cal A}}^{m \dagger}_{f }$, have the same properties with 
respect to the Cartan subalgebra (${\cal S}^{03}, {\cal S}^{12}, {\cal S}^{56}$) or 
with respect to the superposition of (${\cal S}^{03}, {\cal S}^{12}, {\cal S}^{56}$),
manifesting the subgroups $SU(2)\times SU(2)\times U(1)$ or $SU(3)\times U(1)$.  
To point out this fact the same symbols are used in both tables to denote either selfadjoint 
members ($\bigcirc$), or Hermitian conjugated partners with the same quantum 
numbers ($\star \star$, $\bigtriangleup$ , e.t.c.). 

Since the nilpotents and projectors are not the same, the 
algebraic application of 
${}^{II}{\hat{\cal A}}^{m \dagger}_{f }$ on Clifford odd ''basis vectors'' and their
Hermitian conjugated partners differ as well.

\section{``Basis vectors'' in $d=(3+1)$}
\label{basis3+1}

This section is the copy of the one in Ref.~\cite{n2023NPB}. It was suggested by the referee
of Ref.~\cite{n2023NPB},  to illustrate on a simple case of $d=(3+1)$ 
the properties of ``basis vectors'' when describing internal spaces of fermions and bosons
by the Clifford algebra: 
i. The way of constructing the``basis vectors'' for fermions which appear in families 
and for bosons which have no families.
ii. The manifestation of anti-commutativity of the second quantized fermion fields and
commutativity of the second quantized boson fields.
iii. The creation and annihilation operators, described by a tensor product, $*_{T}$,
of the ``basis vectors'' and their Hermitian conjugated partners with the basis in ordinary
space-time. 

%

This section is a short overview of some sections presented in the
article~\cite{n2023MDPI}, equipped by concrete examples of ``basis vectors''
for fermions and bosons in $d=(3+1)$.

\vspace{2mm}

{\bf ``Basis vectors''}

\vspace{2mm}

Let us start by arranging the ``basis vectors'' as a superposition of products of
(operators~\footnote{We repeat that we treat $\gamma^{a}$  as operators,
not as matrices. We write ``basis vectors'' as the superposition of products of $\gamma^{a}$.
If we want to look for a matrix representation of any operator, say $S^{ab}$, we arrange
the ``basis vectors'' into a series and write a matrix of transformations caused by
the operator. However, we do not need to look for the matrix representations of the
operators since we can directly calculate the application of any operators on
``basis vectors''.})
$\gamma^a$, each ``basis vector'' is the eigenvector of all the Cartan subalgebra
members, Eq.~(\ref{cartangrasscliff}). To achieve this, we arrange ``basis vectors'' to
be products of nilpotents and projectors, Eqs.~(\ref{nilproj}, \ref{signature0}), so that
every nilpotent and every projector is the eigenvector of one of the Cartan subalgebra
members.

\begin{small}
Example 1.\\
Let us notice that, for example, two nilpotents anti-commute, while one nilpotent and
one projector (or two projectors) commute due to
Eq.~(\ref{gammatildeantiher0}): \\
$\frac{1}{2} (\gamma^0 - \gamma^3) \frac{1}{2}
(\gamma^1 - i\gamma^2)=- \frac{1}{2} (\gamma^1 - i\gamma^2) \frac{1}{2}
(\gamma^0 - \gamma^3)$, while $\frac{1}{2} (\gamma^0 - \gamma^3) \frac{1}{2}
(1+i \gamma^1 \gamma^2)= \frac{1}{2} (1+i \gamma^1 \gamma^2) \frac{1}{2}
(\gamma^0 - \gamma^3)$.
\end{small}

\vspace{3mm}

In $d=(3+1) $ there are $16 \, (2^{d=4})$ ``eigenvectors" of the Cartan subalgebra
members ($S^{03}, S^{12}$) and (${\bf {\cal S}}^{03}, {\bf {\cal S}}^{12}$) of the
Lorentz algebras $S^{ab}$ and ${\bf {\cal S}}^{ab}$ , Eq.~(\ref{cartangrasscliff}).\\

Half of them are the Clifford odd ``basis vectors'', appearing in two irreducible
representations, in two ``families'' ($2^{\frac{4}{2}-1}$, $f=(1,2)$), each with two
($2^{\frac{4}{2}-1}$, $m=(1,2)$) members, $\hat{b}^{ m \dagger}_{f}$, 
Eq.~(\ref{3+1oddb}). \\
There is a separate group of $2^{\frac{4}{2}-1}\times $$2^{\frac{4}{2}-1} $
(Clifford odd) Hermitian conjugated partners $\hat{b}^{ m}_{f}=
(\hat{b}^{ m \dagger}_{f})^{\dagger}$ appearing in a separate group which is
not reachable by $S^{ab}$, Eq.~(\ref{3+1oddHb}).

There are two separate groups of $2^{\frac{4}{2}-1}\times 2^{\frac{4}{2}-1} $
Clifford even ''basis vectors'', ${}^{i}{\bf {\cal A}}^{m \dagger}_{f}, i=(I, II)$,
the $2^{\frac{4}{2}-1}$ members of each are self-adjoint, the rest have their
Hermitian conjugated partners within the same group, 
Eqs.~(\ref{3+1evenAI}, \ref{3+1evenAII}).\\
All the members of each group are reachable by $S^{ab}$ or $\tilde{S}^{ab}$
from any starting ''basis vector'' ${}^{i}{\bf {\cal A}}^{1\dagger}_{1}$.

\begin{small}
Example 2.\\
$\hat{b}^{ m=1 \dagger}_{f=1}=\stackrel{03}{(+i)}\stackrel{12}{[+]}
(=\frac{1}{2} (\gamma^0 - \gamma^3) \frac{1}{2} (1+i\gamma^1 \gamma^2))$ 
is a Clifford odd ``basis vector'', its Hermitian conjugated partner, 
Eq.~(\ref{gammatildeantiher0}), is 
$\hat{b}^{ m=1 }_{f=1}=\stackrel{03}{(-i)}\stackrel{12}{[+]}(=\frac{1}{2} 
(\gamma^0 +\gamma^3) \frac{1}{2} (1+i\gamma^1 \gamma^2)$, not reachable 
by either  $S^{ab}$ or by $\tilde{S}^{ab}$ from any of two members in any of 
two ``families'' of the group of $\hat{b}^{ m \dagger}_{f}$, presented in 
Eq.~(\ref{3+1oddb}).\\
 ${}^{I}{\bf {\cal A}}^{m=1 \dagger}_{f=1} (=\stackrel{03}{[+i]}\stackrel{12}{[+]}
= \frac{1}{2} (1+\gamma^0 \gamma^3) \frac{1}{2} (1+i\gamma^1 \gamma^2)$ 
is self-adjoint,
${}^{I}{\bf {\cal A}}^{m=2 \dagger}_{f=1} (=\stackrel{03}{(-i)}\stackrel{12}{(-)}
= \frac{1}{2} (\gamma^0 -\gamma^3) (\gamma^1 -i\gamma^2)$. Its Hermitian 
conjugated partner, belonging to the same group, is 
${}^{I}{\bf {\cal A}}^{m=1 \dagger}_{f=2}$  and is reachable from 
${}^{I}{\bf {\cal A}}^{m=1 \dagger}_{f=1}$ by the application of $\tilde{S}^{01}$,
since $\tilde{\gamma}^{0}*_{A}\stackrel{03}{[+i]}=i\stackrel{03}{(+i)}$ and
$\tilde{\gamma}^{1}*_{A}\stackrel{12}{[+]}=i\stackrel{12}{(+)}$.
\end{small}

\vspace{2mm}

{\bf Clifford odd ``basis vectors''}

\vspace{2mm}

Let us first present the Clifford odd anti-commuting ``basis vectors'', appearing in
two ``families'' ${\hat b}^{m \dagger}_{f}$, and their Hermitian conjugated partners
$({\hat b}^{m \dagger}_f)^{\dagger}$. Each member of the two groups is a
product of one nilpotent and one projector. We choose the right-handed
Clifford odd ``basis vectors''~\footnote{
We could choose the left-handed Clifford odd ``basis vectors'' by exchanging
the role of `basis vectors'' and their Hermitian conjugated partners.}. Clifford odd
``basis vectors'' appear in two families, each family has two members~\footnote{
In the case of $d=(1+1)$, we would have one family with one member only, which
must be nilpotent.}. Let us notice that members of each of two families have the 
same quantum numbers ($S^{03}\,,S^{12}$). They distinguish in ``family'' quantum 
numbers ($\tilde{S}^{03}\,,\tilde{S}^{12}$).
\begin{small}
\begin{eqnarray}
\label{3+1oddb}
\begin{array} {ccrr}
f=1&f=2&&\\
\tilde{S}^{03}=\frac{i}{2}, \tilde{S}^{12}=-\frac{1}{2}&
\;\;\tilde{S}^{03}=-\frac{i}{2}, \tilde{S}^{12}=\frac{1}{2}\;\;\; &S^{03}\, &S^{12}\\
\hat{b}^{ 1 \dagger}_{1}=\stackrel{03}{(+i)}\stackrel{12}{[+]}&
\hat{b}^{ 1 \dagger}_{2}=\stackrel{03}{[+i]}\stackrel{12}{(+)}&\frac{i}{2}&
\frac{1}{2}\\
\hat{b}^{ 2 \dagger}_{1}=\stackrel{03}{[-i]}\stackrel{12}{(-)}&
\hat{b}^{ 2 \dagger}_{2}=\stackrel{03}{(-i)}\stackrel{12}{[-]}&-\frac{i}{2}&
-\frac{1}{2}\,.
\end{array}
\end{eqnarray}
\end{small}
We find for their Hermitian conjugated partners 
\begin{small}
\begin{eqnarray}
\label{3+1oddHb}
\begin{array} {ccrr}
S^{03}=- \frac{i}{2}, S^{12}=\frac{1}{2}&
\;\;S^{03}=\frac{i}{2}, S^{12}=-\frac{1}{2}\;\;&\tilde{S}^{03} &\tilde{S}^{12}\\
\hat{b}^{ 1 }_{1}=\stackrel{03}{(-i)}\stackrel{12}{[+]}&
\hat{b}^{ 1 }_{2}=\stackrel{03}{[+i]}\stackrel{12}{(-)}&-\frac{i}{2}&
-\frac{1}{2}\\
\hat{b}^{ 2 }_{1}=\stackrel{03}{[-i]}\stackrel{12}{(+)}&
\hat{b}^{ 2 }_{2}=\stackrel{03}{(+i)}\stackrel{12}{[-]}&\frac{i}{2}&
\frac{1}{2}\,.
\end{array}
\end{eqnarray}
\end{small}
The vacuum state $|\psi_{oc}>$, Eq.~(\ref{vaccliffodd}), on which the Clifford odd
''basis vectors'' apply is equal to:
$|\psi_{oc}>= \frac{1}{\sqrt{2}} (\stackrel{03}{[-i]}\stackrel{12}{[+]}
+\stackrel{03}{[+i]}\stackrel{12}{[+]} )$.

Let us recognize that the Clifford odd ''basis vectors'' anti-commute due to the odd
number of nilpotents, Example 1. And they are orthogonal according to Eqs.~(\ref{graficcliff0},
\ref{graficcliff1}, \ref{graficfollow1}):
$\hat{b}^{ m \dagger}_{f} *_{A} \hat{b}^{ m' \dagger}_{f '}=0$.


\begin{small}
Example 3.\\
According to the vacuum state presented above, one finds that, for example, 
$\hat{b}^{1 \dagger}_1 (=\stackrel{03}{(+i)}\stackrel{12}{[+]})|\psi_{oc}>$
is $\hat{b}^{1 \dagger}_1$ back, since $\stackrel{03}{(+i)}\stackrel{12}{[+]} *_{A}
\stackrel{03}{[-i]}\stackrel{12}{[+]}=\stackrel{03}{(+i)}\stackrel{12}{[+]}$, 
according to Eq.~(\ref{graficcliff0}), while  $\stackrel{03}{(-i)}\stackrel{12}{[+]} *_{A}
\stackrel{03}{[-i]}\stackrel{12}{[+]}=0$ (due to $(\gamma^0 + \gamma^3) 
(1-\gamma^0 \gamma^3) =0$).\\
Let us apply $S^{01}$ and $\tilde{S}^{01}$ on some of the
``basis vectors'' $\hat{b}^{m \dagger}_{f}$, say $\hat{b}^{1 \dagger}_{1}$.\\
When applying $S^{01}=\frac{i}{2}\gamma^0 \gamma^1$ on 
$\frac{1}{2} (\gamma^0 - \gamma^3) \frac{1}{2} (1+ i\gamma^1 \gamma^2)
(\equiv \stackrel{03}{(+i)} \stackrel{12}{[+]})$ we 
get $- \frac{i}{2}  \frac{1}{2} (1-\gamma^0 \gamma^3) 
\frac{1}{2} (\gamma^1-i \gamma^2) (\equiv(- \frac{i}{2} \stackrel{03}{[-i]} 
 \stackrel{12}{(-)})$.\\
 When applying $\tilde{S}^{01}=\frac{i}{2}\tilde{\gamma}^0 \tilde{\gamma}^1$ on 
$\frac{1}{2} (\gamma^0 - \gamma^3) \frac{1}{2} (1+ i\gamma^1 \gamma^2)
(\equiv \stackrel{03}{(+i)} \stackrel{12}{[+]})$ we get, according to 
Eq.~(\ref{tildegammareduced0}), or if using Eq.~(\ref{usefulrel}), 
$- \frac{i}{2}  \frac{1}{2} (1+\gamma^0 \gamma^3) 
\frac{1}{2} (\gamma^1+i \gamma^2) (\equiv(- \frac{i}{2} \stackrel{03}{[+i]} 
 \stackrel{12}{(+)})$.\\
\end{small}


 It then follows, after using Eqs.~(\ref{usefulrel}, \ref{graficcliff0}, \ref{graficcliff1}, 
 \ref{graficfollow1}) or  just the starting relation, Eq.~(\ref{gammatildeantiher0}),
 and taking into account the above concrete evaluations, 
 the relations of Eq.~(\ref{almostDirac}) for our particular case
\begin{small}
\begin{eqnarray}
\label{3+1oddDirac}
\hat{b}^{ m \dagger}_{f}*_{A} |\psi_{oc}>&=&|\psi^m_{f}>\,,\nonumber\\
\hat{b}^{ m }_{f}*_{A} |\psi_{oc}>&=& 0\cdot|\psi_{oc}>\,,\nonumber\\
\{\hat{b}^{ m \dagger}_{f}, \hat{b}^{ m'\dagger}_{f '}\}_{-}*_{A}|\psi_{oc}>&=&0\cdot|\psi_{oc}>\,,
\nonumber\\
\{\hat{b}^{ m }_{f}, \hat{b}^{ m'}_{f '}\}_{-}*_{A}|\psi_{oc}>&=&0\cdot|\psi_{oc}>\,,\nonumber\\
\{\hat{b}^{ m }_{f}, \hat{b}^{ m'\dagger}_{f '}\}_{-}*_{A}|\psi_{oc}>&=&\delta^{m m'}\delta_{f f `}|\psi_{oc}>\,.
\end{eqnarray}
\end{small}
%
The last relation of Eq.~(\ref{3+1oddDirac}) takes into account that each 
``basis vector'' carries  the ``family'' quantum number, determined by 
$\tilde{S}^{ab}$ of the Cartan subalgebra members, Eq.~(\ref{cartangrasscliff}), 
and the appropriate normalization of ``basis vectors'', Eqs.~(\ref{3+1oddb}, \ref{3+1oddHb}).

\vspace{2mm}

{\bf Clifford even ``basis vectors''}

\vspace{2mm}

Besides $2^{\frac{4}{2}-1}\times 2^{\frac{4}{2}-1} $ Clifford odd
``basis vectors'' and the same number of their Hermitian conjugated partners,
Eqs.~(\ref{3+1oddb}, \ref{3+1oddHb}), the Clifford algebra objects offer two
groups of $2^{\frac{4}{2}-1}\times 2^{\frac{4}{2}-1} $ Clifford even ''basis
vectors'', the members of the group ${}^{I}{\bf {\cal A}}^{m \dagger}_{f}$
and ${}^{II}{\bf {\cal A}}^{m \dagger}_{f}$, which have Hermitian conjugated
partners within the same group or are self-adjoint~\footnote {
Let be repeated that ${\bf {\cal S}}^{ab}=S^{ab} + \tilde{S}^{ab} $~\cite{n2022epjc}.}.
We have the group ${}^{I}{\bf {\cal A}}^{m \dagger}_{f}$, $m=(1,2), f=(1,2)$, the
members of which are Hermitian conjugated to each other or are self-adjoint,
\begin{small}
\begin{eqnarray}
\label{3+1evenAI}
\begin{array} {crrcrr}
&{\bf {\cal S}}^{03}&{\bf {\cal S}}^{12}&&{\bf {\cal S}}^{03}&{\bf {\cal S}}^{12}\\
{}^{I}{\bf {\cal A}}^{1 \dagger}_{1}= \stackrel{03}{[+i]}\stackrel{12}{[+]}&0&0&\,,\quad
{}^{I}{\bf {\cal A}}^{1 \dagger}_{2}= \stackrel{03}{(+i)}\stackrel{12}{(+)}&i&1\\
{}^{I}{\bf {\cal A}}^{2 \dagger}_{1}= \stackrel{03}{(-i)}\stackrel{12}{(-)}&-i&-1&\,,\quad
{}^{I}{\bf {\cal A}}^{2 \dagger}_{2}= \stackrel{03}{[-i]}\stackrel{12}{[-]}&0&0\,,
\end{array}
\end{eqnarray}
\end{small}
and the group ${}^{II}{\bf {\cal A}}^{m \dagger}_{f}$, $m=(1,2), f=(1,2)$, the
members of which are either Hermitian conjugated to each other or are self adjoint
\begin{small}
\begin{eqnarray}
\label{3+1evenAII}
\begin{array} {crrcrr}
&{\bf {\cal S}}^{03}&{\bf {\cal S}}^{12}&&{\bf {\cal S}}^{03}&{\bf {\cal S}}^{12}\\
{}^{II}{\bf {\cal A}}^{1 \dagger}_{1}= \stackrel{03}{[+i]}\stackrel{12}{[-]}&0&0&\,,\quad
{}^{II}{\bf {\cal A}}^{1 \dagger}_{2}= \stackrel{03}{(+i)}\stackrel{12}{(-)}&i&-1\\
{}^{II}{\bf {\cal A}}^{2 \dagger}_{1}= \stackrel{03}{(-i)}\stackrel{12}{(+)}&-i&1&\,,\quad
{}^{II}{\bf {\cal A}}^{2 \dagger}_{2}= \stackrel{03}{[-i]}\stackrel{12}{[+]}&0&0\,.
\end{array}
\end{eqnarray}
\end{small}
The Clifford even ``basis vectors'' have no families. The two groups, 
$ {}^{I}{\bf {\cal A}}^{m \dagger}_{f}$ and ${}^{II}{\bf {\cal A}}^{m \dagger}_{f}$ 
(they are not reachable from one another by ${\bf {\cal S}}^{ab}$), are orthogonal (which 
can easily be checked, since  $ \stackrel{ab}{(\pm k)} *_{A}  \stackrel{ab}{(\pm k)}=0$, 
and $ \stackrel{ab}{[\pm k]} *_{A}  \stackrel{ab}{[\mp k]}=0$).
\begin{eqnarray}
\label{3+1AIAIIorth}
{}^{I}{\bf {\cal A}}^{m \dagger}_{f} *_{A} {}^{II}{\bf {\cal A}}^{m' \dagger}_{f `} 
=0, \quad{\rm for \;any } \;(m, m', f, f `)\,.
\end{eqnarray}

\vspace{2mm}

{\bf Application of ${}^{i}{\bf {\cal A}}^{m \dagger}_{f}, i=(I,II)$ on $\hat{b}^{m \dagger}_{f}$}

\vspace{2mm}

Let us demonstrate the application of $ {}^{i}{\bf {\cal A}}^{m \dagger}_{f}, i=(I,II)$,
on the Clifford odd ``basis vectors'' $\hat{b}^{m \dagger}_{f}$, Eqs.~(\ref{calIAb1234gen},
\ref{calbIIA1234gen}), for our particular case $d=(3+1)$ and compare the result with the
result of application $S^{ab}$ and $\tilde{S}^{ab}$ on $\hat{b}^{m \dagger}_{f}$
evaluated above in Example 3. We found, for example, that
$S^{01}(=\frac{i}{2}\gamma^0 \gamma^1) *_{A} \hat{b}^{1 \dagger}_1(=$
$\frac{1}{2} (\gamma^0 - \gamma^3) \frac{1}{2} (1+ i\gamma^1 \gamma^2)
(=\stackrel{03}{(+i)} \stackrel{12}{[+]})=$
$- \frac{i}{2} \frac{1}{2} (1-\gamma^0 \gamma^3)
\frac{1}{2} (\gamma^1-i \gamma^2) (=(- \frac{i}{2} \stackrel{03}{[-i]}
\stackrel{12}{(-i)})=\hat{b}^{2 \dagger}_{1}$.

Applying ${}^{I}{\bf {\cal A}}^{2 \dagger}_{1} (=\stackrel{03}{(-i)}\stackrel{12}{(-)})
*_{A} \, \hat{b}^{1 \dagger}_1(=$
$\stackrel{03}{(+i)} \stackrel{12}{[+]})=-\stackrel{03}{[-i]} \stackrel{12}{(-)}$, which is
$ \hat{b}^{2 \dagger}_1$, presented in Eq.~(\ref{3+1oddb}). We obtain in both cases
the same result, up to the factor $\frac{i}{2}$ (in front of $\gamma^{0} \gamma^{1}$ in
$S^{01}$). In the second case one sees that ${}^{I}{\bf {\cal A}}^{2 \dagger}_{1}$
(carrying ${\cal S}^{03}=-i, {\cal S}^{12}=-1$) transfers these quantum numbers to
$ \hat{b}^{1 \dagger}_1 $ (carrying ${S}^{03}=\frac{i}{2}, {S}^{12}=\frac{1}{2}$)
what results in $ \hat{b}^{2 \dagger}_1 $ (carrying ${S}^{03}=\frac{-i}{2}, {S}^{12}
=\frac{-1}{2}$).

We can check what the application of the rest three
$ {}^{I}{\bf {\cal A}}^{m \dagger}_{f}$,
do when applying on $ \hat{b}^{m \dagger}_f $. The self-adjoint member carrying
${\cal S}^{03}=0, {\cal S}^{12}=0$, either gives $ \hat{b}^{m \dagger}_f $
back, or gives zero, according to Eq.~(\ref{graficcliff0}). The Clifford even ``basis
vectors'', carrying non zero ${\cal S}^{03}$ and ${\cal S}^{12}$ transfer their
internal values to $ \hat{b}^{m \dagger}_f $ or give zero. In all cases
$ {}^{I}{\bf {\cal A}}^{m \dagger}_{f}$ transform a ``family'' member to another
or the same ``family'' member of the same ``family''.

\begin{small}
Example 4.:\\
${}^{I}{\bf {\cal A}}^{1 \dagger}_{1} (=\stackrel{03}{[+i]}\stackrel{12}{[+]}) *_{A}
\,\hat{b}^{1 \dagger}_1(=$ $\stackrel{03}{(+i)} \stackrel{12}{[+]})=\hat{b}^{1 \dagger}_1(=$  
$\stackrel{03}{(+i)} \stackrel{12}{[+]})$\,,\quad
${}^{I}{\bf {\cal A}}^{1 \dagger}_{1} (=\stackrel{03}{[+i]}\stackrel{12}{[+]}) *_{A}
\,\hat{b}^{1 \dagger}_2(=$ $\stackrel{03}{[+i]} \stackrel{12}{(+)})=\hat{b}^{1 \dagger}_2(=$  
$\stackrel{03}{[+i]} \stackrel{12}{(+)})$\,,\\
${}^{I}{\bf {\cal A}}^{2 \dagger}_{1}  (=\stackrel{03}{(-i)}\stackrel{12}{(-)}) *_{A}
\,\hat{b}^{1 \dagger}_2(=$ $\stackrel{03}{[+i]} \stackrel{12}{(+)})=\hat{b}^{2 \dagger}_2(=$  
$\stackrel{03}{(-i)} \stackrel{12}{[-]})$\,,\quad
${}^{I}{\bf {\cal A}}^{2 \dagger}_{1}  (=\stackrel{03}{(-i)}\stackrel{12}{(-)}) *_{A}
\hat{b}^{2 \dagger}_2 (=$  $\stackrel{03}{(-i)} \stackrel{12}{[-]})=0$.\\
\end{small}

\vspace{2mm}

One easily sees that the application of $  {}^{II}{\bf {\cal A}}^{m \dagger}_{f}$ on 
$ \hat{b}^{m' \dagger}_{f `}$ give zero for all $(m,m',f, f ' )$ (due to 
$\stackrel{ab}{[\pm k]} *_{A} \stackrel{ab}{[\mp k]} =0$,  
$\stackrel{ab}{[\pm k]} *_{A} \stackrel{ab}{(\mp k)} = 0$, and similar applications).

We realised in Example 3. that  the application of $\tilde{S}^{01}=
\frac{i}{2}\tilde{\gamma}^0 \tilde{\gamma}^1$ on $\hat{b}^{1 \dagger}_1$
gives $(- \frac{i}{2} \stackrel{03}{[+i]}  \stackrel{12}{(+i)}) = 
-\frac{i}{2} \hat{b}^{1 \dagger}_{2}$.

Let us algebraically,  $*_{A} $, apply $  {}^{II}{\bf {\cal A}}^{2 \dagger}_{1}
(= \stackrel{03}{(-i)} \stackrel{12}{(+)}$), with quantum numbers $({\cal S}^{03}, 
{\cal S}^{12})=(-i,1)$, from the right hand side the Clifford odd ``basis vector'' 
$\hat{b}^{1 \dagger}_{1}$. This application causes the transition of 
$\hat{b}^{1 \dagger}_{1}$ (with quantum numbers $(\tilde{S}^{03}, \tilde{S}^{12})=
(\frac{i}{2}, -\frac{1}{2})$ (see Eq.~(\ref{signature0})) into $\hat{b}^{1 \dagger}_{2}$ 
(with quantum numbers $(\tilde{S}^{03}, \tilde{S}^{12})=(-\frac{i}{2}, \frac{1}{2})$). 
$  {}^{II}{\bf {\cal A}}^{2 \dagger}_{1}$ obviously transfers its quantum numbers to
Clifford odd ``basis vectors'', keeping $m$ unchanged, and changing the ``family'' quantum 
number:
$\hat{b}^{1 \dagger}_{1} *_{A}  {}^{II}{\bf {\cal A}}^{2 \dagger}_{1}=
\hat{b}^{1 \dagger}_{2}$.

\vspace{2mm}

We can conclude: The internal space of the Clifford even ``basis vectors'' has 
properties of the gauge fields of the Clifford odd ``basis vectors''; 
${}^{I}{\bf {\cal A}}^{m \dagger}_{f}$ transform ``family'' members of the Clifford odd 
``basis vectors'' among themselves, keeping the ``family'' quantum number unchanged, 
${}^{II}{\bf {\cal A}}^{m \dagger}_{f}$ transform a particular ``family'' member into 
the same ``family'' member of another ``family''.

\vspace{3mm}

{\bf Creation and annihilation operators}

\vspace{3mm}

To define creation and annihilation operators for fermion and boson fields, we must 
include besides the internal space, the ordinary space, presented in Eq.~(\ref{creatorp}), 
which defines the momentum or coordinate part of fermion and boson fields.

\vspace{2mm}

We define the creation operators for the single particle fermion states as a tensor
product, $*_{T}$, of the Clifford odd ``basis vectors'' and the basis in ordinary space,
Eq.~(\ref{wholespacefermions}):\\
${\bf \hat{b}}^{s \dagger}_{f} (\vec{p}) =
\sum_{m} c^{sm}{}_f (\vec{p}) \,\hat{b}^{\dagger}_{\vec{p}}\,*_{T}\,
\hat{b}^{m \dagger}_{f}$. The annihilation operators are their Hermitian conjugated
partners.

We have seen in Example 1. that Clifford odd ``basis vectors'' (having odd products
of nilpotents) anti-commute. The commuting objects $\hat{b}^{\dagger}_{\vec{p}}$
(multiplying the ``basis vectors'') do not change the Clifford oddness of 
${\bf \hat{b}}^{s \dagger}_{f} (\vec{p})$.
The two Clifford odd objects, ${\bf \hat{b}}^{s \dagger}_{f} (\vec{p})$ and
${\bf \hat{b}}^{s' \dagger}_{f `} (\vec{p'}) $, keep their anti-commutativity,
fulfilling the anti-commutation relations as presented in Eq.~(\ref{Weylpp'comrel}).
Correspondingly we do not need to postulate anti-commutation relations of Dirac.
The Clifford odd ``basis vectors'' in a tensor product with the basis in ordinary
space explain the second quantized postulates for fermion fields.

The Clifford odd ``basis vectors'' contribute for each $\vec{p}$ a finite number
of ${\bf \hat{b}}^{s \dagger}_{f} (\vec{p})$, the ordinary basis offers infinite
possibilities~\footnote{
An infinitesimally small difference between $\vec{p}$ and
$\vec{p'}$ makes two creation operators ${\bf \hat{b}}^{s \dagger}_{f} (\vec{p})$
and ${\bf \hat{b}}^{s \dagger}_{f } (\vec{p'}) $ with the same ``basis vector''
describing the internal space of fermion fields
still fulfilling the anti-commutation relations (as we learn from atomic physics;
two electrons can carry the same spin if they distinguish in the coordinate part of
the state).}.

\vspace{2mm}

Recognizing that internal spaces of fermion fields and their corresponding
boson gauge fields are describable in even dimensional spaces by the Clifford
odd and even ``basis vectors'', respectively, it becomes evidently that when
including the basis in ordinary space, we must take into account
that boson gauge fields have the space index $\alpha$, which describes the
$\alpha$ component of the boson fields in ordinary space.

We multiply, therefore, as presented in Eq.~(\ref{wholespacebosons}),
the Clifford even ``basis vectors'' with the coefficient
${}^{i}{\cal C}^{ m}{}_{f \alpha}$ carrying the space index $\alpha$ so
that the creation operators
${\bf {}^{i}{\hat{\cal A}}^{m \dagger}_{f \alpha}} (\vec{p})=
\hat{b}^{\dagger}_{\vec{p}}\,*_{T}\,
{}^{i}{\cal C}^{ m}{}_{f \alpha}\, {}^{i}{\hat{\cal A}}^{m \dagger}_{f} \,
\,, i=(I,II)$ carry the space index $\alpha$~\footnote{
Requiring the local phase symmetry for the fermion part of the action,
Eq.~(\ref{wholeaction}), would lead to the requirement of the existence of the
boson fields with the space index $\alpha$.}.
The self-adjoint ``basis vectors'', like
($ {}^{i}{\hat{\cal A}}^{1 \dagger}_{1 \alpha},
{}^{i}{\hat{\cal A}}^{2 \dagger}_{2 \alpha}, i=(I,II)$), do not change
quantum numbers of the Clifford odd ``basis vectors'', since they have
internal quantum numbers equal to zero.

In higher dimensional space,
like in $d=(5+1)$, ${}^{I}{\hat{\cal A}}^{1 \dagger}_{3}$, presented in
Table~\ref{Cliff basis5+1even I.}, could represent the internal space of a
photon field, which transfers to, for example, a fermion and anti-fermion
pair with the internal space described by ($\hat{b}^{1 \dagger}_{1}$,
$\hat{b}^{3 \dagger}_{1}$), presented in Table~\ref{oddcliff basis5+1.},
the momentum in ordinary space.

The subgroup structure of $SU(3)$ gauge fields can be recognized in
Fig.~\ref{FigSU3U1even}.

\vspace{3mm}

Properties of the gauge fields $ {}^{i}{\hat{\cal A}}^{m \dagger}_{f \alpha}$
need further studies. 
 


%




In even dimensional spaces, the Clifford odd and even ``basis vectors'', describing internal
spaces of fermion and boson fields, offer the explanation for the second quantized
postulates for fermion and boson fields~[17].


%
\section{Some useful relations in Grassmann and Clifford algebras, needed also in App.~\ref{13+1representation}}
\label{A}

This appendix contains some helpful relations. For more detailed explanations and 
for proofs, the reader is kindly asked to read~\cite{n2023NPB,nh2021RPPNP} 
and the \mbox{references therein.}

 
 For fermions, the operator of handedness $\Gamma^d$ is determined as follows:
  \begin{small}
\begin{eqnarray}
\label{Gamma}
 \Gamma^{(d)}= \prod_a (\sqrt{\eta^{aa}} \gamma^a)  \cdot \left \{ \begin{array}{l l}
 (i)^{\frac{d}{2}} \,, &\rm{ for\, d \,even}\,,\\
 (i)^{\frac{d-1}{2}}\,,&\rm{for \, d \,odd}\,,
  \end{array} \right.
 \end{eqnarray}
 \end{small}
%

The vacuum state for the Clifford odd ''basis vectors'', $|\psi_{oc}>$, is defined as
 \begin{small}
\begin{eqnarray}
\label{vaccliffodd}
|\psi_{oc}>= \sum_{f=1}^{2^{\frac{d}{2}-1}}\,\hat{b}^{m}_{f}{}_{*_A}
\hat{b}^{m \dagger}_{f} \,|\,1\,>\,.
\end{eqnarray}
 \end{small}

Taking into account that the Clifford objects  $\gamma^a$ and $\tilde{\gamma}^a$ fulfil 
relations of Eq.~\ref{gammatildeantiher0}, one obtains beside the relations presented in 
Eq.~(\ref{usefulrel}) the following once
where $i=(I,II)$ denotes  the two groups of Clifford even ``basis vectors'', while
$m$  and $f$ determine membership  of ``basis vectors'' in any of the two groups $I$
or $II$.
\begin{small}
\begin{eqnarray}
\stackrel{ab}{(k)}\stackrel{ab}{(-k)}& =& \eta^{aa} \stackrel{ab}{[k]}\,,\quad 
\stackrel{ab}{(-k)}\stackrel{ab}{(k)} = \eta^{aa} \stackrel{ab}{[-k]}\,,\quad
\stackrel{ab}{(k)}\stackrel{ab}{[k]} =0\,,\quad 
\stackrel{ab}{(k)}\stackrel{ab}{[-k]} =
 \stackrel{ab}{(k)}\,,\quad 
 \nonumber\\
 \stackrel{ab}{(-k)}\stackrel{ab}{[k]} &=& \stackrel{ab}{(-k)}\,,\quad 
\stackrel{ab}{[k]}\stackrel{ab}{(k)}= \stackrel{ab}{(k)}\,,
\quad 
 \stackrel{ab}{[k]}\stackrel{ab}{(-k)} =0\,,\quad 
 \stackrel{ab}{[k]}\stackrel{ab}{[-k]} =0\,,\quad 
 \nonumber\\
\stackrel{ab}{(k)}^{\dagger} &=& \eta^{aa}\stackrel{ab}{(-k)}\,,\quad 
(\stackrel{ab}{(k)})^2 =0\,, \quad \stackrel{ab}{(k)}\stackrel{ab}{(-k)}
=\eta^{aa}\stackrel{ab}{[k]}\,,\nonumber\\
 \stackrel{ab}{[k]}:&=&
 \frac{1}{2}(1+ \frac{i}{k} \gamma^a \gamma^b)\,,\quad \;\,
(\stackrel{ab}{[k]})^2 = \stackrel{ab}{[k]}\,, 
\quad \stackrel{ab}{[k]}\stackrel{ab}{[-k]}=0\,,\nonumber\\
\stackrel{ab}{\tilde{(k)}} \stackrel{ab}{(k)}&=& 0\,,\quad
\stackrel{ab}{\tilde{(k)}} \stackrel{ab}{(-k)}=-i \eta^{aa}\stackrel{ab}{[-k]}
\,,\quad
\stackrel{ab}{\widetilde{(-k)}} \stackrel{ab}{(k)}=-i \eta^{aa}\stackrel{ab}{[k]}
\,,\quad
\stackrel{ab}{\tilde{(k)}} \stackrel{ab}{[k]}= i  \stackrel{ab}{(k)}
\,, \quad \nonumber\\
%
\stackrel{ab}{\tilde{(k)}} \stackrel{ab}{[-k]}&=&0
\,,\quad
\stackrel{ab}{\widetilde{(-k)}} \stackrel{ab}{[k]}=0\,,
\quad
\stackrel{ab}{\widetilde{(-k)}} \stackrel{ab}{[-k]}=i  \stackrel{ab}{(-k)}\,,
\quad \stackrel{ab}{\tilde{[k]}} \stackrel{ab}{(k)}=\stackrel{ab}{(k)}\,,\nonumber\\
\stackrel{ab}{\tilde{[k]}} \stackrel{ab}{(-k)}&=&0\,, \quad
 \stackrel{ab}{\widetilde{[k]}} \stackrel{ab}{[k]}=0\,,\quad 
 \stackrel{ab}{\widetilde{[-k]}} \stackrel{ab}{[k]}=
\stackrel{ab}{[k]}\,,\quad
\stackrel{ab}{\tilde{[k]}}\stackrel{ab}{[-k]}=\stackrel{ab}{[-k]}\,,
\label{graficcliff0}
 \end{eqnarray}
\end{small}
The algebraic multiplication among $\stackrel{ab}{\tilde{(k)}}$ and
$\stackrel{ab}{\tilde{[k]}}$  goes as in the case of $\stackrel{ab}{(k)}$ and
$\stackrel{ab}{[k]}$
\begin{small}
\begin{eqnarray}
\stackrel{ab}{\tilde{(k)}}\stackrel{ab}{\tilde{[k]}}& =& 0\,,\quad
\stackrel{ab}{\tilde{[k]}}\stackrel{ab}{\tilde{(k)}}=  \stackrel{ab}{\tilde{(k)}}\,, 
\quad 
  \stackrel{ab}{\tilde{(k)}}\stackrel{ab}{\tilde{[-k]}} =  \stackrel{ab}{\tilde{(k)}}\,,
\quad \, \stackrel{ab}{\tilde{[k]}}\stackrel{ab}{\tilde{(-k)}} =0\,, \nonumber\\
  \stackrel{ab}{\widetilde{(-k)}}\stackrel{ab}{\tilde{(k)}}&=& \eta^{aa} \stackrel{ab}{[-k]}\,,\quad
 \stackrel{ab}{\widetilde{(-k)}} \stackrel{ab}{\widetilde{[-k]}}= 0\,,\quad 
%
\label{graficcliff1}
\end{eqnarray}
\end{small}
%
One can further find that
\begin{small}
\begin{eqnarray}
\label{graficfollow1}
S^{ac}\stackrel{ab}{(k)}\stackrel{cd}{(k)} &=& -\frac{i}{2} \eta^{aa} \eta^{cc} 
\stackrel{ab}{[-k]}\stackrel{cd}{[-k]}\,, \quad 
S^{ac}\stackrel{ab}{[k]}\stackrel{cd}{[k]} = 
\frac{i}{2} \stackrel{ab}{(-k)}\stackrel{cd}{(-k)}\,,\nonumber\\
S^{ac}\stackrel{ab}{(k)}\stackrel{cd}{[k]} &=& -\frac{i}{2} \eta^{aa}  
\stackrel{ab}{[-k]}\stackrel{cd}{(-k)}\,, \quad
S^{ac}\stackrel{ab}{[k]}\stackrel{cd}{(k)} = \frac{i}{2} \eta^{cc}  
\stackrel{ab}{(-k)}\stackrel{cd}{[-k]}\,.
\end{eqnarray}
\end{small}


%
\section{One family representation of Clifford odd ``basis vectors'' in $d=(13+1)$
}
\label{13+1representation}  

This appendix, is following App.~D of Ref.~\cite{n2023MDPI}, with a short comment
on the corresponding gauge vector and scalar fields and fermion and boson
representations in $d=(14+1)$-dimensional space included.

In even dimensional space $d=(13 +1)$~(\cite{n2022epjc}, App.~A), one irreducible
representation of the Clifford odd ``basis vectors'', analysed from the point of view of
the subgroups $SO(3,1)\times SO(4) $ (included in $SO(7,1)$) and $SO(7,1)\times
SO(6)$ (included in $SO(13,1)$, while $SO(6)$ breaks into $SU(3)\times U(1)$),
contains the Clifford odd ``basis vectors'' describing internal spaces of quarks and
leptons and antiquarks, and antileptons with the quantum numbers assumed by the
{\it standard model} before the electroweak break. Since $SO(4)$ contains two
$SU(2)$ groups, $Y=\tau^{23} + \tau^4$, one irreducible representation includes
the right-handed neutrinos and the left-handed antineutrinos, which are not in the
{\it standard model} scheme.\\

The Clifford even ``basis vectors'', analysed to the same subgroups,
offer the description of the internal spaces of the corresponding vector and scalar
fields, appearing in the {\it standard model} before the electroweak
break~\cite{n2021SQ,n2022epjc}; as explained in Subsect.~\ref{deven}.

For an overview of the properties of the vector and scalar gauge fields in the
{\it spin-charge-family} theory, the reader is invited to see Refs.~%
(\cite{nh2021RPPNP,nd2017} and the references therein). The vector gauge
fields, expressed as the superposition of spin connections and vielbeins, carrying the space
index $m=(0,1,2,3)$, manifest properties of the observed boson fields. The scalar
gauge fields, causing the electroweak break, carry the space index $s=(7,8)$ and
determine the symmetry of mass matrices of quarks and leptons.

In this Table~\ref{Table so13+1.}, one can check the quantum numbers of the
Clifford odd ``basis vectors'' representing quarks and leptons {\it and antiquarks and
antileptons} if taking into account that all the nilpotents and projectors are eigenvectors
of one of the Cartan subalgebra members, ($S^{03}, S^{12}, S^{56}, \dots, $
$S^{13\,14}$), with the eigenvalues $\pm \frac{i}{2}$ for $\stackrel{ab}{(\pm i)}$
and $\stackrel{ab}{[\pm i]}$, and with the eigenvalues $\pm \frac{1}{2}$ for
$\stackrel{ab}{(\pm 1)}$ and $\stackrel{ab}{[\pm 1]}$.

Taking into account that the third component of the weak charge,
$\tau^{13}=\frac{1}{2} (S^{56}-S^{78})$, for the second $SU(2)$ charge,
$\tau^{23}=\frac{1}{2} (S^{56}+ S^{78})$, for the colour charge [$\tau^{33}=
\frac{1}{2} (S^{9\, 10}-S^{11\,12})$ and $\tau^{38}=
\frac{1}{2\sqrt{3}} (S^{9\, 10}+S^{11\,12} - 2 S^{13\,14})$], for the ``fermion
charge'' $\tau^4=-\frac{1}{3} (S^{9\, 10}+S^{11\,12} + S^{13\,14})$, for the hyper
charge $Y=\tau^{23} + \tau^4$, and electromagnetic charge $Q=Y + \tau^{13}$,
one reproduces all the quantum numbers of quarks, leptons, and {\it antiquarks, and
antileptons}. One notices that the $SO(7,1)$ part is the same for quarks and leptons
and the same for antiquarks and antileptons. Quarks distinguish from leptons only in
the colour and ``fermion'' quantum numbers and antiquarks distinguish from antileptons
only in the anti-colour and ``anti-fermion'' quantum numbers.

\vspace{2mm}

In odd dimensional space, $d=(14+1)$, the eigenstates of handedness are the superposition
of one irreducible representation of $SO(13,1)$, presented in Table~\ref{Table so13+1.},
and the one obtained if on each ``basis vector'' appearing in $SO(13,1)$ the operator
$S^{0 \, (14+1)}$ applies, Subsect.~\ref{dodd}, Ref.~\cite{n2023MDPI}.

Let me point out that in addition to the electroweak break of the {\it standard
model} the break at $\ge 10^{16}$ GeV is needed~(\cite{nh2021RPPNP}, and references
therein).
The condensate of the two right-handed neutrinos causes this break
(Ref.~\cite{nh2021RPPNP}, Table 6); it  interacts with all the scalar and vector gauge
fields, except the weak, $U(1), SU(3)$ and the gravitational field in $d=(3+1)$, leaving
these gauge fields massless up to the electroweak break, when the scalar fields, leaving
massless only the electromagnetic, colour and gravitational fields, cause masses of 
fermions and weak bosons.

The theory predicts two groups of four families: To the lower group of four families,
the three so far observed contribute. The theory predicts the symmetry of both groups
to be $SU(2)\times SU(2) \times U(1)$, Ref.~(\cite{nh2021RPPNP}, Sect. 7.3), which
enable to calculate mixing matrices of quarks and leptons for the accurately enough
measured $3\times 3$ sub-matrix of the $4\times 4$ unitary matrix. No sterile neutrinos
are needed, and no symmetry of the mass matrices must be guessed~\cite{gn2013}.

In the literature, one finds a lot of papers trying to reproduce mass matrices
and measured mixing matrices for quarks and leptons~\cite{Fritz,Frogatt,Jarlskog,%
Branco,Harari,Altarelli}.

The stable of the upper four families predicted by the {\it spin-charge-family} theory is 
a candidate for the dark matter, as discussed in Refs.~\cite{gn2009,nh2021RPPNP}. In the
literature, there are several works suggesting candidates for the dark matter
and also for matter/antimatter asymmetry~\cite{Paul,Raby}.

%

\bottomcaption{\label{Table so13+1.}%
\begin{small}
The left-handed ($\Gamma^{(13,1)} = -1$, Eq.~(\ref{Gamma})) irreducible representation
of one family of spinors --- the product of the odd number of nilpotents and of projectors,
which are eigenvectors of the Cartan subalgebra of the $SO(13,1)$ 
group~\cite{n2014matterantimatter,nh02}, manifesting the subgroup $SO(7,1)$ of the
colour charged quarks and antiquarks and the colourless leptons and antileptons ---
is presented.
It contains the left-handed ($\Gamma^{(3,1)}=-1$) weak ($SU(2)_{I}$) charged
($\tau^{13}=\pm \frac{1}{2}$, 
and $SU(2)_{II}$ chargeless ($\tau^{23}=0$ 
quarks and leptons, and the right-handed
($\Gamma^{(3,1)}=1$) weak ($SU(2)_{I}$) chargeless and $SU(2)_{II}$ charged
($\tau^{23}=\pm \frac{1}{2}$) quarks and leptons, both with the spin $ S^{12}$ up
and down ($\pm \frac{1}{2}$, respectively).
Quarks distinguish from leptons only in the $SU(3) \times U(1)$ part: Quarks are triplets
of three colours ($c^i$ $= (\tau^{33}, \tau^{38})$ $ = [(\frac{1}{2},\frac{1}{2\sqrt{3}}),
(-\frac{1}{2},\frac{1}{2\sqrt{3}}), (0,-\frac{1}{\sqrt{3}}) $, 
carrying the "fermion charge" ($\tau^{4}=\frac{1}{6}$). 
The colourless leptons carry the "fermion charge" ($\tau^{4}=-\frac{1}{2}$).
The same multiplet contains also the left handed weak ($SU(2)_{I}$) chargeless and
$SU(2)_{II}$ charged antiquarks and antileptons and the right handed weak
($SU(2)_{I}$) charged and $SU(2)_{II}$ chargeless antiquarks and antileptons.
Antiquarks distinguish from antileptons again only in the $SU(3) \times U(1)$ part:
Antiquarks are anti-triplets carrying the "fermion charge" ($\tau^{4}=-\frac{1}{6}$).
The anti-colourless antileptons carry the "fermion charge" ($\tau^{4}=\frac{1}{2}$).
$Y=(\tau^{23} + \tau^{4})$ is the hyper charge, the electromagnetic charge
is $Q=(\tau^{13} + Y$).
%
\end{small}
}

\tablehead{\hline
i&$$&$|^a\psi_i>$&$\Gamma^{(3,1)}$&$ S^{12}$&
$\tau^{13}$&$\tau^{23}$&$\tau^{33}$&$\tau^{38}$&$\tau^{4}$&$Y$&$Q$\\
\hline
&& ${\rm (Anti)octet},\,\Gamma^{(7,1)} = (-1)\,1\,, \,\Gamma^{(6)} = (1)\,-1$&&&&&&&&& \\
&& ${\rm of \;(anti) quarks \;and \;(anti)leptons}$&&&&&&&&&\\
\hline\hline}
\tabletail{\hline \multicolumn{12}{r}{\emph{Continued on next page}}\\}
\tablelasttail{\hline}
\begin{tiny}
\begin{supertabular}{|r|c||c||c|c||c|c||c|c|c||r|r|}
1&$ u_{R}^{c1}$&$ \stackrel{03}{(+i)}\,\stackrel{12}{[+]}|
\stackrel{56}{[+]}\,\stackrel{78}{(+)}
||\stackrel{9 \;10}{(+)}\;\;\stackrel{11\;12}{[-]}\;\;\stackrel{13\;14}{[-]} $ &1&$\frac{1}{2}$&0&
$\frac{1}{2}$&$\frac{1}{2}$&$\frac{1}{2\,\sqrt{3}}$&$\frac{1}{6}$&$\frac{2}{3}$&$\frac{2}{3}$\\
\hline
2&$u_{R}^{c1}$&$\stackrel{03}{[-i]}\,\stackrel{12}{(-)}|\stackrel{56}{[+]}\,\stackrel{78}{(+)}
||\stackrel{9 \;10}{(+)}\;\;\stackrel{11\;12}{[-]}\;\;\stackrel{13\;14}{[-]}$&1&$-\frac{1}{2}$&0&
$\frac{1}{2}$&$\frac{1}{2}$&$\frac{1}{2\,\sqrt{3}}$&$\frac{1}{6}$&$\frac{2}{3}$&$\frac{2}{3}$\\
\hline
3&$d_{R}^{c1}$&$\stackrel{03}{(+i)}\,\stackrel{12}{[+]}|\stackrel{56}{(-)}\,\stackrel{78}{[-]}
||\stackrel{9 \;10}{(+)}\;\;\stackrel{11\;12}{[-]}\;\;\stackrel{13\;14}{[-]}$&1&$\frac{1}{2}$&0&
$-\frac{1}{2}$&$\frac{1}{2}$&$\frac{1}{2\,\sqrt{3}}$&$\frac{1}{6}$&$-\frac{1}{3}$&$-\frac{1}{3}$\\
\hline
4&$ d_{R}^{c1} $&$\stackrel{03}{[-i]}\,\stackrel{12}{(-)}|
\stackrel{56}{(-)}\,\stackrel{78}{[-]}
||\stackrel{9 \;10}{(+)}\;\;\stackrel{11\;12}{[-]}\;\;\stackrel{13\;14}{[-]} $&1&$-\frac{1}{2}$&0&
$-\frac{1}{2}$&$\frac{1}{2}$&$\frac{1}{2\,\sqrt{3}}$&$\frac{1}{6}$&$-\frac{1}{3}$&$-\frac{1}{3}$\\
\hline
5&$d_{L}^{c1}$&$\stackrel{03}{[-i]}\,\stackrel{12}{[+]}|\stackrel{56}{(-)}\,\stackrel{78}{(+)}
||\stackrel{9 \;10}{(+)}\;\;\stackrel{11\;12}{[-]}\;\;\stackrel{13\;14}{[-]}$&-1&$\frac{1}{2}$&
$-\frac{1}{2}$&0&$\frac{1}{2}$&$\frac{1}{2\,\sqrt{3}}$&$\frac{1}{6}$&$\frac{1}{6}$&$-\frac{1}{3}$\\
\hline
6&$d_{L}^{c1} $&$ - \stackrel{03}{(+i)}\,\stackrel{12}{(-)}|\stackrel{56}{(-)}\,\stackrel{78}{(+)}
||\stackrel{9 \;10}{(+)}\;\;\stackrel{11\;12}{[-]}\;\;\stackrel{13\;14}{[-]} $&-1&$-\frac{1}{2}$&
$-\frac{1}{2}$&0&$\frac{1}{2}$&$\frac{1}{2\,\sqrt{3}}$&$\frac{1}{6}$&$\frac{1}{6}$&$-\frac{1}{3}$\\
\hline
7&$ u_{L}^{c1}$&$ - \stackrel{03}{[-i]}\,\stackrel{12}{[+]}|\stackrel{56}{[+]}\,\stackrel{78}{[-]}
||\stackrel{9 \;10}{(+)}\;\;\stackrel{11\;12}{[-]}\;\;\stackrel{13\;14}{[-]}$ &-1&$\frac{1}{2}$&
$\frac{1}{2}$&0 &$\frac{1}{2}$&$\frac{1}{2\,\sqrt{3}}$&$\frac{1}{6}$&$\frac{1}{6}$&$\frac{2}{3}$\\
\hline
8&$u_{L}^{c1}$&$\stackrel{03}{(+i)}\,\stackrel{12}{(-)}|\stackrel{56}{[+]}\,\stackrel{78}{[-]}
||\stackrel{9 \;10}{(+)}\;\;\stackrel{11\;12}{[-]}\;\;\stackrel{13\;14}{[-]}$&-1&$-\frac{1}{2}$&
$\frac{1}{2}$&0&$\frac{1}{2}$&$\frac{1}{2\,\sqrt{3}}$&$\frac{1}{6}$&$\frac{1}{6}$&$\frac{2}{3}$\\
\hline\hline
\shrinkheight{0.25\textheight}
9&$ u_{R}^{c2}$&$ \stackrel{03}{(+i)}\,\stackrel{12}{[+]}|
\stackrel{56}{[+]}\,\stackrel{78}{(+)}
||\stackrel{9 \;10}{[-]}\;\;\stackrel{11\;12}{(+)}\;\;\stackrel{13\;14}{[-]} $ &1&$\frac{1}{2}$&0&
$\frac{1}{2}$&$-\frac{1}{2}$&$\frac{1}{2\,\sqrt{3}}$&$\frac{1}{6}$&$\frac{2}{3}$&$\frac{2}{3}$\\
\hline
10&$u_{R}^{c2}$&$\stackrel{03}{[-i]}\,\stackrel{12}{(-)}|\stackrel{56}{[+]}\,\stackrel{78}{(+)}
||\stackrel{9 \;10}{[-]}\;\;\stackrel{11\;12}{(+)}\;\;\stackrel{13\;14}{[-]}$&1&$-\frac{1}{2}$&0&
$\frac{1}{2}$&$-\frac{1}{2}$&$\frac{1}{2\,\sqrt{3}}$&$\frac{1}{6}$&$\frac{2}{3}$&$\frac{2}{3}$\\
\hline
11&$d_{R}^{c2}$&$\stackrel{03}{(+i)}\,\stackrel{12}{[+]}|\stackrel{56}{(-)}\,\stackrel{78}{[-]}
||\stackrel{9 \;10}{[-]}\;\;\stackrel{11\;12}{(+)}\;\;\stackrel{13\;14}{[-]}$
&1&$\frac{1}{2}$&0&
$-\frac{1}{2}$&$ - \frac{1}{2}$&$\frac{1}{2\,\sqrt{3}}$&$\frac{1}{6}$&$-\frac{1}{3}$&$-\frac{1}{3}$\\
\hline
12&$ d_{R}^{c2} $&$\stackrel{03}{[-i]}\,\stackrel{12}{(-)}|
\stackrel{56}{(-)}\,\stackrel{78}{[-]}
||\stackrel{9 \;10}{[-]}\;\;\stackrel{11\;12}{(+)}\;\;\stackrel{13\;14}{[-]} $
&1&$-\frac{1}{2}$&0&
$-\frac{1}{2}$&$-\frac{1}{2}$&$\frac{1}{2\,\sqrt{3}}$&$\frac{1}{6}$&$-\frac{1}{3}$&$-\frac{1}{3}$\\
\hline
13&$d_{L}^{c2}$&$\stackrel{03}{[-i]}\,\stackrel{12}{[+]}|\stackrel{56}{(-)}\,\stackrel{78}{(+)}
||\stackrel{9 \;10}{[-]}\;\;\stackrel{11\;12}{(+)}\;\;\stackrel{13\;14}{[-]}$
&-1&$\frac{1}{2}$&
$-\frac{1}{2}$&0&$-\frac{1}{2}$&$\frac{1}{2\,\sqrt{3}}$&$\frac{1}{6}$&$\frac{1}{6}$&$-\frac{1}{3}$\\
\hline
14&$d_{L}^{c2} $&$ - \stackrel{03}{(+i)}\,\stackrel{12}{(-)}|\stackrel{56}{(-)}\,\stackrel{78}{(+)}
||\stackrel{9 \;10}{[-]}\;\;\stackrel{11\;12}{(+)}\;\;\stackrel{13\;14}{[-]} $&-1&$-\frac{1}{2}$&
$-\frac{1}{2}$&0&$-\frac{1}{2}$&$\frac{1}{2\,\sqrt{3}}$&$\frac{1}{6}$&$\frac{1}{6}$&$-\frac{1}{3}$\\
\hline
15&$ u_{L}^{c2}$&$ - \stackrel{03}{[-i]}\,\stackrel{12}{[+]}|\stackrel{56}{[+]}\,\stackrel{78}{[-]}
||\stackrel{9 \;10}{[-]}\;\;\stackrel{11\;12}{(+)}\;\;\stackrel{13\;14}{[-]}$ &-1&$\frac{1}{2}$&
$\frac{1}{2}$&0 &$-\frac{1}{2}$&$\frac{1}{2\,\sqrt{3}}$&$\frac{1}{6}$&$\frac{1}{6}$&$\frac{2}{3}$\\
\hline
16&$u_{L}^{c2}$&$\stackrel{03}{(+i)}\,\stackrel{12}{(-)}|\stackrel{56}{[+]}\,\stackrel{78}{[-]}
||\stackrel{9 \;10}{[-]}\;\;\stackrel{11\;12}{(+)}\;\;\stackrel{13\;14}{[-]}$&-1&$-\frac{1}{2}$&
$\frac{1}{2}$&0&$-\frac{1}{2}$&$\frac{1}{2\,\sqrt{3}}$&$\frac{1}{6}$&$\frac{1}{6}$&$\frac{2}{3}$\\
\hline\hline
17&$ u_{R}^{c3}$&$ \stackrel{03}{(+i)}\,\stackrel{12}{[+]}|
\stackrel{56}{[+]}\,\stackrel{78}{(+)}
||\stackrel{9 \;10}{[-]}\;\;\stackrel{11\;12}{[-]}\;\;\stackrel{13\;14}{(+)} $ &1&$\frac{1}{2}$&0&
$\frac{1}{2}$&$0$&$-\frac{1}{\sqrt{3}}$&$\frac{1}{6}$&$\frac{2}{3}$&$\frac{2}{3}$\\
\hline
18&$u_{R}^{c3}$&$\stackrel{03}{[-i]}\,\stackrel{12}{(-)}|\stackrel{56}{[+]}\,\stackrel{78}{(+)}
||\stackrel{9 \;10}{[-]}\;\;\stackrel{11\;12}{[-]}\;\;\stackrel{13\;14}{(+)}$&1&$-\frac{1}{2}$&0&
$\frac{1}{2}$&$0$&$-\frac{1}{\sqrt{3}}$&$\frac{1}{6}$&$\frac{2}{3}$&$\frac{2}{3}$\\
\hline
19&$d_{R}^{c3}$&$\stackrel{03}{(+i)}\,\stackrel{12}{[+]}|\stackrel{56}{(-)}\,\stackrel{78}{[-]}
||\stackrel{9 \;10}{[-]}\;\;\stackrel{11\;12}{[-]}\;\;\stackrel{13\;14}{(+)}$&1&$\frac{1}{2}$&0&
$-\frac{1}{2}$&$0$&$-\frac{1}{\sqrt{3}}$&$\frac{1}{6}$&$-\frac{1}{3}$&$-\frac{1}{3}$\\
\hline
20&$ d_{R}^{c3} $&$\stackrel{03}{[-i]}\,\stackrel{12}{(-)}|
\stackrel{56}{(-)}\,\stackrel{78}{[-]}
||\stackrel{9 \;10}{[-]}\;\;\stackrel{11\;12}{[-]}\;\;\stackrel{13\;14}{(+)} $&1&$-\frac{1}{2}$&0&
$-\frac{1}{2}$&$0$&$-\frac{1}{\sqrt{3}}$&$\frac{1}{6}$&$-\frac{1}{3}$&$-\frac{1}{3}$\\
\hline
21&$d_{L}^{c3}$&$\stackrel{03}{[-i]}\,\stackrel{12}{[+]}|\stackrel{56}{(-)}\,\stackrel{78}{(+)}
||\stackrel{9 \;10}{[-]}\;\;\stackrel{11\;12}{[-]}\;\;\stackrel{13\;14}{(+)}$&-1&$\frac{1}{2}$&
$-\frac{1}{2}$&0&$0$&$-\frac{1}{\sqrt{3}}$&$\frac{1}{6}$&$\frac{1}{6}$&$-\frac{1}{3}$\\
\hline
22&$d_{L}^{c3} $&$ - \stackrel{03}{(+i)}\,\stackrel{12}{(-)}|\stackrel{56}{(-)}\,\stackrel{78}{(+)}
||\stackrel{9 \;10}{[-]}\;\;\stackrel{11\;12}{[-]}\;\;\stackrel{13\;14}{(+)} $&-1&$-\frac{1}{2}$&
$-\frac{1}{2}$&0&$0$&$-\frac{1}{\sqrt{3}}$&$\frac{1}{6}$&$\frac{1}{6}$&$-\frac{1}{3}$\\
\hline
23&$ u_{L}^{c3}$&$ - \stackrel{03}{[-i]}\,\stackrel{12}{[+]}|\stackrel{56}{[+]}\,\stackrel{78}{[-]}
||\stackrel{9 \;10}{[-]}\;\;\stackrel{11\;12}{[-]}\;\;\stackrel{13\;14}{(+)}$ &-1&$\frac{1}{2}$&
$\frac{1}{2}$&0 &$0$&$-\frac{1}{\sqrt{3}}$&$\frac{1}{6}$&$\frac{1}{6}$&$\frac{2}{3}$\\
\hline
24&$u_{L}^{c3}$&$\stackrel{03}{(+i)}\,\stackrel{12}{(-)}|\stackrel{56}{[+]}\,\stackrel{78}{[-]}
||\stackrel{9 \;10}{[-]}\;\;\stackrel{11\;12}{[-]}\;\;\stackrel{13\;14}{(+)}$&-1&$-\frac{1}{2}$&
$\frac{1}{2}$&0&$0$&$-\frac{1}{\sqrt{3}}$&$\frac{1}{6}$&$\frac{1}{6}$&$\frac{2}{3}$\\
\hline\hline
25&$ \nu_{R}$&$ \stackrel{03}{(+i)}\,\stackrel{12}{[+]}|
\stackrel{56}{[+]}\,\stackrel{78}{(+)}
||\stackrel{9 \;10}{(+)}\;\;\stackrel{11\;12}{(+)}\;\;\stackrel{13\;14}{(+)} $ &1&$\frac{1}{2}$&0&
$\frac{1}{2}$&$0$&$0$&$-\frac{1}{2}$&$0$&$0$\\
\hline
26&$\nu_{R}$&$\stackrel{03}{[-i]}\,\stackrel{12}{(-)}|\stackrel{56}{[+]}\,\stackrel{78}{(+)}
||\stackrel{9 \;10}{(+)}\;\;\stackrel{11\;12}{(+)}\;\;\stackrel{13\;14}{(+)}$&1&$-\frac{1}{2}$&0&
$\frac{1}{2}$ &$0$&$0$&$-\frac{1}{2}$&$0$&$0$\\
\hline
27&$e_{R}$&$\stackrel{03}{(+i)}\,\stackrel{12}{[+]}|\stackrel{56}{(-)}\,\stackrel{78}{[-]}
||\stackrel{9 \;10}{(+)}\;\;\stackrel{11\;12}{(+)}\;\;\stackrel{13\;14}{(+)}$&1&$\frac{1}{2}$&0&
$-\frac{1}{2}$&$0$&$0$&$-\frac{1}{2}$&$-1$&$-1$\\
\hline
28&$ e_{R} $&$\stackrel{03}{[-i]}\,\stackrel{12}{(-)}|
\stackrel{56}{(-)}\,\stackrel{78}{[-]}
||\stackrel{9 \;10}{(+)}\;\;\stackrel{11\;12}{(+)}\;\;\stackrel{13\;14}{(+)} $&1&$-\frac{1}{2}$&0&
$-\frac{1}{2}$&$0$&$0$&$-\frac{1}{2}$&$-1$&$-1$\\
\hline
29&$e_{L}$&$\stackrel{03}{[-i]}\,\stackrel{12}{[+]}|\stackrel{56}{(-)}\,\stackrel{78}{(+)}
||\stackrel{9 \;10}{(+)}\;\;\stackrel{11\;12}{(+)}\;\;\stackrel{13\;14}{(+)}$&-1&$\frac{1}{2}$&
$-\frac{1}{2}$&0&$0$&$0$&$-\frac{1}{2}$&$-\frac{1}{2}$&$-1$\\
\hline
30&$e_{L} $&$ - \stackrel{03}{(+i)}\,\stackrel{12}{(-)}|\stackrel{56}{(-)}\,\stackrel{78}{(+)}
||\stackrel{9 \;10}{(+)}\;\;\stackrel{11\;12}{(+)}\;\;\stackrel{13\;14}{(+)} $&-1&$-\frac{1}{2}$&
$-\frac{1}{2}$&0&$0$&$0$&$-\frac{1}{2}$&$-\frac{1}{2}$&$-1$\\
\hline
31&$ \nu_{L}$&$ - \stackrel{03}{[-i]}\,\stackrel{12}{[+]}|\stackrel{56}{[+]}\,\stackrel{78}{[-]}
||\stackrel{9 \;10}{(+)}\;\;\stackrel{11\;12}{(+)}\;\;\stackrel{13\;14}{(+)}$ &-1&$\frac{1}{2}$&
$\frac{1}{2}$&0 &$0$&$0$&$-\frac{1}{2}$&$-\frac{1}{2}$&$0$\\
\hline
32&$\nu_{L}$&$\stackrel{03}{(+i)}\,\stackrel{12}{(-)}|\stackrel{56}{[+]}\,\stackrel{78}{[-]}
||\stackrel{9 \;10}{(+)}\;\;\stackrel{11\;12}{(+)}\;\;\stackrel{13\;14}{(+)}$&-1&$-\frac{1}{2}$&
$\frac{1}{2}$&0&$0$&$0$&$-\frac{1}{2}$&$-\frac{1}{2}$&$0$\\
\hline\hline
33&$ \bar{d}_{L}^{\bar{c1}}$&$ \stackrel{03}{[-i]}\,\stackrel{12}{[+]}|
\stackrel{56}{[+]}\,\stackrel{78}{(+)}
||\stackrel{9 \;10}{[-]}\;\;\stackrel{11\;12}{(+)}\;\;\stackrel{13\;14}{(+)} $ &-1&$\frac{1}{2}$&0&
$\frac{1}{2}$&$-\frac{1}{2}$&$-\frac{1}{2\,\sqrt{3}}$&$-\frac{1}{6}$&$\frac{1}{3}$&$\frac{1}{3}$\\
\hline
34&$\bar{d}_{L}^{\bar{c1}}$&$\stackrel{03}{(+i)}\,\stackrel{12}{(-)}|\stackrel{56}{[+]}\,\stackrel{78}{(+)}
||\stackrel{9 \;10}{[-]}\;\;\stackrel{11\;12}{(+)}\;\;\stackrel{13\;14}{(+)}$&-1&$-\frac{1}{2}$&0&
$\frac{1}{2}$&$-\frac{1}{2}$&$-\frac{1}{2\,\sqrt{3}}$&$-\frac{1}{6}$&$\frac{1}{3}$&$\frac{1}{3}$\\
\hline
35&$\bar{u}_{L}^{\bar{c1}}$&$ - \stackrel{03}{[-i]}\,\stackrel{12}{[+]}|\stackrel{56}{(-)}\,\stackrel{78}{[-]}
||\stackrel{9 \;10}{[-]}\;\;\stackrel{11\;12}{(+)}\;\;\stackrel{13\;14}{(+)}$&-1&$\frac{1}{2}$&0&
$-\frac{1}{2}$&$-\frac{1}{2}$&$-\frac{1}{2\,\sqrt{3}}$&$-\frac{1}{6}$&$-\frac{2}{3}$&$-\frac{2}{3}$\\
\hline
36&$ \bar{u}_{L}^{\bar{c1}} $&$ - \stackrel{03}{(+i)}\,\stackrel{12}{(-)}|
\stackrel{56}{(-)}\,\stackrel{78}{[-]}
||\stackrel{9 \;10}{[-]}\;\;\stackrel{11\;12}{(+)}\;\;\stackrel{13\;14}{(+)} $&-1&$-\frac{1}{2}$&0&
$-\frac{1}{2}$&$-\frac{1}{2}$&$-\frac{1}{2\,\sqrt{3}}$&$-\frac{1}{6}$&$-\frac{2}{3}$&$-\frac{2}{3}$\\
\hline
37&$\bar{d}_{R}^{\bar{c1}}$&$\stackrel{03}{(+i)}\,\stackrel{12}{[+]}|\stackrel{56}{[+]}\,\stackrel{78}{[-]}
||\stackrel{9 \;10}{[-]}\;\;\stackrel{11\;12}{(+)}\;\;\stackrel{13\;14}{(+)}$&1&$\frac{1}{2}$&
$\frac{1}{2}$&0&$-\frac{1}{2}$&$-\frac{1}{2\,\sqrt{3}}$&$-\frac{1}{6}$&$-\frac{1}{6}$&$\frac{1}{3}$\\
\hline
38&$\bar{d}_{R}^{\bar{c1}} $&$ - \stackrel{03}{[-i]}\,\stackrel{12}{(-)}|\stackrel{56}{[+]}\,\stackrel{78}{[-]}
||\stackrel{9 \;10}{[-]}\;\;\stackrel{11\;12}{(+)}\;\;\stackrel{13\;14}{(+)} $&1&$-\frac{1}{2}$&
$\frac{1}{2}$&0&$-\frac{1}{2}$&$-\frac{1}{2\,\sqrt{3}}$&$-\frac{1}{6}$&$-\frac{1}{6}$&$\frac{1}{3}$\\
\hline
39&$ \bar{u}_{R}^{\bar{c1}}$&$\stackrel{03}{(+i)}\,\stackrel{12}{[+]}|\stackrel{56}{(-)}\,\stackrel{78}{(+)}
||\stackrel{9 \;10}{[-]}\;\;\stackrel{11\;12}{(+)}\;\;\stackrel{13\;14}{(+)}$ &1&$\frac{1}{2}$&
$-\frac{1}{2}$&0 &$-\frac{1}{2}$&$-\frac{1}{2\,\sqrt{3}}$&$-\frac{1}{6}$&$-\frac{1}{6}$&$-\frac{2}{3}$\\
\hline
40&$\bar{u}_{R}^{\bar{c1}}$&$\stackrel{03}{[-i]}\,\stackrel{12}{(-)}|\stackrel{56}{(-)}\,\stackrel{78}{(+)}
||\stackrel{9 \;10}{[-]}\;\;\stackrel{11\;12}{(+)}\;\;\stackrel{13\;14}{(+)}$
&1&$-\frac{1}{2}$&
$-\frac{1}{2}$&0&$-\frac{1}{2}$&$-\frac{1}{2\,\sqrt{3}}$&$-\frac{1}{6}$&$-\frac{1}{6}$&$-\frac{2}{3}$\\
\hline\hline
41&$ \bar{d}_{L}^{\bar{c2}}$&$ \stackrel{03}{[-i]}\,\stackrel{12}{[+]}|
\stackrel{56}{[+]}\,\stackrel{78}{(+)}
||\stackrel{9 \;10}{(+)}\;\;\stackrel{11\;12}{[-]}\;\;\stackrel{13\;14}{(+)} $
&-1&$\frac{1}{2}$&0&
$\frac{1}{2}$&$\frac{1}{2}$&$-\frac{1}{2\,\sqrt{3}}$&$-\frac{1}{6}$&$\frac{1}{3}$&$\frac{1}{3}$\\
\hline
42&$\bar{d}_{L}^{\bar{c2}}$&$\stackrel{03}{(+i)}\,\stackrel{12}{(-)}|\stackrel{56}{[+]}\,\stackrel{78}{(+)}
||\stackrel{9 \;10}{(+)}\;\;\stackrel{11\;12}{[-]}\;\;\stackrel{13\;14}{(+)}$
&-1&$-\frac{1}{2}$&0&
$\frac{1}{2}$&$\frac{1}{2}$&$-\frac{1}{2\,\sqrt{3}}$&$-\frac{1}{6}$&$\frac{1}{3}$&$\frac{1}{3}$\\
\hline
43&$\bar{u}_{L}^{\bar{c2}}$&$ - \stackrel{03}{[-i]}\,\stackrel{12}{[+]}|\stackrel{56}{(-)}\,\stackrel{78}{[-]}
||\stackrel{9 \;10}{(+)}\;\;\stackrel{11\;12}{[-]}\;\;\stackrel{13\;14}{(+)}$
&-1&$\frac{1}{2}$&0&
$-\frac{1}{2}$&$\frac{1}{2}$&$-\frac{1}{2\,\sqrt{3}}$&$-\frac{1}{6}$&$-\frac{2}{3}$&$-\frac{2}{3}$\\
\hline
44&$ \bar{u}_{L}^{\bar{c2}} $&$ - \stackrel{03}{(+i)}\,\stackrel{12}{(-)}|
\stackrel{56}{(-)}\,\stackrel{78}{[-]}
||\stackrel{9 \;10}{(+)}\;\;\stackrel{11\;12}{[-]}\;\;\stackrel{13\;14}{(+)} $
&-1&$-\frac{1}{2}$&0&
$-\frac{1}{2}$&$\frac{1}{2}$&$-\frac{1}{2\,\sqrt{3}}$&$-\frac{1}{6}$&$-\frac{2}{3}$&$-\frac{2}{3}$\\
\hline
45&$\bar{d}_{R}^{\bar{c2}}$&$\stackrel{03}{(+i)}\,\stackrel{12}{[+]}|\stackrel{56}{[+]}\,\stackrel{78}{[-]}
||\stackrel{9 \;10}{(+)}\;\;\stackrel{11\;12}{[-]}\;\;\stackrel{13\;14}{(+)}$
&1&$\frac{1}{2}$&
$\frac{1}{2}$&0&$\frac{1}{2}$&$-\frac{1}{2\,\sqrt{3}}$&$-\frac{1}{6}$&$-\frac{1}{6}$&$\frac{1}{3}$\\
\hline
46&$\bar{d}_{R}^{\bar{c2}} $&$ - \stackrel{03}{[-i]}\,\stackrel{12}{(-)}|\stackrel{56}{[+]}\,\stackrel{78}{[-]}
||\stackrel{9 \;10}{(+)}\;\;\stackrel{11\;12}{[-]}\;\;\stackrel{13\;14}{(+)} $
&1&$-\frac{1}{2}$&
$\frac{1}{2}$&0&$\frac{1}{2}$&$-\frac{1}{2\,\sqrt{3}}$&$-\frac{1}{6}$&$-\frac{1}{6}$&$\frac{1}{3}$\\
\hline
47&$ \bar{u}_{R}^{\bar{c2}}$&$\stackrel{03}{(+i)}\,\stackrel{12}{[+]}|\stackrel{56}{(-)}\,\stackrel{78}{(+)}
||\stackrel{9 \;10}{(+)}\;\;\stackrel{11\;12}{[-]}\;\;\stackrel{13\;14}{(+)}$
 &1&$\frac{1}{2}$&
$-\frac{1}{2}$&0 &$\frac{1}{2}$&$-\frac{1}{2\,\sqrt{3}}$&$-\frac{1}{6}$&$-\frac{1}{6}$&$-\frac{2}{3}$\\
\hline
48&$\bar{u}_{R}^{\bar{c2}}$&$\stackrel{03}{[-i]}\,\stackrel{12}{(-)}|\stackrel{56}{(-)}\,\stackrel{78}{(+)}
||\stackrel{9 \;10}{(+)}\;\;\stackrel{11\;12}{[-]}\;\;\stackrel{13\;14}{(+)}$
&1&$-\frac{1}{2}$&
$-\frac{1}{2}$&0&$\frac{1}{2}$&$-\frac{1}{2\,\sqrt{3}}$&$-\frac{1}{6}$&$-\frac{1}{6}$&$-\frac{2}{3}$\\
\hline\hline
49&$ \bar{d}_{L}^{\bar{c3}}$&$ \stackrel{03}{[-i]}\,\stackrel{12}{[+]}|
\stackrel{56}{[+]}\,\stackrel{78}{(+)}
||\stackrel{9 \;10}{(+)}\;\;\stackrel{11\;12}{(+)}\;\;\stackrel{13\;14}{[-]} $ &-1&$\frac{1}{2}$&0&
$\frac{1}{2}$&$0$&$\frac{1}{\sqrt{3}}$&$-\frac{1}{6}$&$\frac{1}{3}$&$\frac{1}{3}$\\
\hline
50&$\bar{d}_{L}^{\bar{c3}}$&$\stackrel{03}{(+i)}\,\stackrel{12}{(-)}|\stackrel{56}{[+]}\,\stackrel{78}{(+)}
||\stackrel{9 \;10}{(+)}\;\;\stackrel{11\;12}{(+)}\;\;\stackrel{13\;14}{[-]} $&-1&$-\frac{1}{2}$&0&
$\frac{1}{2}$&$0$&$\frac{1}{\sqrt{3}}$&$-\frac{1}{6}$&$\frac{1}{3}$&$\frac{1}{3}$\\
\hline
51&$\bar{u}_{L}^{\bar{c3}}$&$ - \stackrel{03}{[-i]}\,\stackrel{12}{[+]}|\stackrel{56}{(-)}\,\stackrel{78}{[-]}
||\stackrel{9 \;10}{(+)}\;\;\stackrel{11\;12}{(+)}\;\;\stackrel{13\;14}{[-]} $&-1&$\frac{1}{2}$&0&
$-\frac{1}{2}$&$0$&$\frac{1}{\sqrt{3}}$&$-\frac{1}{6}$&$-\frac{2}{3}$&$-\frac{2}{3}$\\
\hline
52&$ \bar{u}_{L}^{\bar{c3}} $&$ - \stackrel{03}{(+i)}\,\stackrel{12}{(-)}|
\stackrel{56}{(-)}\,\stackrel{78}{[-]}
||\stackrel{9 \;10}{(+)}\;\;\stackrel{11\;12}{(+)}\;\;\stackrel{13\;14}{[-]}  $&-1&$-\frac{1}{2}$&0&
$-\frac{1}{2}$&$0$&$\frac{1}{\sqrt{3}}$&$-\frac{1}{6}$&$-\frac{2}{3}$&$-\frac{2}{3}$\\
\hline
53&$\bar{d}_{R}^{\bar{c3}}$&$\stackrel{03}{(+i)}\,\stackrel{12}{[+]}|\stackrel{56}{[+]}\,\stackrel{78}{[-]}
||\stackrel{9 \;10}{(+)}\;\;\stackrel{11\;12}{(+)}\;\;\stackrel{13\;14}{[-]} $&1&$\frac{1}{2}$&
$\frac{1}{2}$&0&$0$&$\frac{1}{\sqrt{3}}$&$-\frac{1}{6}$&$-\frac{1}{6}$&$\frac{1}{3}$\\
\hline
54&$\bar{d}_{R}^{\bar{c3}} $&$ - \stackrel{03}{[-i]}\,\stackrel{12}{(-)}|\stackrel{56}{[+]}\,\stackrel{78}{[-]}
||\stackrel{9 \;10}{(+)}\;\;\stackrel{11\;12}{(+)}\;\;\stackrel{13\;14}{[-]} $&1&$-\frac{1}{2}$&
$\frac{1}{2}$&0&$0$&$\frac{1}{\sqrt{3}}$&$-\frac{1}{6}$&$-\frac{1}{6}$&$\frac{1}{3}$\\
\hline
55&$ \bar{u}_{R}^{\bar{c3}}$&$\stackrel{03}{(+i)}\,\stackrel{12}{[+]}|\stackrel{56}{(-)}\,\stackrel{78}{(+)}
||\stackrel{9 \;10}{(+)}\;\;\stackrel{11\;12}{(+)}\;\;\stackrel{13\;14}{[-]} $ &1&$\frac{1}{2}$&
$-\frac{1}{2}$&0 &$0$&$\frac{1}{\sqrt{3}}$&$-\frac{1}{6}$&$-\frac{1}{6}$&$-\frac{2}{3}$\\
\hline
56&$\bar{u}_{R}^{\bar{c3}}$&$\stackrel{03}{[-i]}\,\stackrel{12}{(-)}|\stackrel{56}{(-)}\,\stackrel{78}{(+)}
||\stackrel{9 \;10}{(+)}\;\;\stackrel{11\;12}{(+)}\;\;\stackrel{13\;14}{[-]} $&1&$-\frac{1}{2}$&
$-\frac{1}{2}$&0&$0$&$\frac{1}{\sqrt{3}}$&$-\frac{1}{6}$&$-\frac{1}{6}$&$-\frac{2}{3}$\\
\hline\hline
57&$ \bar{e}_{L}$&$ \stackrel{03}{[-i]}\,\stackrel{12}{[+]}|
\stackrel{56}{[+]}\,\stackrel{78}{(+)}
||\stackrel{9 \;10}{[-]}\;\;\stackrel{11\;12}{[-]}\;\;\stackrel{13\;14}{[-]} $ &-1&$\frac{1}{2}$&0&
$\frac{1}{2}$&$0$&$0$&$\frac{1}{2}$&$1$&$1$\\
\hline
58&$\bar{e}_{L}$&$\stackrel{03}{(+i)}\,\stackrel{12}{(-)}|\stackrel{56}{[+]}\,\stackrel{78}{(+)}
||\stackrel{9 \;10}{[-]}\;\;\stackrel{11\;12}{[-]}\;\;\stackrel{13\;14}{[-]}$&-1&$-\frac{1}{2}$&0&
$\frac{1}{2}$ &$0$&$0$&$\frac{1}{2}$&$1$&$1$\\
\hline
59&$\bar{\nu}_{L}$&$ - \stackrel{03}{[-i]}\,\stackrel{12}{[+]}|\stackrel{56}{(-)}\,\stackrel{78}{[-]}
||\stackrel{9 \;10}{[-]}\;\;\stackrel{11\;12}{[-]}\;\;\stackrel{13\;14}{[-]}$&-1&$\frac{1}{2}$&0&
$-\frac{1}{2}$&$0$&$0$&$\frac{1}{2}$&$0$&$0$\\
\hline
60&$ \bar{\nu}_{L} $&$ - \stackrel{03}{(+i)}\,\stackrel{12}{(-)}|
\stackrel{56}{(-)}\,\stackrel{78}{[-]}
||\stackrel{9 \;10}{[-]}\;\;\stackrel{11\;12}{[-]}\;\;\stackrel{13\;14}{[-]} $&-1&$-\frac{1}{2}$&0&
$-\frac{1}{2}$&$0$&$0$&$\frac{1}{2}$&$0$&$0$\\
\hline
61&$\bar{\nu}_{R}$&$\stackrel{03}{(+i)}\,\stackrel{12}{[+]}|\stackrel{56}{(-)}\,\stackrel{78}{(+)}
||\stackrel{9 \;10}{[-]}\;\;\stackrel{11\;12}{[-]}\;\;\stackrel{13\;14}{[-]}$&1&$\frac{1}{2}$&
$-\frac{1}{2}$&0&$0$&$0$&$\frac{1}{2}$&$\frac{1}{2}$&$0$\\
\hline
62&$\bar{\nu}_{R} $&$ - \stackrel{03}{[-i]}\,\stackrel{12}{(-)}|\stackrel{56}{(-)}\,\stackrel{78}{(+)}
||\stackrel{9 \;10}{[-]}\;\;\stackrel{11\;12}{[-]}\;\;\stackrel{13\;14}{[-]} $&1&$-\frac{1}{2}$&
$-\frac{1}{2}$&0&$0$&$0$&$\frac{1}{2}$&$\frac{1}{2}$&$0$\\
\hline
63&$ \bar{e}_{R}$&$\stackrel{03}{(+i)}\,\stackrel{12}{[+]}|\stackrel{56}{[+]}\,\stackrel{78}{[-]}
||\stackrel{9 \;10}{[-]}\;\;\stackrel{11\;12}{[-]}\;\;\stackrel{13\;14}{[-]}$ &1&$\frac{1}{2}$&
$\frac{1}{2}$&0 &$0$&$0$&$\frac{1}{2}$&$\frac{1}{2}$&$1$\\
\hline
64&$\bar{e}_{R}$&$\stackrel{03}{[-i]}\,\stackrel{12}{(-)}|\stackrel{56}{[+]}\,\stackrel{78}{[-]}
||\stackrel{9 \;10}{[-]}\;\;\stackrel{11\;12}{[-]}\;\;\stackrel{13\;14}{[-]}$&1&$-\frac{1}{2}$&
$\frac{1}{2}$&0&$0$&$0$&$\frac{1}{2}$&$\frac{1}{2}$&$1$\\
\hline
\end{supertabular}
\end{tiny}

\vspace {3mm}


%

%
\section*{Acknowledgments} 

The author thanks Department of Physics, FMF, University of Ljubljana, Society of Mathematicians, Physicists and Astronomers of Slovenia,  for supporting the research on the {\it spin-charge-family} theory by offering the room and computer facilities and Matja\v z Breskvar of Beyond Semiconductor for donations, in particular for the annual workshops entitled "What comes beyond the standard models".

\vspace{3mm}



\begin{thebibliography}{99}
\bibitem{norma93} N. Manko\v c Bor\v stnik, "Spinor and vector representations in four dimensional Grassmann
              space", {\it J. of Math. Phys.} {\bf 34} (1993) 3731-3745, 
              ''Unification of spin and charges 
            in Grassmann space?'', hep-th 9408002, IJS.TP.94/22, 
            Mod. Phys. Lett.{\bf A (10)} No.7 (1995) 587-595; 
\bibitem{n2019PRD} N. S. Manko\v c Bor\v stnik, "New way of second 
             quantized theory of  fermions with either Clifford or Grassmann coordinates
             and  {\it spin-charge-family} theory " [arXiv:1802.05554v4, arXiv:1902.10628], 
             N. S. Manko\v c Bor\v stnik, ''How Clifford algebra can help understand second  
             quantization of fermion and boson fields'', 
             [arXiv: 2210.06256. physics.gen-ph V1] . 
\bibitem{pikanorma} A. Bor\v stnik,  N.S.  Manko\v c Bor\v stnik, ''Left and
              right handedness of fermions and bosons'', 
              J. of Phys. G:  Nucl. Part. 
              Phys.{\bf 24}(1998)963-977, hep-th/9707218. 
\bibitem{pikanorma2005} A. Bor\v stnik Bra\v ci\v c, N. S. Manko\v c Bor\v stnik, 
             ''On the origin of  families of fermions and their mass matrices'', hep-ph/0512062,  
              Phys Rev. {\bf D 74} 073013-28  (2006).
             
\bibitem{n2023NPB} N. S. Manko\v c Bor\v stnik, ''How Clifford algebra helps understand second quantized quarks and
  leptons and corresponding vector and scalar boson fields, opening a new step
  beyond the standard model'', Reference: NUPHB 994 (2023) 116326 , 
  [arXiv: 2210.06256, physics.gen-ph V2].\\ 
\bibitem{n2023MDPI}  N. S. Manko\v c Bor\v stnik,  ''Clifford odd and even objects in even and odd 
  dimensional spaces'',   Symmetry 2023,15,818-12-V2 94818, https:doi.org/10.3390/sym15040818,
 [arxiv.org/abs/2301.04466] , https://www.mdpi.com/2073-8994/15/4/818  Manuscript ID: symmetry-2179313.
\bibitem{Geor} H. Georgi, in {\it Particles and Fields} (edited by C. E. Carlson), A.I.P., 1975; Google Scholar.
\bibitem{FritzMin} H. Fritzsch and P. Minkowski, {\it Ann. Phys.} {\bf 93} (1975) 193. 
\bibitem{PatiSal} J. Pati and A. Salam, {\it Phys.Rev.} {\bf D 8} (1973) 1240. 
\bibitem{GeorGlas} H. Georgy and S.L. Glashow, {\it Phys. Rev. Lett.}  {\bf 32 } (1974) 438.
\bibitem{Cho}  Y. M. Cho, {\it J. Math. Phys.} {\bf 16}  (1975) 2029.
%
%
\bibitem{n2014matterantimatter}  N.S. Manko\v c Bor\v stnik, 
"Matter-antimatter asymmetry in the {\it spin-charge-family} theory", 
{\it Phys. Rev.} {\bf D 91} (2015) 065004  [arXiv:1409.7791].
\bibitem{JMP2013} N.S. Manko\v c Bor\v stnik N S, "The spin-charge-family theory is explaining the  
origin of families, of the Higgs and the Yukawa couplings", {\it J. of Modern Phys.} 
{\bf 4} (2013) 823 [arXiv:1312.1542].  
\bibitem{nh2021RPPNP}   N. S. Manko\v c Bor\v stnik, H. B. Nielsen,
 "How does Clifford algebra 
            show the way to the second quantized fermions with unified spins, 
            charges and families, and with vector and scalar gauge fields beyond 
            the {\it standard model}", Progress in Particle and Nuclear Physics,
           http://doi.org/10.1016.j.ppnp.2021.103890 . 
\bibitem{DN2018}  "Representations in Grassmann space and 
            fermion degrees of freedom", Proceedings  to  the $20^{th}$ Workshop "What comes 
             beyond the standard models", Bled, 9-17 of July, 2017, Ed. N.S. Manko\v c Bor\v stnik, H.B. Nielsen,
              D. Lukman, DMFA  Zalo\v zni\v stvo, Ljubljana, December 2017         
           [arxiv:1805.06318, arXiv:1806.01629]  
\bibitem{pikan2006}  A. Bor\v stnik Bra\v ci\v c, N. S. Manko\v c Bor\v stnik, 
 ''On the origin of families of fermions and their mass matrices'', hep-ph/0512062,  
Phys. Rev. {\bf D 74}  073013-28  (2006).




\bibitem{nh02}  N.S. Manko\v c Bor\v stnik, H.B.F. Nielsen, {\it J. of Math. Phys.} {\bf 43}, 
5782 (2002) [arXiv:hep-th/0111257],
``How to generate families of spinors'',
{\it J. of Math. Phys.} {\bf 44} 4817 (2003) [arXiv:hep-th/0303224].
\bibitem{nd2017} N.S. Manko\v c Bor\v stnik, D. Lukman, "Vector and scalar gauge 
fields with respect to $d=(3+1)$ in Kaluza-Klein theories and in 
the {\it spin-charge-family theory}", {\it Eur. Phys. J. C} {\bf 77} (2017) 231.
\bibitem{nh2018}  N.S. Manko\v c Bor\v stnik and H.B.  Nielsen, "Why nature made a choice of 
Clifford and not  Grassmann coordinates",   Proceedings  to  the $20^{th}$ Workshop "What comes 
beyond the standard models", Bled, 9-17 of July, 2017, Ed. N.S. Manko\v c Bor\v stnik, H.B. Nielsen,
D. Lukman, DMFA  Zalo\v zni\v stvo, Ljubljana, December 2017, p. 89-120 
[arXiv:1802.05554v1v2]. 
\bibitem{2020PartIPartII}   N.S. Manko\v c Bor\v stnik, H.B.F. Nielsen, 
"Understanding the second  quantization of fermions in Clifford and in Grassmann 
space",
             ``New way of second quantization of fermions ---  Part I and Part II 
             proceedings 
             [arXiv:2007.03517, arXiv:2007.03516], "New way of second 
             quantized theory of  fermions with either Clifford or Grassmann coordinates
             and  {\it spin-charge-family} theory " 
             [arXiv:1802.05554v4,arXiv:1902.10628]. 
             
 

\bibitem{n2021SQ}    N. S. Manko\v c Bor\v stnik, ''How do Clifford algebras show 
             the way to the second quantized fermions with unified spins, charges and 
             families, and to the corresponding second quantized vector and scalar 
             gauge field '', Proceedings  to  the $24^{rd}$ Workshop 
             "What comes    beyond the standard models", 5 - 11 of July, 2021, 
             Ed. N.S. Manko\v c Bor\v stnik, H.B. Nielsen, D. Lukman, A. Kleppe, DMFA  
             Zalo\v zni\v stvo,  Ljubljana, December 2021, [arXiv:2112.04378] .  
 %
 \bibitem{n2022epjc}  N. S. Manko\v c Bor\v stnik, ''How Clifford algebra can help understand 
           second  quantization of fermion and boson fields'', 
             [arXiv: 2210.06256. physics.gen-ph V1] , 
\bibitem{nIARD2022} N. S. Manko\v c Bor\v stnik, 
''Clifford odd and even objects 
            offer description of internal space of fermions and bosons, respectively, 
            opening new insight into the second quantization of fields'', 
            The $13^{th}$ Bienal Conference on Classical and Quantum Relativistic 
            Dynamics of Particles and Fields IARD 2022, Prague, $6 - 9$ June,
             [http://arxiv.org/abs/2210.07004] .
\bibitem{2020PartIPartII}   N.S. Manko\v c Bor\v stnik, H.B.F. Nielsen, 
``Understanding the second  quantization of fermions in Clifford and in Grassmann 
space: {\it New way of second quantization of fermions---Part I and Part II}. 
              (2020), [arXiv:2007.03516],    [arXiv:1802.05554v4,  arXiv:1902.10628].  

 %
\bibitem{Dirac} P.A.M. Dirac {\it Proc. Roy. Soc. (London)}, {\bf A 117} (1928) 610.
\bibitem{BetheJackiw} H.A. Bethe, R.W. Jackiw, "Intermediate quantum mechanics",
New York : W.A. Benjamin, 1968.
\bibitem{Weinberg} S. Weinberg, "The quantum theory of fields", Cambridge, 
Cambridge University Press, 2015.
%
%
%
%
%
\bibitem{nh2017} N.S. Manko\v c Bor\v stnik, H.B.F. Nielsen, "The spin-charge-family theory 
             offers understanding of the triangle anomalies cancellation in the standard model",
             {\it Fortschritte der Physik, Progress of Physics} (2017) 1700046.
 \bibitem{normaJMP2015} N.S. Manko\v c Bor\v stnik, "The explanation for the origin of the 
Higgs  scalar and for the Yukawa couplings by the {\it spin-charge-family} theory", 
 {\it J. of Mod. Physics} {\bf 6} (2015) 2244-2274, http://dx.org./10.4236/jmp.2015.615230
              [arXiv:1409.4981].
%
%
\bibitem{NHD} D. Lukman, N.S. Manko\v c Bor\v stnik and H.B. Nielsen,
"An effective two dimensionality cases bring a new hope to the Kaluza-Klein-like theories", 
{\em New J. Phys.} 13:103027, 2011.
%
 \bibitem{KaluzaKlein}  T. Kaluza, ''On the unification problem in Physics'', {\it Sitzungsber. d. Berl. Acad.}
 (1918) 204, O. Klein, ''Quantum theory and five-dimensional relativity'', {\it Zeit. Phys.} {\bf 37}(1926) 895. 
\bibitem{Witten} E. Witten, ''Search for realistic Kaluza-Klein theory'',{\it  Nucl. Phys.} {\bf B 186} (1981) 412.
\bibitem{Duff} M. Duff, B. Nilsson, C. Pope, {\it Phys. Rep.} {\bf C 130} (1984)1,
M. Duff, B. Nilsson, C. Pope, N. Warner, {\it Phys. Lett.} {\bf B 149} (1984) 60.
\bibitem{App} T. Appelquist, H. C. Cheng, B. A. Dobrescu, {\it Phys. Rev.} {\bf D 64}
(2001) 035002.
\bibitem{SapTin}  M. Saposhnikov, P. TinyakovP 2001 {\it Phys. Lett.} {\bf B 515} (2001) 442
 [arXiv:hep-th/0102161v2].
\bibitem{Wetterich} C. Wetterich,{\it Nucl. Phys.} {\bf B 253} (1985) 366.
\bibitem{zelenaknjiga} The authors of the works presented in {\it An introduction to Kaluza-Klein 
theories}, Ed. by H. C. Lee, World Scientific, Singapore 1983.
\bibitem{mil} M. Blagojevi\' c,   {\em Gravitation and gauge symmetries},  IoP Publishing, Bristol 2002.
 %
 \bibitem{mdn2006} M. Breskvar, D. Lukman, N. S. Manko\v c Bor\v stnik, 
                     ''On the Origin of Families of Fermions and Their Mass Matrices\,---\,%
Approximate Analyses of Properties of Four Families Within Approach 
Unifying Spins and Charges", 
                     Proceedings to the $9^{\rm th}$ Workshop ''What Comes Beyond the Standard 
                     Models'', Bled, Sept. 16 - 26, 2006,  Ed. by Norma Manko\v c Bor\v stnik, 
		     Holger Bech Nielsen, Colin Froggatt, Dragan Lukman, DMFA Zalo\v zni\v stvo, 
                     Ljubljana December 2006, p.25-50, hep-ph/0612250.
%
\bibitem{gmdn2007} G. Bregar, M. Breskvar, D. Lukman, N.S. Manko\v c Bor\v stnik,
                    "Families of Quarks and Leptons and Their Mass Matrices", 
                    Proceedings to the $10^{th}$ international workshop ''What Comes Beyond 
		    the Standard Model'', 17 -27 of July, 2007, Ed. Norma Manko\v c  
		    Bor\v stnik, Holger Bech Nielsen, Colin Froggatt, Dragan Lukman,
		    DMFA  Zalo\v zni\v stvo, Ljubljana December 2007, 
		    p.53-70, hep-ph/0711.4681.
%
\bibitem{gmdn2008} G. Bregar, M. Breskvar, D. Lukman, N.S. Manko\v c Bor\v stnik, 
                     "Predictions for four families by the Approach unifying spins and charges"
                     {\it New J. of Phys.} {\bf 10} (2008) 093002,
                     hep-ph/0606159, hep/ph-07082846.
%
              %
\bibitem{gn2009} G. Bregar, N.S. Manko\v c Bor\v stnik, "Does dark matter consist of baryons 
	        of new stable family quarks?", {\it Phys. Rev. D } {\bf 80}, 083534 (2009), 1-16.
%
\bibitem{gn2013}  G. Bregar, N.S. Manko\v c Bor\v stnik, "Can we predict the fourth family masses 
              for quarks and leptons?", Proceedings (arxiv:1403.4441) to the 16 th Workshop "What comes beyond the 
              standard models", Bled, 14-21 of July, 2013, Ed. N.S. Manko\v c Bor\v stnik, 
              H.B. Nielsen, D. Lukman, DMFA  Zalo\v zni\v stvo, Ljubljana December 2013, p. 31-51, 
              http://arxiv.org/abs/1212.4055.
%
\bibitem{gn2014} G. Bregar, N.S. Manko\v c Bor\v stnik, "The new experimental data for the quarks 
           mixing matrix are in better agreement with the {\it spin-charge-family} theory predictions",               
              Proceedings to 
              the $17^th$ Workshop "What comes beyond the standard models", Bled, 20-28 of July, 2014, 
              Ed. N.S. Manko\v c Bor\v stnik, H.B. Nielsen, D. Lukman, DMFA  Zalo\v zni\v stvo, 
              Ljubljana December 2014, p.20-45 [ arXiv:1502.06786v1] [arxiv:1412.5866].  	        
%
\bibitem{nm2015} N.S. Manko\v c Bor\v stnik, M. Rosina, "Are superheavy stable quark 
              clusters viable   candidates  for the dark matter?",
              International Journal of Modern Physics D (IJMPD) {\bf 24} (No. 13) (2015) 1545003.               
 \bibitem{MPavsic}  Pav\v si\v c, M. \emph{The Landscape of Theoretical Physics: Global View};
             van der Merwe, A., Ed.; Kluwer Academic Publishers:  New York, NY, USA, 2001.
\bibitem{MP2017} Pav\v  si\v c, M. Quantized fields \' a la Clifford and unification. \emph{arXiv} \textbf{2017},
 [arXiv:1707.05695].  
 %
\bibitem{NA2017} A. Hernandez-Galeana and N.S. Manko\v c Bor\v stnik, "The symmetry 
             of $4 \times 4$ mass matrices predicted by the 
             {\it spin-charge-family} theory --- $SU(2) \times SU(2) \times U(1)$ ---  
              remains in all  loop corrections", Proceedings  to  the $21^{st}$ Workshop "What comes 
             beyond the standard models", 23 of June - 1 of July, 2018, Ed. N.S. Manko\v c Bor\v stnik, 
            H.B. Nielsen, D. Lukman, DMFA  Zalo\v zni\v stvo, Ljubljana, December 2017         
           [arXiv:1902.02691, arXiv:1902.10628].
 %
 \bibitem{datanew} A. Ceccucci (CERN), Z. Ligeti (LBNL), Y. Sakai (KEK), 
Particle Data Group, Aug. 29, 2014 
[http://pdg.lbl.gov/2014/reviews/rpp2014-rev-ckm-matrix.pdf].
%
\bibitem{PDG2020} Review of Particle,
Particle Data Group, P.A. Zyla, R.M. Barnett, J. Beringer, O. Dahl, D.A. Dwyer, 
D.E. Groom, C -J. Lin, K.S. Lugovsky, E. Pianori ...., Author Notes, 
Progress of Theoretical and Experimental Physics, Volume 2020, Issue 8, August 2020, 083C01,
https://doi.org/10.1093/ptep/ptaa104, 14 August 2020. 
%
 \bibitem{normaJMP2015} N.S. Manko\v c Bor\v stnik, "The explanation for the origin of the 
Higgs  scalar and for the Yukawa couplings by the {\it spin-charge-family} theory", 
 {\it J. of Mod. Physics} {\bf 6} (2015) 2244-2274, http://dx.org./10.4236/jmp.2015.615230
              [arXiv:1409.4981].  
%
\bibitem{nh2008}  N.S. Manko\v c Bor\v stnik, H.B. Nielsen, ``Particular boundary condition 
                     ensures that a fermion in d=1+5, compactified on a finite disk, manifests 
                     in d=1+3 as massless spinor with a charge 1/2, mass protected and 
                     chirally coupled to the gauge field'',  hep-th/0612126, arxiv:0710.1956, 
                      {\it Phys.  Lett.} {\bf B} 663, Issue 3, 22 May 2008, Pages 265-269.  
%
\bibitem{NHDJMP}  160. D. Lukman, N.S. Manko\v c Bor\v stnik, H.B. Nielsen, "An effective two 
               dimensionality" cases bring a new hope to the Kaluza-Klein-like theories'', 
                              http://arxiv.org/abs/1001.4679v5, 
                            {\it New J. Phys.} {\bf 13} (2011) 103027, 1-25. 
 %
 \bibitem{NHD2010} D. Lukman, N.S. Manko\v c Bor\v stnik, H.B. Nielsen,   
              "Families of Spinors in $d=(1+5)$ Compactified on an Infinite Disc with 
the Zweibein Which Makes a Disc Curved on $S^2$ and a Possibility for Masslessness",  
 Proceedings to the 
                  $13^{th}$ international workshop ''What Comes Beyond 
		                the Standard Models'', 13 -23 of July, 2010, Ed. N.S. Manko\v c Bor\v stnik, 
                H.B. Nielsen, D. Lukman, DMFA  Zalo\v zni\v stvo, Ljubljana December 2010, 
                 p. 193-202,  arXiv:1012.0224. \\
%
\bibitem{ND2012Disk}  D. Lukman and N.S. Manko\v c Bor\v stnik, "Spinor states on a curved infinite disc 
                   with non-zero spin-connection fields", http://arxiv.org/abs/1205.1714, 
                   {\it J. Phys. A:  Math. Theor.} {\bf 45} (2012) 465401 (19pp). \\
%
\bibitem{Rita2022}	R. Bernabei, P. Belli, A. Bussolotti, V. Caracciolo, R. Cerulli, N. Ferrari, A.
Leoncini, V. Merlo, F. Montecchia, F. Cappella, A. d’Angelo, A. Incicchitti, A.
Mattei, C.J. Dai, X.H. Ma, X.D. Sheng, Z.P. Ye, ``New and recent results, and perspectives from
 DAMA/LIBRA–phase2, [ arXiv:2209.00882 ], Proceedings  to  the $25^{rd}$ Workshop 
 "What comes beyond the standard models", 6 - 12 of 
July, 2022,   Ed. N.S. Manko\v c Bor\v stnik, H.B. Nielsen, A. Kleppe, DMFA  
             Zalo\v zni\v stvo,  Ljubljana, December 2022, [arXiv:2303.17040 in physics.gen-ph     ].
             The direct link is: http://bsm.fmf.uni-lj.si/bled2022bsm/talks/bled22.pd.  \\
%
 \bibitem{prd2018} N.S. Manko\v c Bor\v stnik, H.B.F. Nielsen, "New way of second 
             quantized theory of  fermions with either Clifford or Grassmann coordinates
             and  {\it spin-charge-family} theory " 
             [arXiv:1802.05554v4,arXiv:1902.10628].              
                  
%
%
%
\bibitem{string} Hawking, Stephen W. (28 February 2006). The Theory of Everything: The Origin 
and Fate of the Universe. Phoenix Books; Special Anniversary. ISBN 978-1-59777-508-3.
%
\bibitem{Fritz} H. Fritzsch,  Weak-interaction mixing in the six-quark theory.
Fritzsch, H. \emph{Phys. Lett.} {\bf 1978}, \emph{73B}, 317. %
\bibitem{Frogatt} Frogatt, C.D.; Nielsen, H.B.  Hierarchy of quark masses, cabibbo angles 
and CP violation. \emph{Nucl. Phys.} {\bf 1979}, \emph{B147}, 277. 
%
\bibitem{Jarlskog} Jarlskog, C. Commutator of the Quark Mass Matrices in the Standard Electroweak Model and a Measure of Maximal CP Nonconservation.
\emph{Phys. Rev. Lett.} {\bf 1985}, \emph{55}, 1039. 
\bibitem{Branco} Gustavo C. ; Branco  C.; LavouraL.
Ansatz for the quark mass matrices allowing for a high top-quark mass.%
\emph{Phys. Rev.}  {\bf 1991}, \emph{ D 44, R582(R) }. 
\bibitem{Harari} Harari, H.; Nir, Y. B-anti B mixing and relations among quark masses, 
angles and phases.
\emph{Phys. Lett.} {\bf 1987}, \emph{B195}, 586. 
\bibitem{FRI10} Stech, B. Are the neutrino masses and mixings closely related to the
masses and mixings of quarks?
 \emph{Phys. Lett.} {\bf 1997}, \emph{B403}, 114. 
\bibitem{Altarelli} Altarelli, G.;  Feruglio, F.
Models of neutrino masses and mixings.
 \emph{New J. Phys.} \textbf{2004}, \emph{6}, 106. 
%
%
%
%
\bibitem{Raby} Mohapatra, R.N. A unified solution to the big problems of the
standard model.
 \emph{arXiv} \text{2022} arXiv:2207.10619.
%
\bibitem{Paul} Frampton, P. Predictions of additional baryons and mesons. \emph{arXiv} \textbf{2022}, 
 [arXiv:2209.05349].
 %
 [arxiv:2211.09579. ] 
\end{thebibliography}
\end{document}